\documentclass[aps,pra,onecolumn,showpacs,showkeys,groupedaddress,nobalancelastpage,nofootinbib,floatfix,superscriptaddress,longbibliography]{revtex4-2}
\usepackage{graphics}
\usepackage{graphicx}
\usepackage{subfigure}
\usepackage{dcolumn}
\usepackage{bm}
\usepackage{comment} 
\usepackage{xcolor}
\usepackage{bookmark}
\usepackage{tensor}
\usepackage{amsmath}
\usepackage{amssymb}
\usepackage{dsfont}
\usepackage{float}
\usepackage[utf8]{inputenc}
\usepackage{romannum}
\usepackage{moreverb}
\usepackage{subfigure}
\usepackage{hyperref}
\usepackage[capitalize]{cleveref}

\makeatletter

\def\p@subsection{}
\makeatother

\begin{document}
\pagenumbering{arabic}
\newcommand{\ket}[1]{\left| #1 \right\rangle}
\newcommand{\bra}[1]{\left\langle #1 \right|}
\newcommand{\braket}[1]{\left\langle #1 \right\rangle} 
\newcommand{\expect}[1]{\langle #1 \rangle}
\newcommand{\ketbra}[2]{\mathinner{|{#1}\rangle\langle{#2}|}} 
\newcommand{\braopket}[3]{\langle{#1}|{#2}|{#3}\rangle}
\newcommand{\beq}{\begin{equation}}
\newcommand{\eeq}{\end{equation}}
\newcommand{\bea}{\begin{align}}
\newcommand{\eea}{\end{align}}
\newcommand{\bq}{\begin{quote}}
\newcommand{\eq}{\end{quote}}
\newcommand{\oem}{\color{blue}}
\newcommand{\blk}{\color{black}}
 \newtheorem{thm}{Theorem}
 \newtheorem{cor}[thm]{Corollary}
 \newtheorem{lem}[thm]{Lemma}
 \newtheorem{prop}[thm]{Proposition}
 \newtheorem{defn}[thm]{Definition}
 \newtheorem{rem}[thm]{Remark}

\pagestyle{plain}
\title{\bf Quantum Estimation of the Stokes Vector Rotation for a General Polarimetric Transformation}
\author{ Ali Pedram}
\email{apedram19@ku.edu.tr}
\affiliation{Department of Physics, Koç University, Istanbul, Sarıyer 34450, Türkiye}
\author{ Vira R. Besaga}
\email{vira.besaga@uni-jena.de}
\affiliation{Institute of Applied Physics, Abbe Center of Photonics, Friedrich-Schiller-Universität Jena, 07745 Jena, Germany}
\author{ Lea Gassab}
\email{lgassab20@ku.edu.tr}
\affiliation{Department of Physics, Koç University, Istanbul, Sarıyer 34450, Türkiye}
\author{ Frank Setzpfandt}
\email{f.setzpfandt@uni-jena.de}
\affiliation{Institute of Applied Physics, Abbe Center of Photonics, Friedrich-Schiller-Universität Jena, 07745 Jena, Germany}
\affiliation{Fraunhofer Institute for Applied Optics and Precision Engineering IOF, 07745 Jena, Germany}
\author{Özgür E. Müstecaplıoğlu}
\email{omustecap@ku.edu.tr}
\affiliation{Department of Physics, Koç University, Istanbul, Sarıyer 34450, Türkiye}
\affiliation{TÜBİTAK Research Institute for Fundamental Sciences, 41470 Gebze, Türkiye}
\affiliation{Faculty of Engineering and Natural Sciences, Sabancı University, Tuzla 34956, Türkiye}
\pagebreak

\begin{abstract}
Classical polarimetry is a well-established discipline with diverse applications across different branches of science. The burgeoning interest in leveraging quantum resources to achieve highly sensitive measurements has spurred researchers to elucidate the behavior of polarized light within a quantum mechanical framework, thereby fostering the development of a quantum theory of polarimetry. In this work, drawing inspiration from polarimetric investigations in biological tissues, we investigate the precision limits of polarization rotation angle estimation about a known rotation axis, in a quantum polarimetric process, comprising three distinct quantum channels. The rotation angle to be estimated is induced by the retarder channel on the Stokes vector of the probe state. The diattenuator and depolarizer channels, acting on the probe state, can be thought of as effective noise processes. We explore the precision constraints inherent in quantum polarimetry by evaluating the quantum Fisher information for probe states of significance in quantum metrology, namely NOON, Kings of Quantumness, and Coherent states. The effects of the noise channels as well as their ordering is analyzed on the estimation error of the rotation angle to characterize practical and optimal quantum probe states for quantum polarimetry. Furthermore, we propose an experimental framework tailored for NOON state quantum polarimetry, aiming to bridge theoretical insights with empirical validation.
\end{abstract}
\maketitle


\section{Introduction}
\label{sec:intro}
Polarization is a property of a propagating electromagnetic wave which quantifies the direction of the oscillations of the electric field. This property of the electromagnetic waves is exploited in a diverse set of technological applications in materials science~\cite{nesse2004introduction,B201314M,doi:10.1146/annurev.ms.11.080181.000525,Losurdo2009,1178170}, astronomy~\cite{tinbergen_1996,Snik2013}, medical sciences~\cite{He2021,doi:https://doi.org/10.1002/9781119011804.ch7,10.1117/1.3652896,Ramella-Roman_2020,Gramatikov2014}, and quantum information~\cite{RevModPhys.79.135}. Interaction with a medium can alter the polarization state of light. A simple framework for calculating the transformation of polarized light by different optical elements is Jones matrix calculus~\cite{hecht2017optics}. Here, the polarized light is modeled using a $2\times1$ vector and the optical elements causing the polarized light to undergo linear transformations is modeled by $2\times2$ matrices. However, Jones calculus is only applicable for fully polarized states. A more general framework is the Mueller matrix calculus, which can be used to model partially polarized states and depolarizing transformations as well~\cite{hecht2017optics}. In Mueller calculus the polarization states are represented by the Stokes vectors which are $4\times1$ vectors and optical elements are modeled using $4\times4$ Mueller matrices.\\

Optical polarimetry is a field of study and a set of techniques concerning measurement and interpretation of the polarization information. Studying the polarization of light before and after it interacts (transmission, reflection, or scattering) with a medium, one can infer several optical and geometric properties of the sample. Based on the optical properties of the material under study different polarimetry techniques might be used, such as transmission polarimetry or ellipsometry. Mueller polarimeters comprehensively measure the change of polarization state of light providing 16 elements of the Mueller matrix of a sample~\cite{Azzam:16}. The non-invasive nature and high precision of polarimetry makes it suitable for studying sensitive samples. Ellipsometry is widely used for many applications including measurement of the refractive index and thickness of thin films~\cite{doi:10.1146/annurev.ms.11.080181.000525}. Polarimetry techniques can also be used to measure the polarization parameters of the biological samples. For instance, the thickness of the retinal nerve fiber layer (RNFL) can be used to diagnose glaucoma~\cite{Gramatikov2014,BUENO20003791,knighton2002,Benoit:01,Huang:03}. The basic principle is that, RNFL is birefringent, therefore it can be modeled by a linear retarder which rotates the Stokes vector of the incoming probe state. Estimation of this rotation angle yields an estimate of the thickness of the RNFL~\cite{knighton2002,Benoit:01,Huang:03}.\\

With growing interest in quantum metrology, there have been attempts to utilize quantum resources for precise measurement of physical parameters. Estimation of rotation angles is also of a great interest in quantum communication for studies in alignment of the reference frames of the communicating parties\cite{PhysRevLett.86.4160,PhysRevLett.87.167901,PhysRevA.96.022310,PhysRevLett.87.257903,PhysRevLett.93.180503,PhysRevA.78.052333,Safranek_2015,Koochakie2021}. Interestingly, there is a straightforward link between classical polarization optics and quantum mechanics of two level systems~\cite{PhysRevA.76.032323}. Other than rotation sensing, there are also attempts to utilize quantum resources for specific polarimetric tasks~\cite{Yoon_2020,Rudnicki:20,jarzyna2021quantum}. Furthermore, quantum correlations of the probe state have been utilized to enhance the measurement precision in polarimetry tasks~\cite{Hannonen:20,Magnitskiy:20,Magnitskiy:22,Restuccia:2022,Vega:2021,https://doi.org/10.1002/qute.202400059}. Goldberg et al.~\cite{PhysRevResearch.2.023038,Goldberg:21,Goldberg_thesis,GOLDBERG2022185} have studied changes in quantum polarization and have consolidated a quantum mechanical framework for polarimetry.\\

Several works have been carried out investigating the ultimate sensitivity limit and possible improvement for measurement of the optical properties of samples. It is shown that the NOON states and Kings of Quantumness (anticoherent states) yield Heisenberg limit (HL) scaling for rotation sensing~\cite{PhysRevA.95.052125,Bouchard:17,PhysRevA.78.052333,PhysRevA.98.032113,Martin2020optimaldetectionof,ZGoldberg_2021,PhysRevApplied.20.024052}. Kings of Quantumness are states which have vanishing expectation values for their Stokes operators and they are called Kings or anticoherent states of order $t$, if all higher moments of the Stokes operators up to $t$ are isotropic. More information on these states is provided in the appendix. For measurement of optical losses in a bosonic channel it is shown that Fock and two-mode squeezed vacuum (TMSV) states are optimal, however their error scaling doesn't surpass the standard quantum limit (SQL)~\cite{PhysRevA.79.040305,PhysRevLett.112.120405,PhysRevA.77.053807,PhysRevA.89.023845,Hayat:99,PhysRevLett.59.2555,PhysRevA.104.052615,JAKEMAN1986219,Losero2018,PhysRevLett.98.160401,PhysRevLett.121.230801,PhysRevApplied.16.044031}. Studies have shown that for sensing the depolarization parameter HL scaling is not possible. However, using correlated probes and ancilla-assisted schemes increases the estimation accuracy~\cite{PhysRevA.66.022308,PhysRevA.92.032324,Frey_2011,PhysRevA.72.052334}. In general, it was argued that no parameter which is encoded on the density matrix by a non-unitary operator or a convex combination of unitary operators can be estimated beyond shot noise limit~\cite{4655455}. At the same time, subsequent studies demonstrated that employing controlled and error-corrected methodologies in metrology and environmental monitoring could potentially reinstate the HL in specific cases~\cite{PhysRevLett.112.150802,Zhou2018,Albarelli2018restoringheisenberg,PhysRevA.101.032347}.\\

Based on the framework developed in~\cite{PhysRevResearch.2.023038} and inspired by the studies in vision and applications of metrology in biological systems~\cite{He2021,doi:https://doi.org/10.1002/9781119011804.ch7,10.1117/1.3652896,Ramella-Roman_2020,Gramatikov2014,BUENO20003791,knighton2002,Benoit:01,Huang:03,TAYLOR20161,PhysRevResearch.4.033060}, we aim to assess the feasibility of using quantum polarimetry for estimation of rotation angle due to a birefringent medium, modeled by a linear retarder, in presence of diattenuation and depolarization. Studies on biological tissues have shown that, although the polar decomposition of Mueller matrix~\cite{Lu:96} in tissue polarimetry can yield reliable polarization properties in classical regime, this decomposition does not correspond to the underlying physical reality~\cite{doi:https://doi.org/10.1002/9781119011804.ch7,https://doi.org/10.1002/jbio.200810040,GAO2021106735}. Furthermore, experimental evidence suggests that optical quantum entanglement can survive while photons are being transmitted through highly scattering biological tissues~\cite{Shi:2016,Huang_2020,Lum:21,https://doi.org/10.1002/jbio.201800096}. Therefore, it is crucial to investigate the precision limits in quantum polarimetry considering non-commutativity of the elementary components of the Mueller matrix, which implies considering different composition orders of the quantum polarization channels.\\

This manuscript is organized as follows. In~\cref{sec:polarimetry} we introduce concepts in quantum polarimetry such as Stokes operators and their transformations due to polarization channels. The basic concepts in quantum metrology, i.e. quantum Fisher information (QFI) and quantum Cramer-Rao bound (QCRB) are introduced in~\cref{sec:metrology}. In~\cref{sec:results} the results of our calculations for QFI using NOON, Kings of Quantumness~\cite{PhysRevA.95.052125,Bouchard:17,PhysRevA.78.052333,PhysRevA.98.032113,Martin2020optimaldetectionof,ZGoldberg_2021,PhysRevApplied.20.024052} and coherent states as probe states are given. In~\cref{sec:expimp} a possible experimental implementation is discussed and finally, in~\cref{sec:conclusion} we present our conclusions. In the appendix we present a brief explanation on Majorana representation of quantum polarization states and quantum concepts in optical polarization and introduce Kings of Quantumness and the concept of anticoherence.

\section{Quantum Description of Polarized Light and Polarization Transformations}
\label{sec:polarimetry}

\begin{figure}[htbp!]
    \centering
    \subfigure[]{\includegraphics[width=0.49\textwidth]{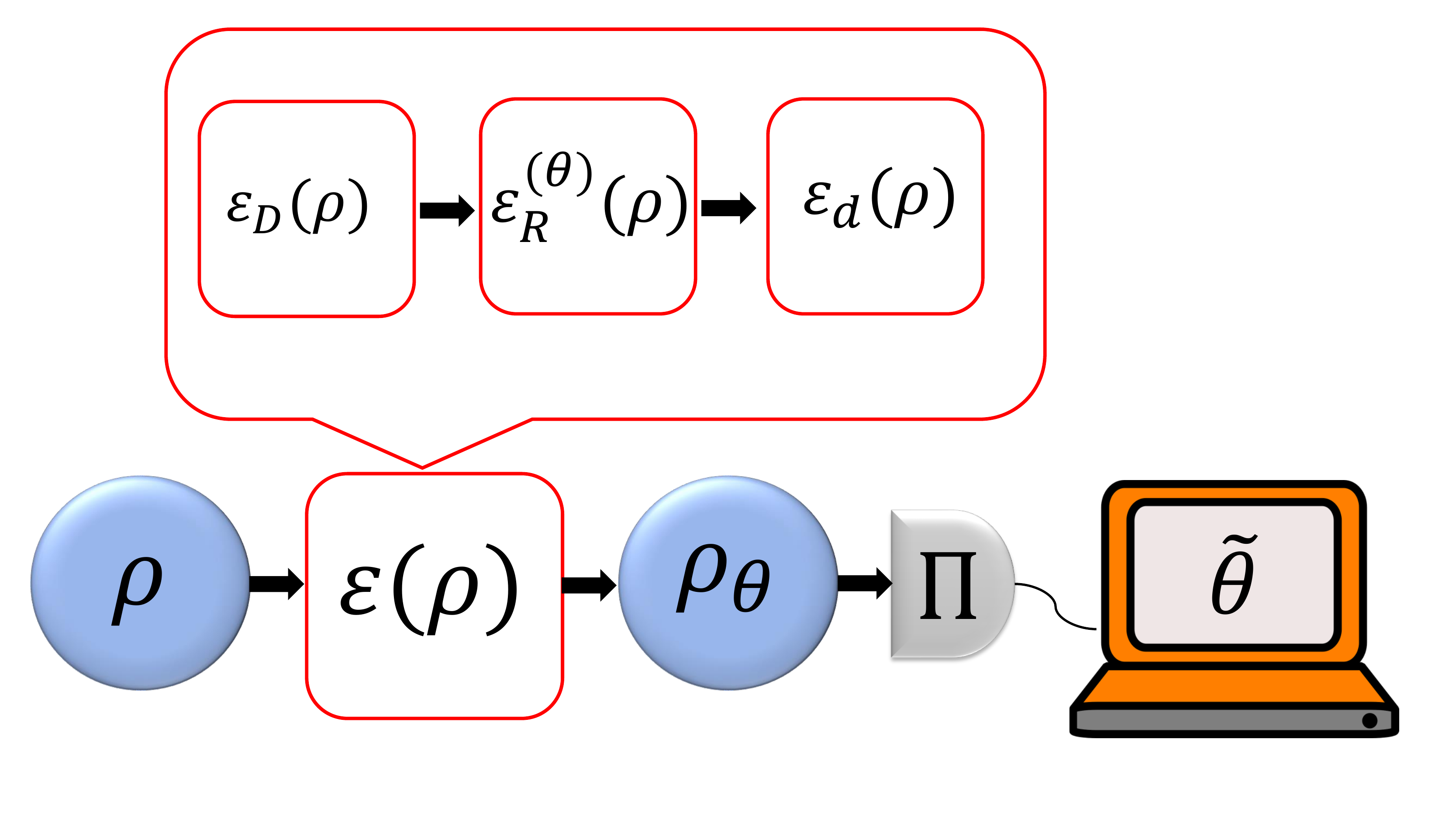}} 
    \subfigure[]{\includegraphics[width=0.49\textwidth]{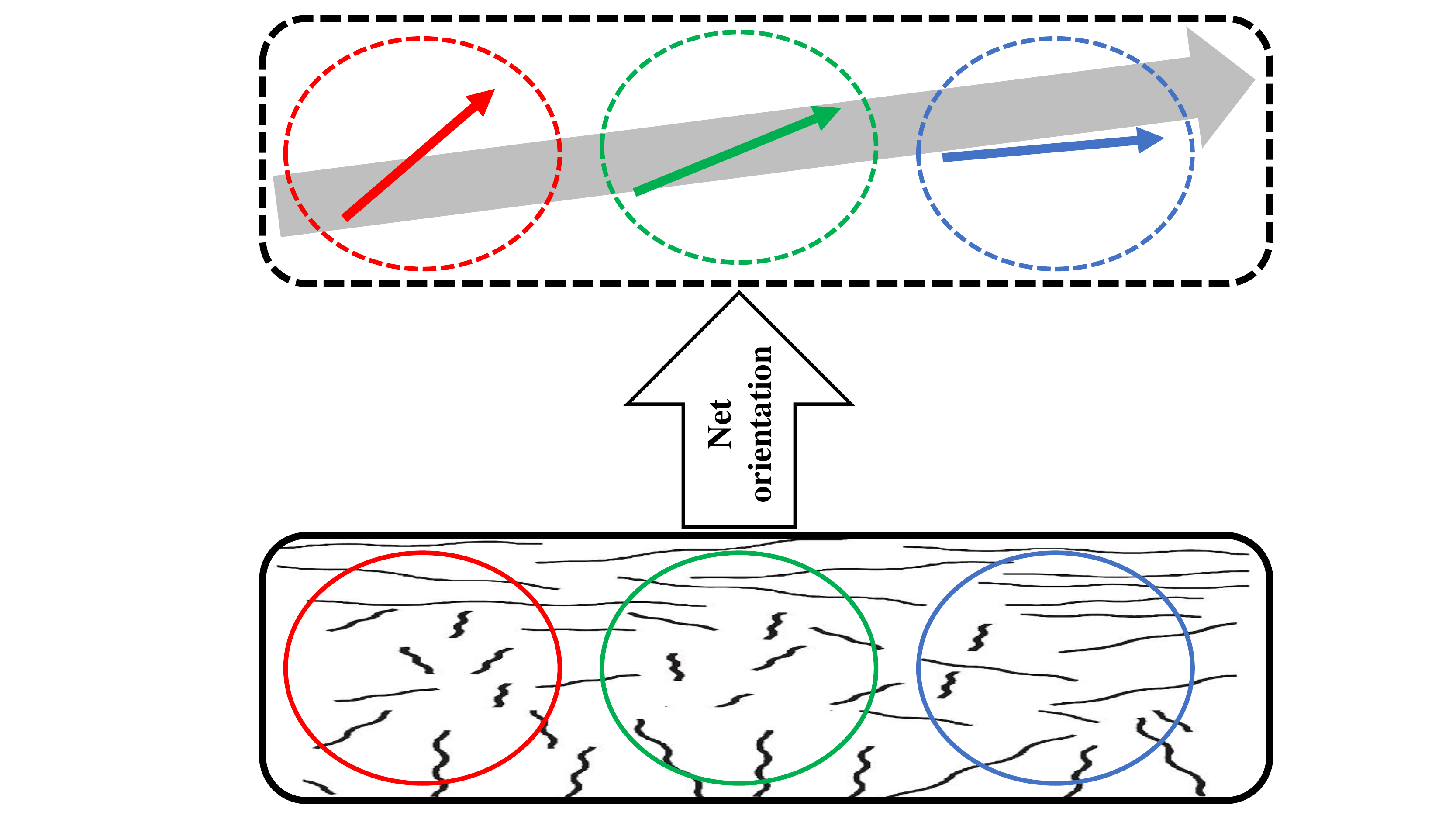}} 
    \caption{(a) A schematic representation of the general polarization channel. In this figure the order of the action of the diattunator, retarder and depolarizer channels follows the standard Lu-Chipman form. (b) Distribution of fibers in an inhomogeneous biological sample and overall rotation axis experienced by the probe. The large gray arrow represents the net rotation axis for the entire sample and the smaller colored arrows in dashed circles represents the local net rotation axis for the respective fiber distribution at each specific subsection of the sample.}
    \label{schematicpolchan}
\end{figure}

In a quantum optical setting, each polarization mode can be thought of as a harmonic oscillator. One can write a general pure state of light by acting the creation operators of the horizontal and vertical polarization modes on the vacuum state.
\begin{equation}\label{polarizationstate}
\begin{aligned}
  \ket{\Psi}=\sum_{n_1,n_2}c_{n_1,n_2}\ket{n_1,n_2},\\
  \ket{n_1,n_2}\equiv \frac{\hat{a}^{\dagger}\vphantom{a}{}^{n_1}\hat{b}^{\dagger}\vphantom{a}{}^{n_2}}{\sqrt{n_1!n_2!}}\ket{0,0}.
\end{aligned}
\end{equation}
We take $\hat{a}$ and $\hat{b}$ to be the annihilation operators of the horizontal and vertical polarization modes respectively. These operators satisfy the bosonic commutation relations. 
\begin{equation}\label{bosoncom}
  [\hat{a}_i,\hat{a}_j^{\dagger}]=\delta_{ij},\quad [\hat{a}_i,\hat{a}_j]=[\hat{a}_i^{\dagger},\hat{a}_j^{\dagger}]=0, \quad \hat{a}_i\in\{\hat{a},\hat{b}\}.
\end{equation}
Using the field operators of the horizontal and vertical polarization modes we can define the Stokes operators as
\begin{equation}\label{stokesops}
\begin{aligned}
  \hat{S}_{0}=\left(\hat{a}^{\dagger}\hat{a}+\hat{b}^{\dagger}\hat{b}\right)/2,\quad\hat{S}_{1}=\left(\hat{a}^{\dagger}\hat{b}+\hat{b}^{\dagger}\hat{a}\right)/2, \\
  \hat{S}_{2}=-\text{i}\left(\hat{a}^{\dagger}\hat{b}-\hat{b}^{\dagger}\hat{a}\right)/2,\quad\hat{S}_{3}=\left(\hat{a}^{\dagger}\hat{a}-\hat{b}^{\dagger}\hat{b}\right)/2.
\end{aligned}
\end{equation}
The Stokes operators are quantum generalizations of the Stokes parameters, which are promoted to an operator status. These operators obey the following $\mathfrak{su}(2)$ algebraic relations,
\begin{equation}\label{stokescom}
\begin{aligned}
  \left[\hat{S}_{i},\hat{S}_{j}\right]=\text{i}\sum_{k=1}^{3}\epsilon_{ijk}\hat{S}_{k},\\ \hat{S}_{1}^{2}+\hat{S}_{2}^{2}+\hat{S}_{3}^{2}=\hat{S}_{0}\left(\hat{S}_{0}+1\right).
\end{aligned}
\end{equation}
Similar to the classical case, one can use the Stokes operators to define a semiclassical degree of polarization (DOP)~\cite{Goldberg:21}:
\begin{equation}\label{dop}
  \mathbb{P}_{sc}=\frac{\lvert\langle\hat{\mathbf{S}}\rangle\rvert}{\langle\hat{S}_{0}\rangle}=\frac{\sqrt{\langle\hat{S}_{1}\rangle^{2}+\langle\hat{S}_{2}\rangle^{2}+\langle\hat{S}_{3}\rangle^{2}}}{\langle\hat{S}_{0}\rangle}\,.
\end{equation}
Here, $\langle \hat{S}_i \rangle = \text{Tr}[\hat{\rho}\hat{S}_i]$ denotes the expectation value of the operator $\hat{S}_i$ and $\hat{\mathbf{S}}$ is the normalized Stokes vector. However, in general for quantum states, this definition for the DOP fails to capture their polarization behavior and one needs to take into account so called ``hidden polarization" or higher order polarization to which higher moments of the Stokes operators contribute~\cite{KLYSHKO1992349,doi:10.1080/09500349908231279,PhysRevA.66.013806,PhysRevA.72.033813,LUIS2007173,PhysRevLett.105.153602,BJORK20104440,PhysRevA.75.053806,PhysRevA.84.045804,10.1103/PhysRevA.87.043821,PhysRevA.87.043814,Sanchez-Soto_2013}. In quantum mechanics, a general transformation of an operator is described by a completely positive trace preserving (CPTP) dynamical map which is also dubbed as a quantum channel. Kraus' theorem states that any quantum channel can be described by a set of Kraus operators $\{K_l\}$ such that,
\begin{equation}\label{stokeskraus}
  \hat{S}_{\mu}\to\sum_{l}\hat{K}_{l}^{\dagger}\hat{S}_{\mu}\hat{K}_{l}.
\end{equation}
For the expectation values of the Stokes operators to transform according to the classical Mueller description, the following condition must be satisfied:
\begin{equation}\label{stokesmueller}
  \langle S_{\mu}\rangle\to\sum_{\nu=0}^{3}M_{\mu\nu}\langle S_{\nu}\rangle.
\end{equation}
For both \cref{stokeskraus} and \cref{stokesmueller} to hold, we can impose the condition that the Stokes operators should transform in the same way as the Stokes parameters~\cite{PhysRevResearch.2.023038}, thus
\begin{equation}\label{krausmueller}
  \sum_{l}\hat{K}_{l}^{\dagger}\hat{S}_{\mu}\hat{K}_{l}=\sum_{\nu}M_{\mu\nu}\hat{S}_{\nu}.
\end{equation}
It has also been shown that if the Mueller matrix is not singular, one can decompose it into a product of three elementary matrix factors with well-defined polarimetric properties, a retarder, a diattenuator, and a depolarizer~\cite{Lu:96}, using the standard Lu-Chipman decomposition as
\begin{equation}\label{muellerdecomp}
  M = M_d M_R M_D.
\end{equation}
Here, $M_D$, $M_R$ and $M_d$ are the elementary Mueller matrices for the diattenuation, retardation and depolarization components respectively. Firstly, it must be noted that this decomposition is made solely for interpreting the polarimetric data and extracting physical parameters more conveniently by capturing the effective polarization transformation and doesn't necessarily describe the actual physical underlying process. This is specially the case for biological tissues
for which the depolarization, diattenuation and retardation effects are likely to occur simultaneously~\cite{doi:https://doi.org/10.1002/9781119011804.ch7,https://doi.org/10.1002/jbio.200810040,GAO2021106735}. Secondly, due to the fact that in general the matrix product is non-commutative, the decomposition order given in \cref{muellerdecomp} is not the only decomposition that one can make out of the original Mueller matrix. Based on the optical characteristics of the sample, there exist a multitude of ways to break down the total Mueller matrix into a combination of the elementary ones, of which \cref{muellerdecomp} is only a special case~\cite{doi:https://doi.org/10.1002/9781119011804.ch7,Lu:96,https://doi.org/10.1002/jbio.200810040,GAO2021106735,Morio:04,Gil:13,Ossikovski:07,Ossikovski:12,Ortega-Quijano:11,https://doi.org/10.1002/pssa.200777793}. Based on \cref{krausmueller} we can describe the quantum mechanical polarization channels as~\cite{PhysRevResearch.2.023038}
\begin{equation}\label{qchdecomp}
  \varepsilon(\hat{\rho}) = \varepsilon_d \circ \varepsilon_R \circ \varepsilon_D(\hat{\rho}),
\end{equation}
in which $\varepsilon_D$, $\varepsilon_R$ and $\varepsilon_d$ are the diattenuator, retarder and depolarizer channels, respectively. A schematic representation of this channel is given in \cref{schematicpolchan}(a).
The retarder channel rotates the Stokes vector which also acts as a unitary rotation on the density matrix~\cite{PhysRevA.98.032113,PhysRevResearch.2.023038,Goldberg_thesis,GOLDBERG2022185}:
\begin{equation}\label{ch_retarder}
  \varepsilon_R(\hat{\rho}) = \hat{R}(\theta,\mathbf{n}) \hat{\rho} \hat{R}^{\dagger}(\theta,\mathbf{n}).
\end{equation}
Here, the operator $\hat{R}$ is given by
\begin{equation}\label{rotation}
  \hat{R}(\theta,\mathbf{n}) = \text{exp}(i\theta \hat{\mathbf{S}}.\mathbf{n}),
\end{equation}
in which $\mathbf{n}=(\Theta,\Phi)$ is the rotation axis with and $\theta$ is the angle of rotation. In composite, anisotropic media such as biological tissues, retardation phenomena may arise from various factors such as reflection, refraction at interfaces, or propagation through the medium. Notably, these materials harbor distributed fibers, such as collagen fibers, exhibiting non-uniform distribution throughout the sample. While initial models often assumed uniformity in the refractive index difference between ordinary and extraordinary axes, and the uniform direction of the extraordinary axis across the sample, recent investigations have begun to challenge this simplification~\cite{doi:https://doi.org/10.1002/9781119011804.ch7}. The orientation of the retarder axis aligns with the prevailing direction of these fibers within a specific volume of the sample~\cite{doi:https://doi.org/10.1002/9781119011804.ch7}. However, during local probing of the sample, fiber orientations may locally fluctuate across different subsections, as illustrated in~\cref{schematicpolchan}(b). Moreover, reconstructing three-dimensional images from various sections to ascertain fiber direction introduces inherent noise susceptibility, leading to inevitable uncertainties in inferred orientations~\cite{https://doi.org/10.1002/mrm.10331,Bancelin:14,Menzel:17,10.1117/12.2548935,https://doi.org/10.1002/advs.202303381}. Consequently, among the retarder's three parameters, two primarily dictate the resultant fiber orientation—thus determining the rotation axis, albeit subject to uncertainty arising from the aforementioned factors—while the remaining parameter quantifies the degree of rotation, facilitating estimation of sample thickness. Our focus lies solely on accurately estimating the rotation angle.

The diattenuator channel can be modeled by a sequential application of rotation operators in an enlarged Hilbert space containing two ancillary modes and tracing out the ancillary modes. It is given by~\cite{PhysRevResearch.2.023038,Goldberg_thesis,GOLDBERG2022185}
\begin{equation}\label{ch_diattenuator}
  \varepsilon_D(\hat{\rho}) = \text{Tr}_{v_1,v_2}[\hat{U}_D (\hat{\rho} \otimes |\text{vac}\rangle_{v_1,v_2}{}_{v_1,v_2}\langle\text{vac}|) \hat{U}_D^{\dagger}],
\end{equation}
in which $v_1$ and $v_2$ are the ancillary modes and the operator $\hat{U}_D$ is
\begin{equation}\label{op_diattenuator}
\begin{split}
  \hat{U}_D = \hat{R}_{a,b}^{\dagger}(0,\beta,\gamma)\hat{R}_{b,v_2}(0,-2\text{cos}^{-1}(0,\sqrt{r},0))\\
  \hat{R}_{a,v_1}(0,-2\text{cos}^{-1}(0,\sqrt{q},0))\hat{R}_{a,b}(0,\beta,\gamma).
\end{split}
\end{equation}
Here, $\hat{R}_{a,b}$ denotes a rotation operator between modes $a$ and $b$ parametrized by the Euler angles $(0,\beta,\gamma)$. $\hat{R}_{a,v_1}$ and $\hat{R}_{b,v_2}$ are rotations between modes $a$ and $v_1$ and $b$ and $v_2$ correspondingly with their respective attenuation parameters $q$ and $r$. For the case of depolarization, one can give a classical-like decomposition of a state into a convex combination of the completely polarized state and the isotropic state~\cite{PhysRevResearch.2.023038}.
\begin{equation}\label{ch_depolarizer}
  \varepsilon_d^{(c)}(\hat{\rho}) = p\hat{\rho}_{pol} + (1-p)\sum_{N}p_N\frac{\hat{\mathcal{I}}_N}{N+1},
\end{equation}
in which $p$ is the depolarization parameter and $p_N$ is the weight factor in each photon number subspace. However, this decomposition doesn't always work for the case of quantum states and also, such a convex combination can only describe single photon states and classical states undergoing isotropic depolarization~\cite{GOLDBERG2022185,Goldberg_thesis}. Within the context of quantum optics, polarization is $\mathfrak{su}(2)$-invariant and an unpolarized state is defined as a state which remains invariant under any linear polarization transformation. A valid quantum optical depolarization model must preserve the invariant subspaces of the optical field and the steady-state in each of them must be a diagonal state~\cite{Klimov:06,PhysRevA.77.033853,PhysRevA.88.052120}. We can construct a quantum depolarization channel by a convex combination of different rotation operators~\cite{Goldberg_thesis,GOLDBERG2022185}.
\begin{equation}\label{simp_depolarizer}
  \varepsilon_d^{(r)}(\hat{\rho}) = \sum_{i=1}^{n} P(i) \hat{R}_i\hat{\rho}\hat{R}_i^{\dagger}.
\end{equation}
In~\cref{simp_depolarizer}, $P(i)$ is the probability for the rotation operator $\hat{R}_i$ and the sum of all the probabilities must add up to one. Depolarization can be caused by the interaction of light with an ensemble of atoms due to random rotations induced by the interaction on the photonic state~\cite{Klimov:06,PhysRevA.77.033853,PhysRevA.88.052120}. The channel given in~\cref{simp_depolarizer} is a simple model for such a physical process~\cite{Goldberg_thesis}. However, there are also models for depolarization of quantum states based on master equations~\cite{Klimov:06,PhysRevA.77.033853,PhysRevA.88.052120}, two of which are presented in~\cref{dep_me_iso} and~\cref{dep_me_aniso}.
\begin{align}
  \mathcal{L}_d^{(i)}(\hat{\rho})=\frac{d\hat{\rho}}{dt} = \nu_i (\mathcal{D}(\hat{S}_1)\hat{\rho}+\mathcal{D}(\hat{S}_2)\hat{\rho}+\mathcal{D}(\hat{S}_3)\hat{\rho}),\label{dep_me_iso}\\
  \mathcal{L}_d^{(a)}(\hat{\rho})=\frac{d\hat{\rho}}{dt} = \nu_{a_0} \mathcal{D}(\hat{S}_0)\hat{\rho}+\nu_{a}(\mathcal{D}(\hat{S}_2)\hat{\rho}+\mathcal{D}(\hat{S}_3)\hat{\rho}).\label{dep_me_aniso}
\end{align}
In~\cref{dep_me_iso} and~\cref{dep_me_aniso}, the expression $\mathcal{D}(\hat{A})\hat{\rho}\equiv 2\hat{A}\hat{\rho}\hat{A}^{\dagger}-\hat{A}^{\dagger}\hat{A}\hat{\rho}-\hat{\rho}\hat{A}^{\dagger}\hat{A}$ denotes the dissipator superoperator with the jump operator $\hat{A}$, while $\mathcal{L}_d^{(i)}$ and $\mathcal{L}_d^{(a)}$ represent the Lindblad generators governing their respective Markovian depolarization dynamics. Consequently, we establish a connection between these generators and their associated channels as $\varepsilon_{d_t}^{(x)}(\hat{\rho})=\text{exp}(\mathcal{L}_d^{(x)}t)\hat{\rho}(0)$, where $x=i,a$. It is important to differentiate between the two depolarization master equations: in~\cref{dep_me_aniso}, unlike~\cref{dep_me_iso}, the depolarization rates along the three axes $\hat{S}_1$, $\hat{S}_2$ and $\hat{S}_3$ are not equal. Moreover, it's worth noting that the dissipator $\mathcal{L}(\hat{S}_0)$ in~\cref{dep_me_aniso} holds no relevance for depolarization dynamics. Accordingly, the master equations described in~\cref{dep_me_iso} and~\cref{dep_me_aniso} are commonly referred to as the isotropic and anisotropic depolarization channels, respectively. The master equation in~\cref{dep_me_iso} is also recognized as the $\mathfrak{su}(2)$-invariant depolarization in literature~\cite{PhysRevA.88.052120}. 

While the isotropic depolarization, as delineated by the Markovian master equation in~\cref{dep_me_iso}, exhibits a more intricate dynamic behavior compared to the semi-classical phenomenological channel outlined in~\cref{simp_depolarizer}~\cite{PhysRevResearch.4.013178}, it inherently assumes that the probability distribution governing the rotation vector (encompassing both rotation direction and angle) is solely contingent upon the magnitude of said vector. Consequently, transformations with non-negligible probability primarily involve small magnitudes of the rotation vector, hence implying small rotation angles. Consequently, the isotropic depolarization channel can be conceptualized as a composite of numerous small rotation operators. On the other hand, the dynamics governing anisotropic depolarization are derived within the weak coupling limit, thereby implying small rotation angles, and are subject to the rotating wave approximation and Markovianity assumption~\cite{Klimov:06,PhysRevA.77.033853}. A thorough examination of isotropic and anisotropic depolarization dynamics concerning spin states is undertaken in~\cite{PhysRevResearch.4.013178}.
\section{Quantum Parameter Estimation and QCRB}
\label{sec:metrology}
In this section we introduce some of the fundamental results for quantum single parameter estimation. Assume that through a process, the parameter $\theta$ gets encoded on a quantum state. For a parameterized state the lower bound for the variance of any unbiased estimator $\tilde{\theta}$ of the parameter, is given by the QCRB~\cite{HELSTROM1967101,1054108,Helstrom1969,PhysRevLett.72.3439,doi:10.1142/S0219749909004839,doi:10.1116/1.5119961}
\begin{equation}\label{qcrb}
  (\Delta\tilde{\theta})^2 \geq \frac{1}{\nu\mathcal{F}_Q(\theta)}.
\end{equation}
Here, $\mathcal{F}_Q$ is the QFI and $\nu$ is the number of measurement repetitions. QFI is a generalization of the classical Fisher information (CFI) which is defined as the expectation value of the squared derivative of the logarithm of the probability distribution. For a pure state QFI is defined as,
\begin{equation}\label{pureqfi}
  \mathcal{F}_Q(\theta) = 4[\langle \partial_{\theta}\psi|\partial_{\theta}\psi \rangle - |\langle \partial_{\theta}\psi|\psi \rangle|^2].
\end{equation}
For a mixed state, one can define QFI using the symmetric logarithmic derivative (SLD) operator as
\begin{equation}\label{mixedqf}
  \mathcal{F}_Q(\theta) = \text{Tr}[\hat{\rho} \hat{L}_{\theta}^2].
\end{equation}
SLD is implicitly defined by the following equation
\begin{equation}\label{sld}
  \partial_{\theta}\hat{\rho} = \frac{\hat{L}_{\theta}\hat{\rho} + \hat{\rho} \hat{L}_{\theta}}{2}.
\end{equation}
By expressing $\hat{L}_{\theta}$ in the eigenbasis of $\hat{\rho}$, one can write QFI as
\begin{equation}\label{mixedqfieigen}
  \mathcal{F}_Q(\theta) = 2\sum_{k,l}\frac{|\langle k|\partial_{\theta}\hat{\rho}|l\rangle|^2}{\lambda_k + \lambda_l}.
\end{equation}
Here, $k$, $l$ and $\lambda_k$, $\lambda_l$ are the eigenvalues and eigenvectors of $\hat{\rho}$.
\section{Results}
\label{sec:results}
In this section we numerically calculate the QFI for the rotation angle $\theta$ in order to determine its ultimate measurement precision. In all of our calculations the rotation angle to be estimated is $\theta = \pi/10$, which falls into the range of small angles for which the Kings of Quantumness yield optimal sensitivity~\cite{Martin2020optimaldetectionof} for every value of $S$ considered in this study. The probe states that have been considered for this task are the coherent state, NOON states and Kings of Quantumness. 
\begin{figure}[htbp!]
    \centering
    \subfigure[]{\includegraphics[width=0.49\textwidth]{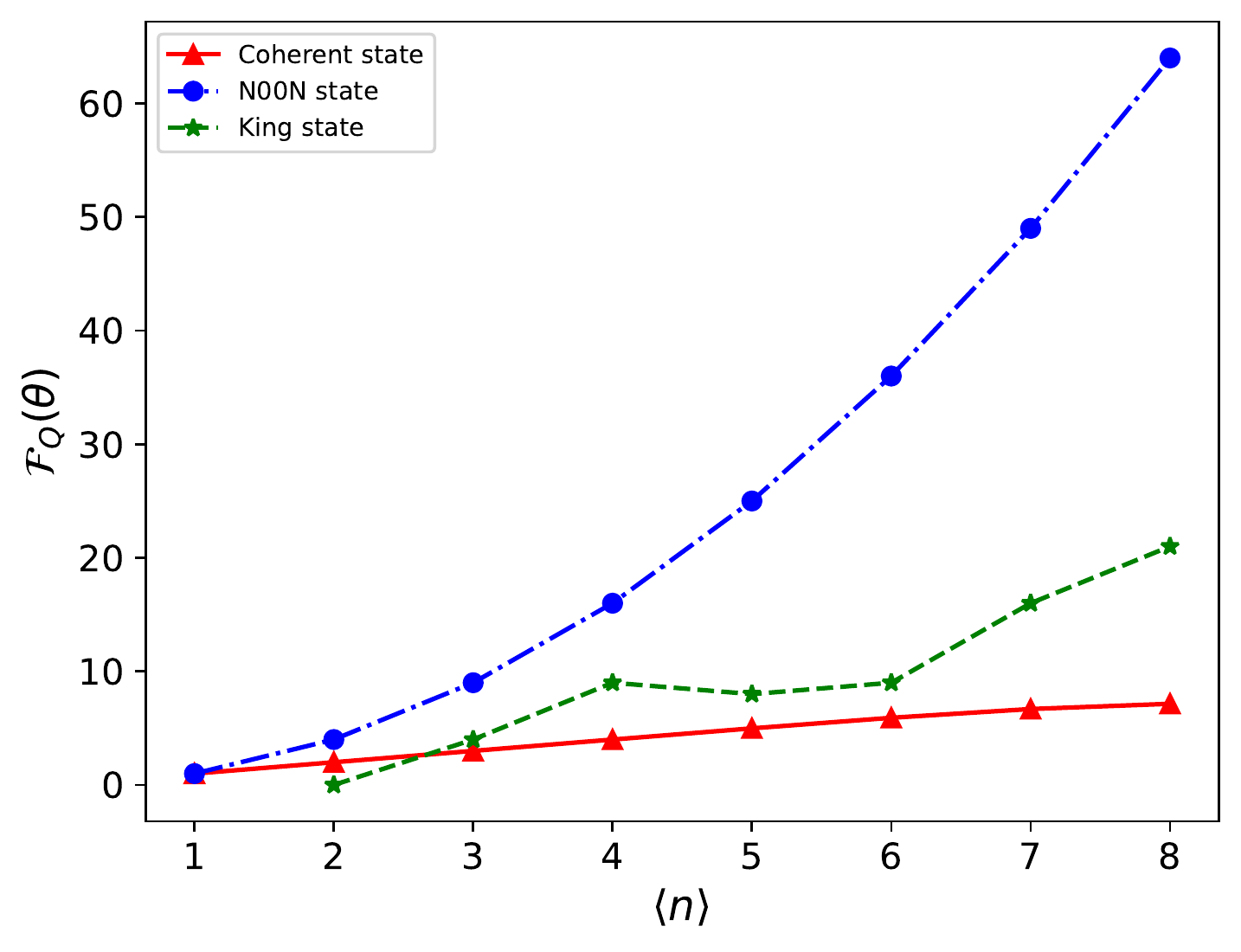}} 
    \subfigure[]{\includegraphics[width=0.49\textwidth]{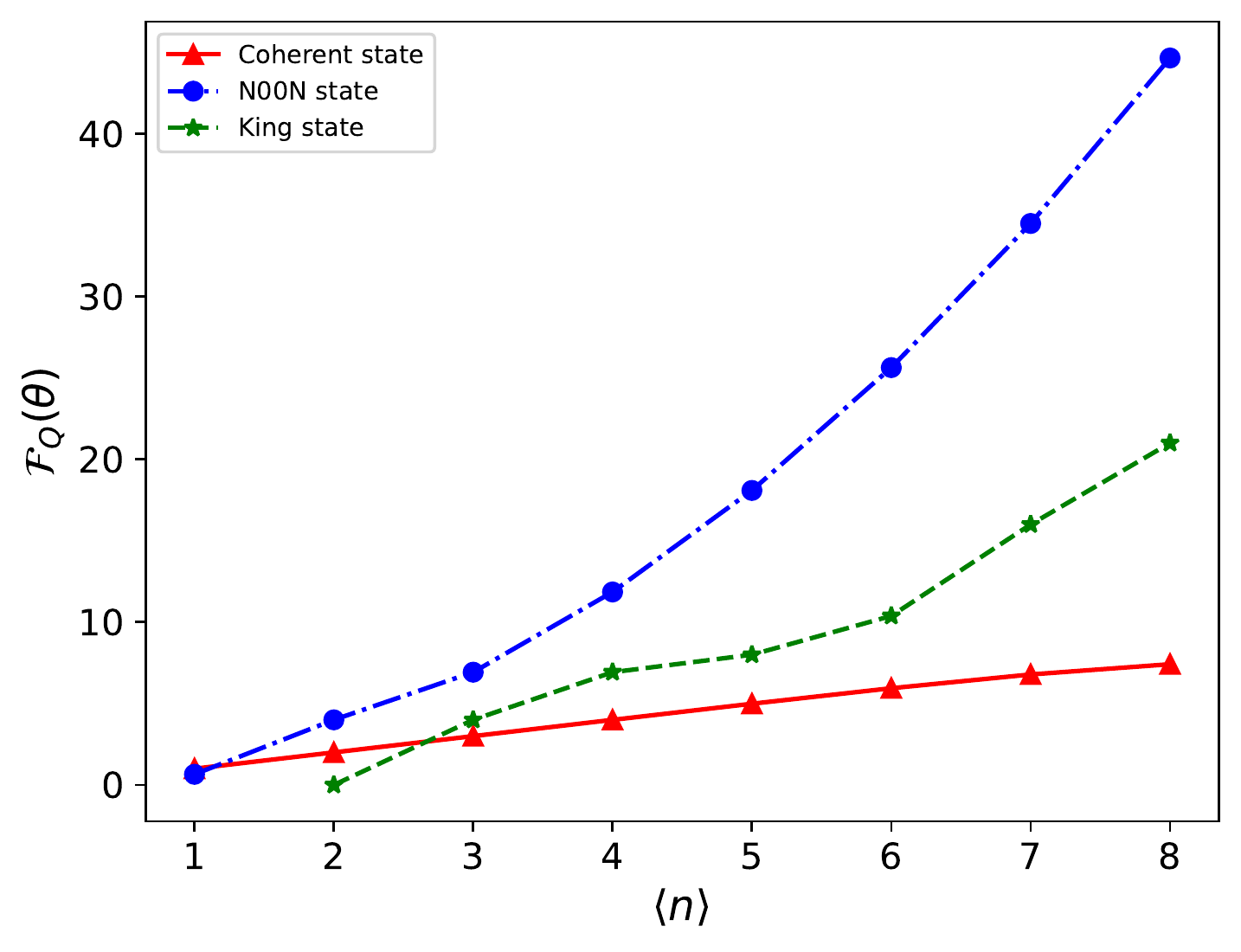}} 
    \caption{QFI for rotation angle (no depolarization and diattenuation) vs average photon number. Calculations are for coherent state (solid line with triangular data points in red), NOON state (dash dotted line with circular data points in blue) and Kings of Quantumness (dashed green line with star datapoints) for (a) $(\Theta,\Phi,\theta)=(0,0,\pi/10)$ (b) $(\Theta,\Phi,\theta)=(\pi/5,0,\pi/10)$}.
    \label{qfi_noiseless}
\end{figure} 
For estimation of small rotation angles, coherent state saturates SQL, NOON state saturates HL as long as the rotation axis is not orthogonal to $(\Theta,\Phi)=(0,0)$ axis (maximum sensitivity occurs for $(\Theta,\Phi)=(0,0)$ axis), and anticoherent states saturate HL for all rotation axes~\cite{PhysRevA.98.032113}. Because of the dependence of the precision of NOON state on the rotation axis, the optimal strategy is to align the sample such that $(\Theta,\Phi)=(0,0)$ axis corresponds to the net orientation direction of the fibers. However, as explained earlier, in realistic cases there will be an uncertainty in the direction of this axis. The Kings of Quantumness utilized in our calculations are taken from~\cite{Martin2020optimaldetectionof}. For all optical modes, the cut-off dimension for Hilbert space is taken to be 12. In our calculations we have used the open source QuTiP~\cite{JOHANSSON20121760,JOHANSSON20131234} and QuanEstimation~\cite{PhysRevResearch.4.043057} packages for Python. As a reference, our numerical results for the QFI in case of no depolarization and diattenuation is presented in~\cref{qfi_noiseless}.

In~\cref{qfi_noiseless}(a) we consider a perfect alignment of the optimal axis for the NOON state ($(\Theta,\Phi)=(0,0)$) and the retarder axis and in~\cref{qfi_noiseless}(b) we consider an uncertainty in the retarder axis which creates a misalignment between the two directions. Comparing~\cref{qfi_noiseless}(a) and~\cref{qfi_noiseless}(b) we see that, as expected, the QFI for NOON state reduces when the rotation axis is not aligned along its optimal axis, although it scales quadratically with the average photon number in both cases. QFI for coherent state scales linearly with $\braket{n}$ regardless of the rotation direction. The Kings of Quantumness (referred to as the King state in the figures) result in a larger value for QFI compared to the coherent state except for $S=1$ ($\braket{n}=2$) which corresponds to the bi-photon state (which is 1-anticoherent). For both of the rotation axes, NOON state achieves a higher sensitivity than the Kings of Quantumness. In~\cref{qfi_noiseless}(a) we observe non-monotonicity in QFI for Kings of Quantumness at $S=5/2$ ($\braket{n}=5$). This is not surprising if we consider the fact that optimal rotation sensors must be 2-anticoherent but for $S=5/2$ the state is only 1-anticoherent~\cite{Martin2020optimaldetectionof}. However, in~\cref{qfi_noiseless}(b) the QFI for Kings of quantumness is monotonic with $\braket{n}$ and the precision limits for $\theta$ are slightly different as compared to~\cref{qfi_noiseless}(a). It is instructive to assess the performance of these states by considering the effect of each of the noise channels on the estimation precision of the rotation angle. Therefore, in \cref{subsec:retdep} we will study the effect of the depolarization channel on the QFI for the rotation angle and in \cref{subsec:retdepdia} the joint effect of depolarization and diattenuation is considered. In both of these cases the ordering of the implementation of the channels is also considered. Based on the polarimetric studies in biological tissues~\cite{https://doi.org/10.1002/mrm.10331,Bancelin:14,Menzel:17,10.1117/12.2548935,https://doi.org/10.1002/advs.202303381}, in all subsequent calculations we will take the retarder axis to be $(\Theta,\Phi)=(\pi/5,0)$ with a misalignment of $\pi/5$ with the optimal axis for the NOON state. Using the terminology common in the classical polarimetry, we will call $\varepsilon_{\text{for}}(\hat{\rho}) = \varepsilon_d \circ \varepsilon_R \circ \varepsilon_D(\hat{\rho})$ as the ``forward" decomposition and $\varepsilon_{\text{rev}}(\hat{\rho}) = \varepsilon_D \circ \varepsilon_R \circ \varepsilon_d(\hat{\rho})$ as the ``reverse" decomposition.
\subsection{Noisy Rotation Sensing with Depolarization}
\label{subsec:retdep}
In this section we calculate the QFI for the rotation angle induced by the retarder, in both forward and reverse processes without considering the effect of diattenuation, focusing only on depolarization. We have considered 3 depolarization channels, namely the isotropic, anisotropic and using convex combination of rotations. Since in the master equation for the anisotropic depolarization~\cref{dep_me_aniso}, the dissipator $\mathcal{D}(\hat{S}_0)$ is irrelevant for depolarization~\cite{PhysRevA.88.052120}, we set $\nu_{a_0}=0$ for the master equation. For implementing the depolarization channel based on convex combination of rotations we have assumed a range of equally spaced $n_r$ rotation angles between $\eta_{min}$ and $\eta_{max}$ with rotation probabilities in~\cref{simp_depolarizer} given by a normal distribution with mean at zero rotation angle. One can choose which rotation angles to include and their probabilities by changing the range of angles considered and the standard deviation $\sigma_r$ for the normal distribution respectively.
\begin{figure}[htbp!]
    \centering
    \subfigure[]{\includegraphics[width=0.49\textwidth]{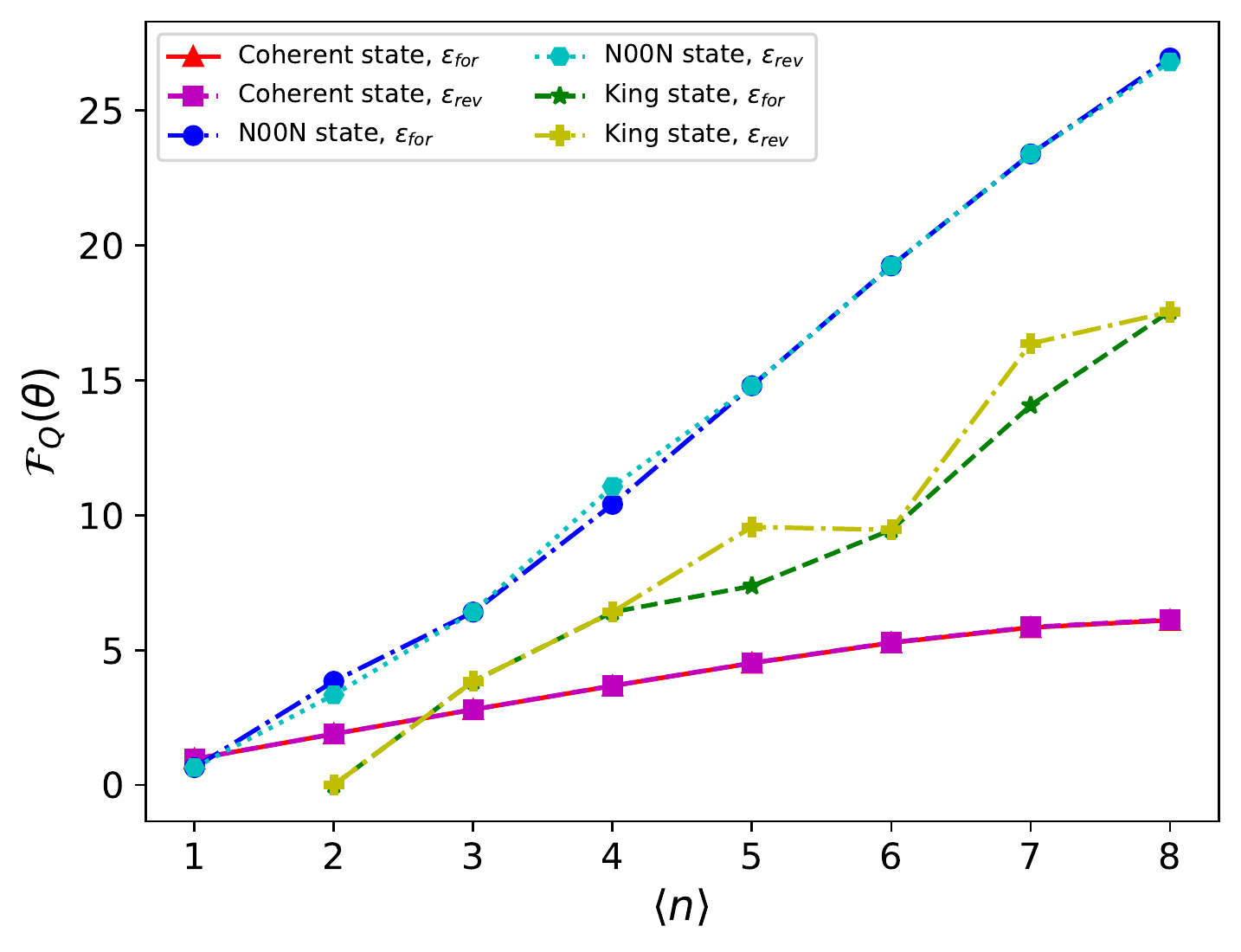}} 
    \subfigure[]{\includegraphics[width=0.49\textwidth]{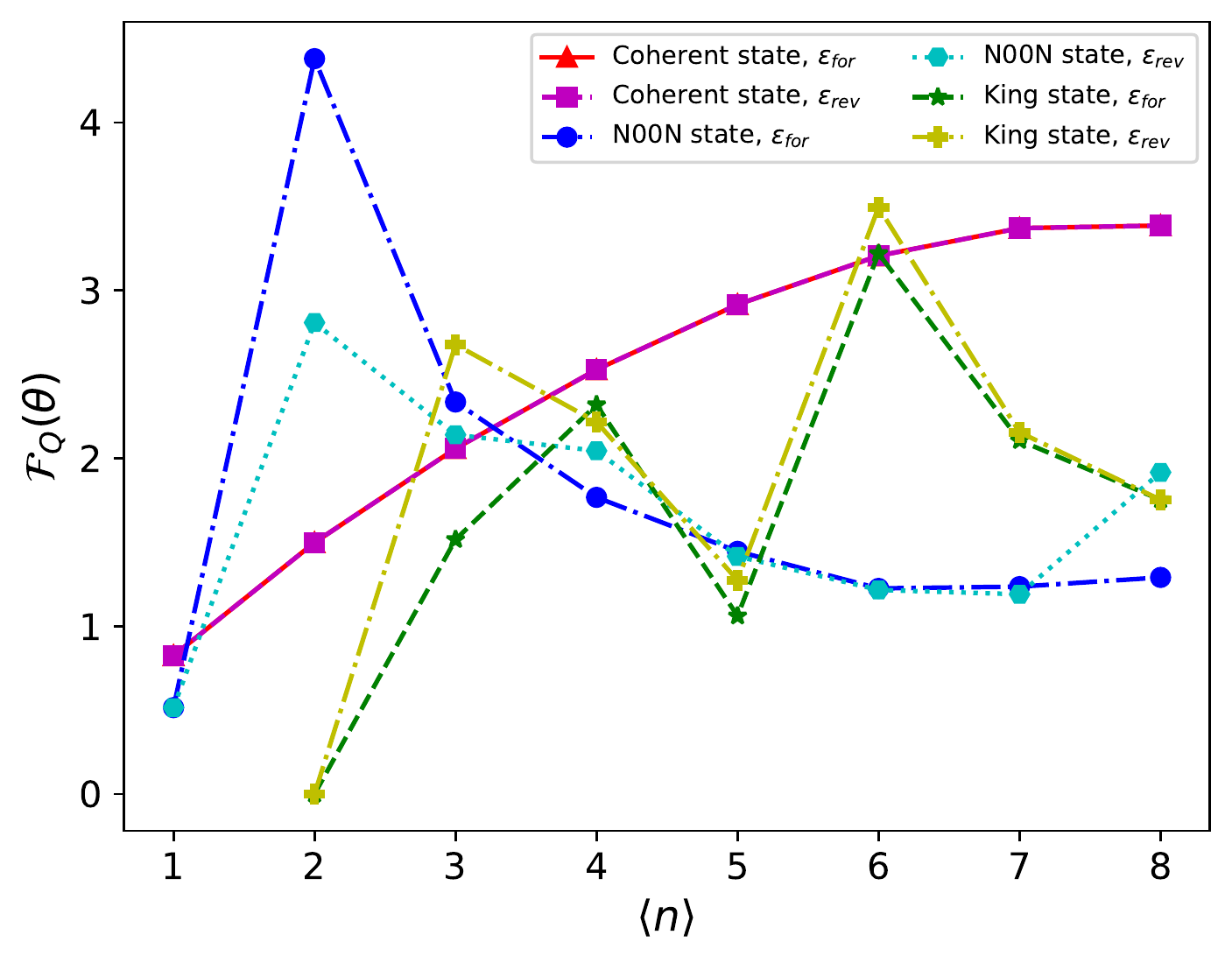}} 
    \caption{QFI for rotation angle in presence of isotropic depolarization noise vs average photon number. Calculations are for coherent state in forward (solid line with triangular data points in red) and reverse (dashed line with square data points in magenta) processes, NOON state in forward (dash dotted line with circular data points in blue) and reverse (dotted line with triangular data points in cyan) processes and Kings of Quantumness in forward (dashed green line with star datapoints) and reverse (dash-dotted yellow line with plus shaped datapoints) processes for (a) $\nu_it = 0.003$ (b) $\nu_it = 0.03$.}
    \label{qfi_depnoise_iso}
\end{figure}
\begin{figure}[htbp!]
    \centering
    \subfigure[]{\includegraphics[width=0.49\textwidth]{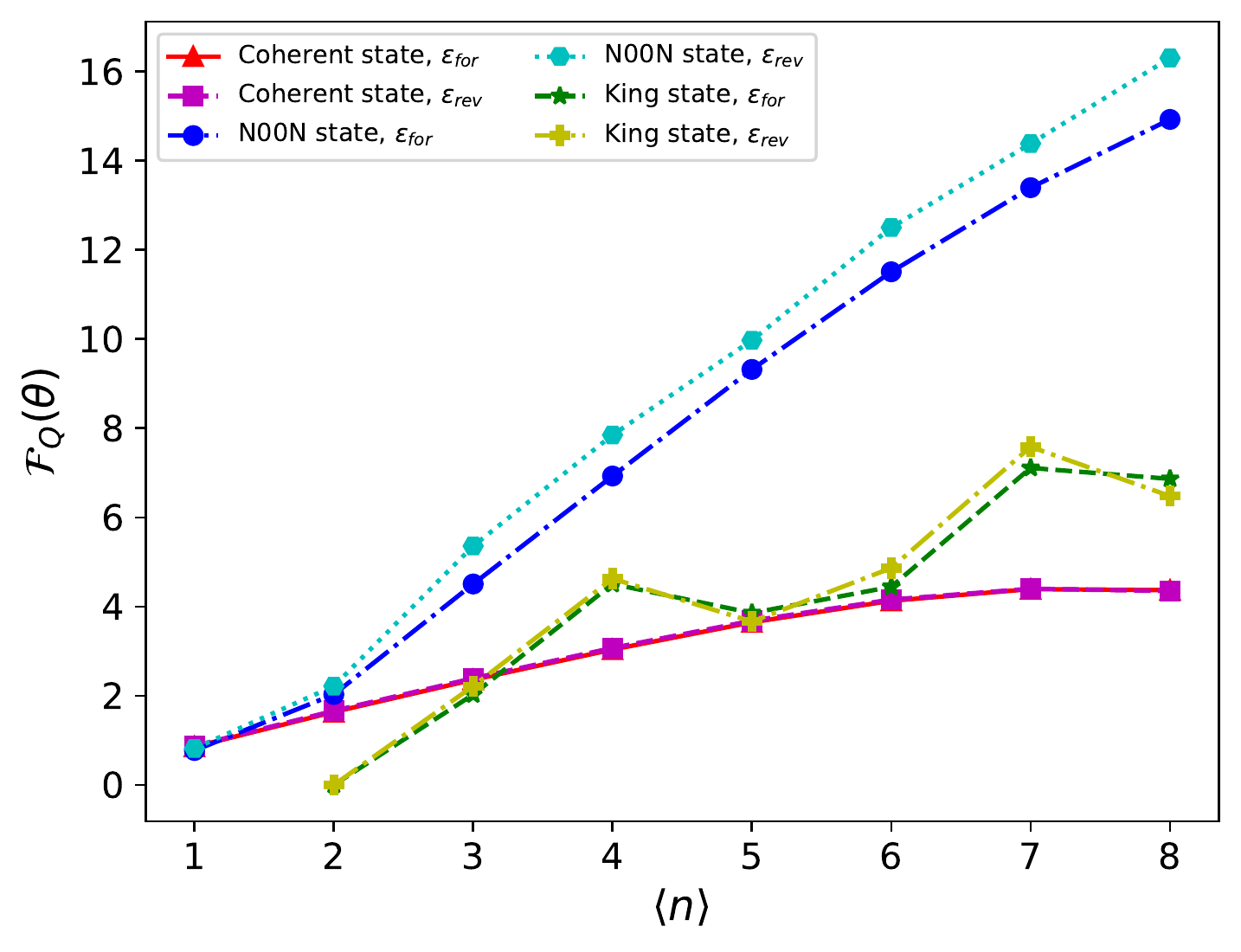}} 
    \subfigure[]{\includegraphics[width=0.49\textwidth]{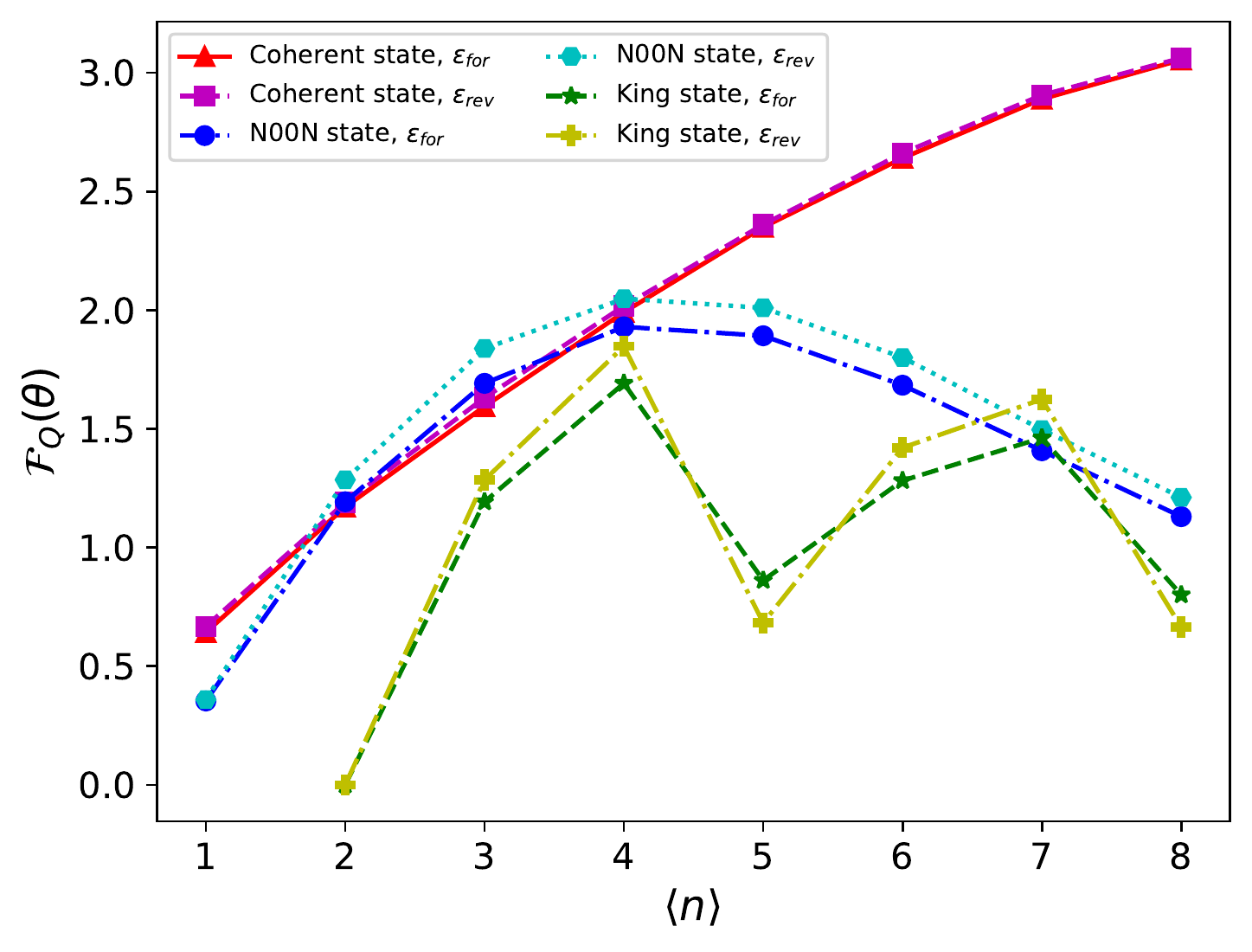}} 
    \caption{QFI for rotation angle in presence of anisotropic depolarization noise vs average photon number. Calculations are for coherent state in forward (solid line with triangular data points in red) and reverse (dashed line with square data points in magenta) processes, NOON state in forward (dash dotted line with circular data points in blue) and reverse (dotted line with triangular data points in cyan) processes and Kings of Quantumness in forward (dashed green line with star datapoints) and reverse (dash-dotted yellow line with plus shaped datapoints) processes for (a) $\nu_at = 0.05$ (b) $\nu_at = 0.15$. }
    \label{qfi_depnoise_ani}
\end{figure}
\begin{figure}[htbp!]
    \centering
    \subfigure[]{\includegraphics[width=0.49\textwidth]{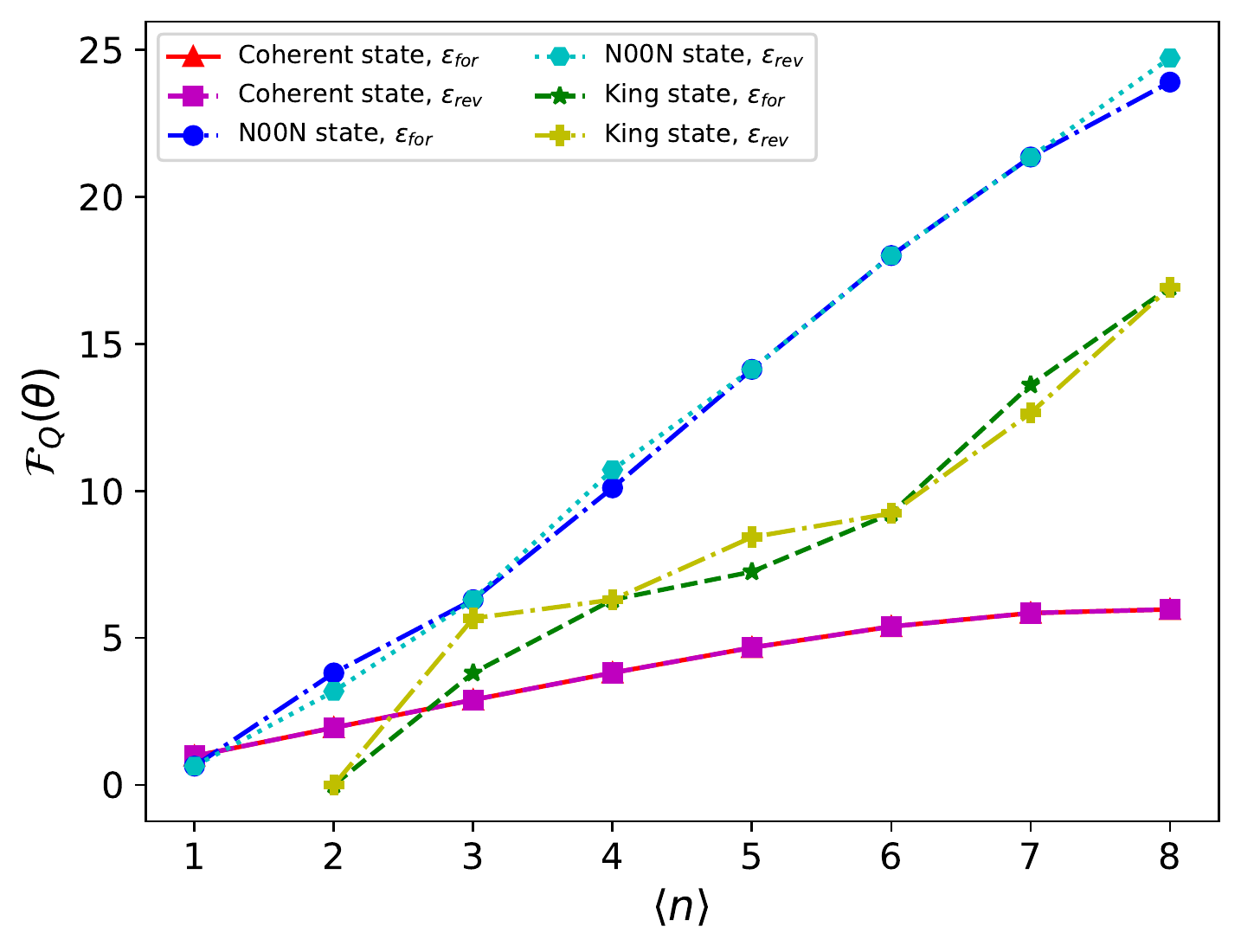}} 
    \subfigure[]{\includegraphics[width=0.49\textwidth]{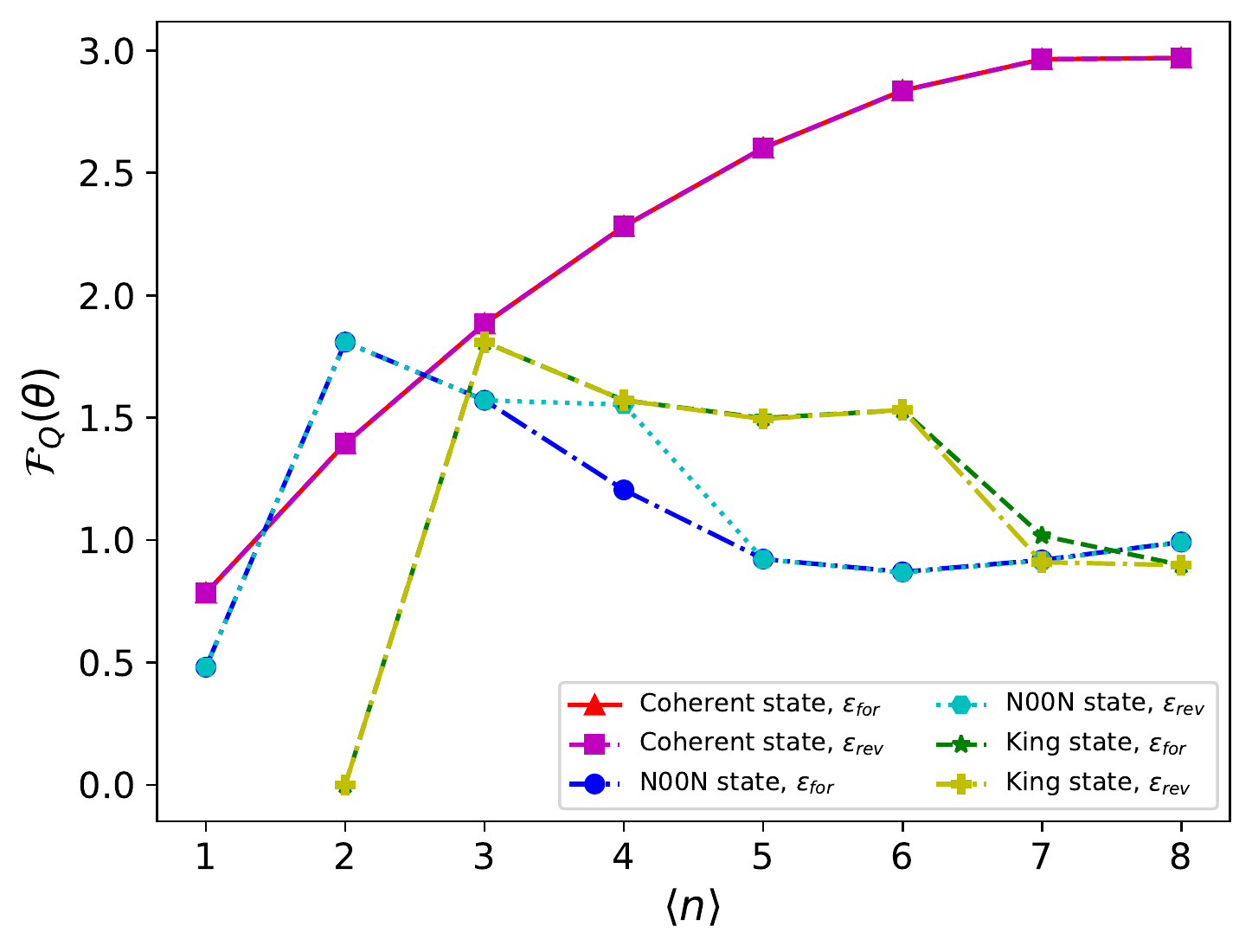}} 
    \caption{QFI for rotation angle in presence depolarization noise based on convex combination of rotations vs average photon number. Calculations are for coherent state in forward (solid line with triangular data points in red) and reverse (dashed line with square data points in magenta) processes, NOON state in forward (dash dotted line with circular data points in blue) and reverse (dotted line with triangular data points in cyan) processes and Kings of Quantumness in forward (dashed green line with star data points) and reverse (dash-dotted yellow line with plus shaped datapoints) processes for (a) $[\eta_{min},\eta_{max}] = [-\pi/8,\pi/8]$, $\sigma_r = \pi/32$, $n_r = 6$ (b) $[\eta_{min},\eta_{max}] = [-\pi,\pi]$, $\sigma_r = \pi/8$, $n_r = 20$.}
    \label{qfi_depnoise_rr}
\end{figure}
To make sure that no rotation axis or decomposition order is preferred, we took an average of the results given by all 6 permutations of the convex combination of $n_r$ rotations using three rotation operators $\hat{R}_1 = \text{exp}(-i\phi \hat{S}_1)$, $\hat{R}_2 = \text{exp}(-i\phi \hat{S}_2)$ and $\hat{R}_3 = \text{exp}(-i\phi \hat{S}_3)$. We stress that our choice of depolarization channel, although is physically motivated, is only a specific one among an infinite number of choices that are possible based on~\cref{simp_depolarizer}. For all depolarization channels we considered low and high depolarization regimes in our calculations. The results for isotropic, anisotropic and random rotation depolarization channels are presented in~\cref{qfi_depnoise_iso},~\cref{qfi_depnoise_ani} and~\cref{qfi_depnoise_rr} respectively.\\

It is clear that for all of the considered cases, introducing depolarization greatly reduces the measurement sensitivity of the retardation angle. Reduction in QFI is more dramatic for larger depolarization values. Additionally for all depolarization channels there is no difference in terms of metrological power between forward and reverse decomposition for coherent state. The isotropic depolarization channel is $\mathfrak{su}(2)$-invariant and as discussed in the derivation of master equation~\cref{dep_me_iso} in~\cite{PhysRevA.88.052120} different ordering of rotation and depolarization channels should yield the same result. Additionally, as explained earlier, the depolarization channel based on convex combinations of rotations is chosen such that the effects of decomposition order and choice of rotation axis for depolarization is minimized. However, from~\cref{qfi_depnoise_iso} and~\cref{qfi_depnoise_rr} we see that there are differences between forward and reverse decomposition for NOON state and Kings of Quantumness, the difference being larger for higher depolarization. These differences can be present due to a few reasons. First of all, due to numerical nature of our study, the results inevitably contain numerical errors. Also, from the explicit relation given for $\nu_{iso}$ in~\cite{PhysRevA.88.052120} we see that for larger values of $\nu_{iso}$ either the time derivative of the probability distribution for the rotation vectors comprising the depolarization channel or the amplitude of the rotation vectors must be large. The latter condition weakens the $\mathfrak{su}(2)$-invariance condition because for perfect $\mathfrak{su}(2)$-invariance to hold the rotations must be infinitesimally small.

Interestingly, in~\cref{qfi_depnoise_ani}(a) we see that for the anisotropic depolarization channel, the metrological performance for the NOON state and Kings of Quantumness is slightly better for the case of reverse decomposition. Inspecting~\cref{qfi_depnoise_ani}(b) we see that this advantage still holds for a larger amount of depolarization. It is also clear from our results that for all depolarization channels considered, NOON state and Kings of Quantumness lose their metrological power faster when depolarization is applied, compared with the coherent state. This confirms the common understanding that quantum states are more susceptible to noise compared to the classical states. Needless to say that for stronger depolarization than what we considered in this study, the coherent state results in a larger QFI than the quantum probe states for all average photon numbers. Another general feature underlying all depolarization channels is that the QFI vs $\braket{n}$ is no longer monotonic for all states except the coherent state, specially for large depolarization. This might be rather surprising considering the fact that usually the QFI scales monotonically with the number of resources. However, as discussed in~\cite{PhysRevA.88.052120} and~\cite{PhysRevResearch.4.013178}, different state multipoles have different decay rates and states with a larger $S$ values experience a larger decay. 

\subsection{Noisy Rotation Sensing with Depolarization and Diattenuation}
\label{subsec:retdepdia}
\begin{figure}[htbp!]
    \centering
    \subfigure[]{\includegraphics[width=0.32\textwidth]{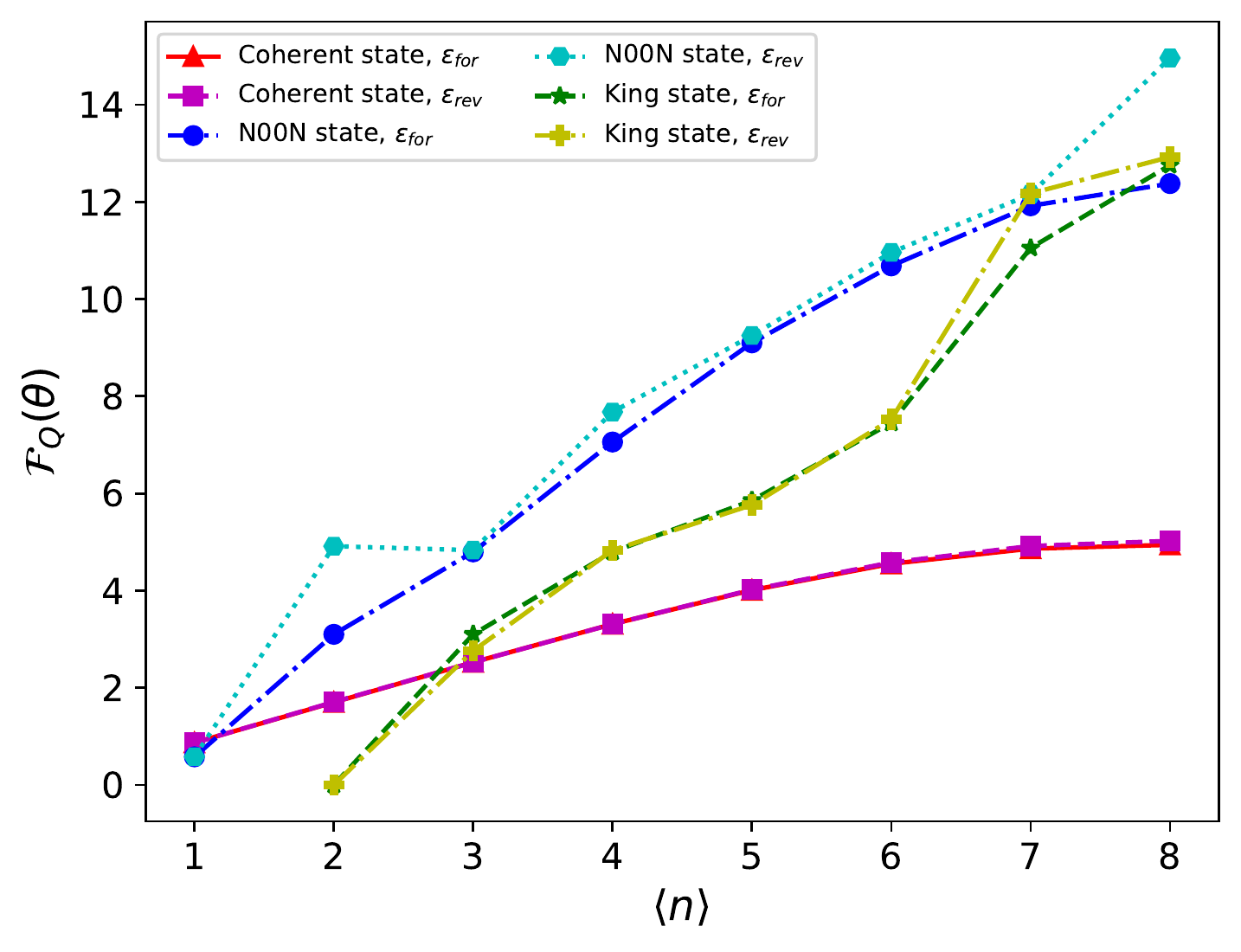}} 
    \subfigure[]{\includegraphics[width=0.32\textwidth]{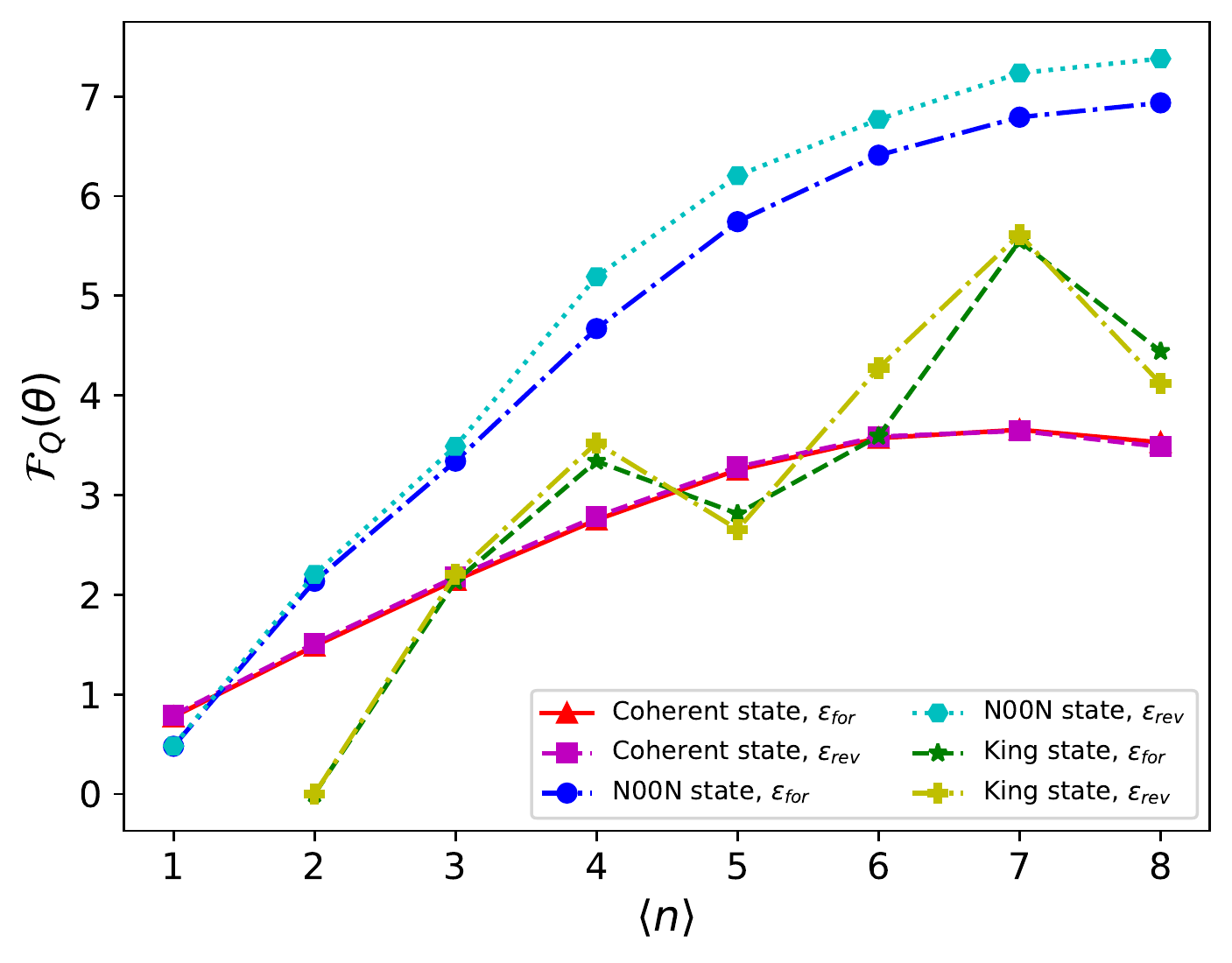}} 
    \subfigure[]{\includegraphics[width=0.32\textwidth]{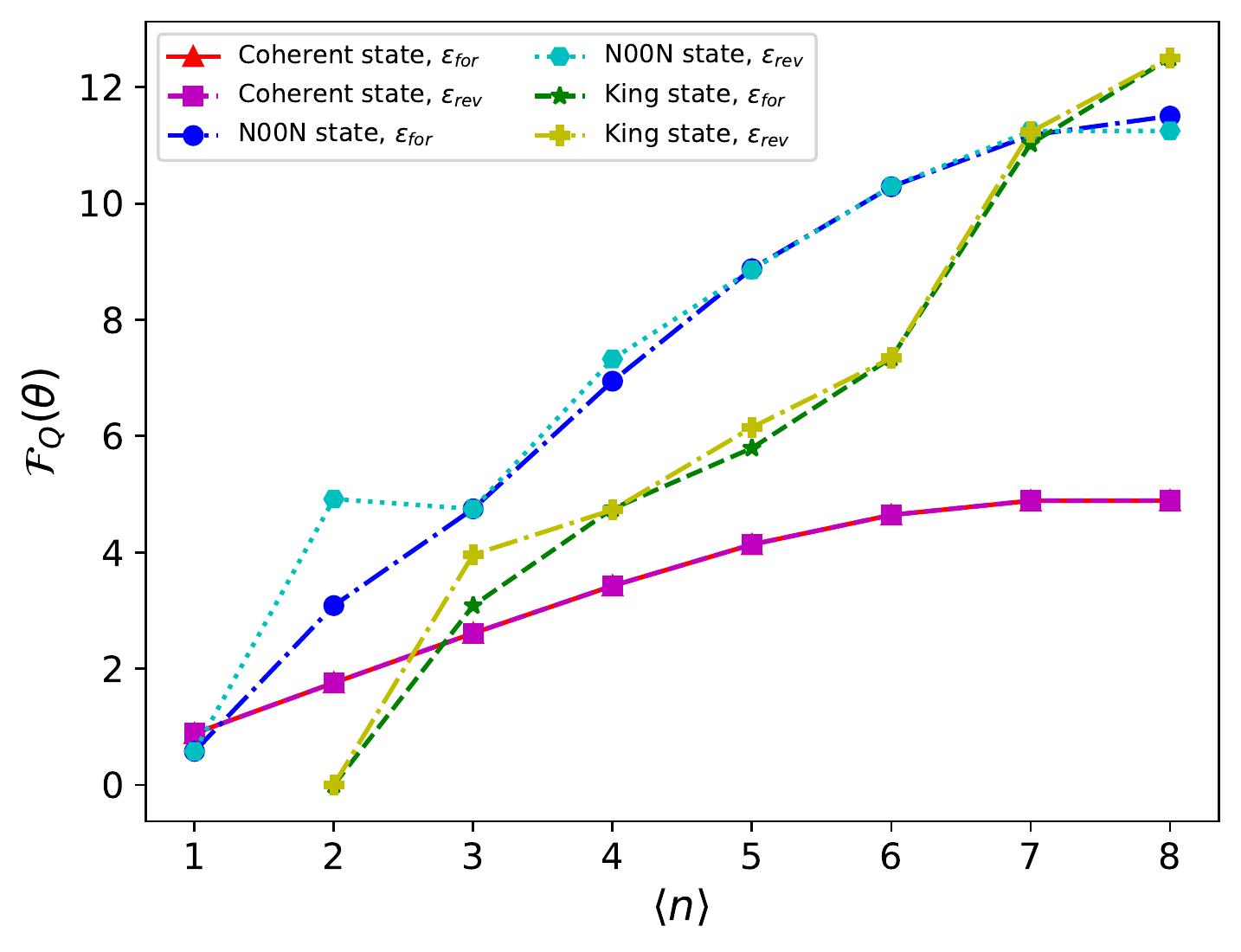}} 
  \caption{QFI for rotation angle in presence of depolarization and diattenuaion noise vs average photon number. Calculations are for coherent state in forward (solid line with triangular data points in red) and reverse (dashed line with square data points in magenta) processes, NOON state in forward (dash dotted line with circular data points in blue) and reverse (dotted line with triangular data points in cyan) processes and Kings of Quantumness in forward (dashed green line with star datapoints) and reverse (dash-dotted yellow line with plus shaped datapoints) processes for $q=r=0.9$ and (a) isotropic depolarization with $\nu_it = 0.003$ (b) anisotropic depolarization with $\nu_at = 0.05$ (c) depolarization based on convex combination of rotations with $[\eta_{min},\eta_{max}] = [-\pi/8,\pi/8]$, $\sigma_r = \pi/32$, $n_r = 6$.}\label{qfi_depdianoise}
\end{figure}

Finally, we calculate the QFI for the rotation angle vs the average photon number of the probe states considering both depolarization and diattenuation channels in forward and reverse decomposition channels. The results are shown in~\cref{qfi_depdianoise}. We have only included the results for the weak depolarization case for each depolarization channel considered in~\cref{subsec:retdep} to see wether the quantum advantage present in this regime still persists after implementation of the new loss channel. The parameters of rotation are identical with the ones chosen in the previous subsection and diattenuation parameters are $q=r=0.9$. 
The general behavior of the QFI with the number of photons is similar to the case where diattenuation is not considered. However, as expected, for all cases it is evident that due to implementation of the diattenuation channel the QFI for each $\braket{n}$ shrinks even further. Comparing~\cref{qfi_depdianoise} with the results given in the previous subsection it is clear that the diattenuation channel diminishes the metrological power of the NOON state more than it does for Kings of Quantumness. Furthermore, the relative advantage of the NOON state in reverse decomposition as compared with the forward decomposition still remains after inclusion of diattenuation. From~\cref{qfi_depdianoise}(a) and~\cref{qfi_depdianoise}(c) it is clear that for larger photon numbers Kings of Quantumness surpass the NOON state in terms of precision and from~\cref{qfi_depdianoise}(b) we see that for Kings of Quantumness at $\braket{n}=5$ (which corresponds to a 1-anticoherent state) the QFI is smaller than that of the coherent state.\\

So far we have compared the precision for estimation of the rotation angle between the three states under various settings. Although the QFI quantifies the ultimate precision bound, achieving the precision specified by the QFI might be experimentally challenging. Therefore it is instructive to briefly discuss the possible achievable precision limits under readily available experimental measurement operators. For this purpose we present the result for the simplest case of a probe state with a single photon ($\braket{n}=1$). Considering the POVMs corresponding to the Pauli operators $\hat{\sigma}_x$, $\hat{\sigma}_y$ and $\hat{\sigma}_z$, which can be easily implemented experimentally, the Fisher information for the NOON and coherent states become $2$ and $2/e$ respectively. Therefore, it is possible to extract the quantum advantage in this simple case using experimentally accessible measurement operators. Extension of this methodology to the higher excitation subspaces of the probe field is tedious but straightforward. Additionally, to gain maximal information, one should utilize symmetric informationally complete (SIC) POVMs to perform the measurements. However, experimental feasibility of such SIC-POVMs and complete characterization of precision based on experimentally easy to implement measurement operators is out of the scope of this work.

\section{Experimental Implementation of NOON State Polarimetry}
\label{sec:expimp}
In addition to theoretical investigations~\cite{Belsley:2022}, several recent experimental examples of applications of NOON states, for instance, for scattering-based probing of delicate samples like an atomic spin ensemble~\cite{Wolfgramm:2013} and highly sensitive measurement of protein concentration via the refractive index change~\cite{Peng:2023}, support the motivation of our study. Aiming at experimental validation of the theoretical findings described in the previous sections and at testing the feasibility to enhance the polarimetric sensing accuracy in presence of depolarization and diattenuation channels, we conceive a series of measurements. Here, we follow the previous works on realization and characterization of the two-photon and higher-order NOON states~\cite{Yurke:1986,Kuzmich:1998,Sanaka:2004,Israel:2012,Israel:2014}.

\begin{figure}[htbp!] 
\centering
\includegraphics[width=.5\textwidth]{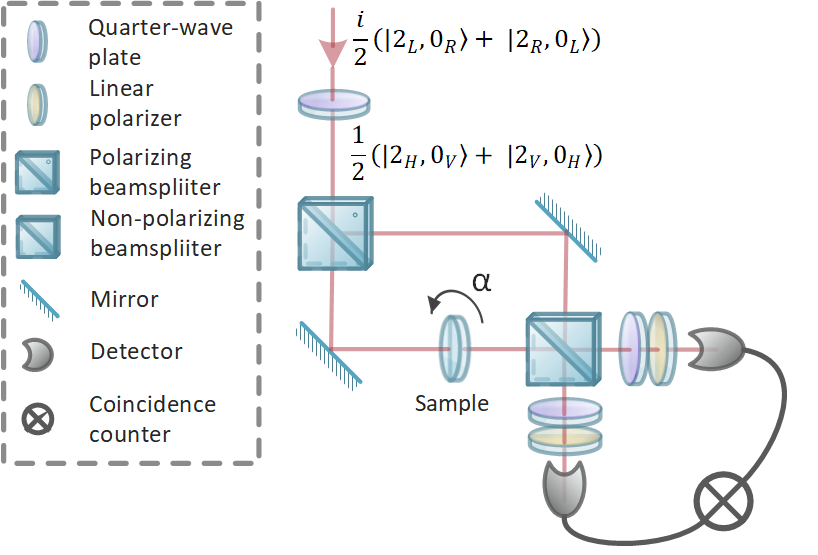}
\caption{A possible optical arrangement for test experiments on polarimetric accuracy enhancement with NOON states. A state $i/\sqrt{2}(|2_L,0_R\rangle + |2_R,0_L\rangle)$ is generated within an SPDC event of type-II phase matching (not depicted). This state is translated into $1/\sqrt{2}(|2_H,0_V\rangle + |0_V,2_H\rangle)$ by a quarter-wave plate and enters a Mach-Zehnder interferometer with a sample in one of the channels. The Hong-Ou-Mandel interference at the last non-polarizing beamsplitter takes place. Coincidence counts between the two outputs of the beamsplitter are measured.}
\label{fig:expSchemeNOON}
\end{figure}

Experimentally a NOON state can be both prepared and detected by exploiting the Hong-Ou-Mandel (HOM) effect~\cite{Hong:1987}. Alternatively, in case of a two-photon NOON state, the preparation can be realized by the spontaneous parametric down-conversion (SPDC) process. Under conditions of type-II phase matching, a correlated pair of the vertically (V) and horizontally (H) polarized photons $|HV\rangle|HV\rangle$ is generated by SPDC. This state can be regarded as a NOON state of the form $i/\sqrt{2}(|2_L,0_R\rangle + |2_R,0_L\rangle)$, where L and R stand for the left- and right-circular polarizations, respectively \cite{Adamson:2007,Wolfgramm:2010}. The projection of this state to the (H,V) basis can be obtained by a quarter-wave plate (QWP), oriented with its fast axis at $45^{\circ}$ to the incoming polarization orientations. Such a strategy is followed in Fig~\ref{fig:expSchemeNOON}, where a possible optical arrangement for test experiments is represented. 

After the two-photon NOON state is prepared, it is split by a polarizing beam splitter into two paths. One of them is regarded as reference channel, and the other is used for introducing the object under study. The two beams are then merged with a non-polarizing beam splitter. At both outputs of the latter a polarization state analyzer consisting of a QWP and a linear polarizer (LP) is implemented to perform projective measurements with the help of single photon detectors. The coincidences between the two channels for state reconstruction via the quantum state tomography~\cite{James:2001,Adamson:2007} or HOM interference visibility measurements are counted with the time tagging device.

In order to check the feasibility of the method for practical applications like for the already mentioned eye retina diagnostics, first technical samples of controllable thickness and retardance should be used. For the first case, a possible option would be a series of transparent homogeneous thin films of different thicknesses. For the second case, the retardance estimation under minimal impacts of diattenuation and depolarization, fixed (QWP, half-wave plate) and variable (Soleil-Babinet compensator) retarders at different orientations (angle $\alpha$) of their optical axis can be used. It is expected that successful demonstration of such experiments with technical samples could be later directly translated for method feasibility checks with biological specimen probing.

\section{Conclusion}
\label{sec:conclusion}
In conclusion, we studied the precision limits for single parameter estimation of the rotation angle of the Stokes vector in a polarimetric setup, within the framework of quantum polarimetry. We have incorporated the effects of depolarization and diattenuation channels within our numerical calculations and considered different orders of implementations of these channels. For the sake of completeness, we have utilized three different depolarization channels found in the literature in our calculations. All three depolarization models are consistent with the notion of polarization in quantum optics. The results show that, for our chosen sets of parameters and operators, QFI greatly diminishes upon implementation of the depolarization channel. For depolarization based on convex combination of rotations and the isotropic depolarization, there exists a minimal difference between forward and reverse decomposition in the final result for the QFI for weak depolarization and the difference is slightly more pronounced in the strong depolarization regime. However, for the anisotropic depolarization, after applying the depolarization channel, the reverse and forward decompositions yield different behavior with the average number of photons for Kings of Quantumness and the NOON state. The scaling of coherent state remains monotonic with the average number of photons and both decompositions yield the same result. However, for NOON state and Kings of Quantumness QFI becomes non-monotonic with the average photon number in the strong depolarization regime. This non-monotonicity manifests itself in all three depolarization models used in our calculations and therefore strongly suggests that there exists a law of diminishing returns for extracting information about the rotation parameter with respect to the number of excitations in the probe field in the higher range of the depolarization regime considered in this study. As expected, implementing diattenuation decreases QFI for all cases. However, our results demonstrate that metrological power of the NOON state diminishes more than that of Kings of Quantumness such that for larger average photon numbers the latter yields a higher QFI. Therefore our results suggest that, although HL scaling is lost after applying the noise channels, one can potentially utilize low photon number NOON states to gain a quantum advantage in polarimetric retardation measurements on a lossy medium. Possible experimental procedure for carrying out such a task is proposed.
\acknowledgments
We gratefully acknowledge financial support from the Scientific and Technological Research Council of Türkiye (TÜBİTAK), grant No. 120F200. This work has been funded by the European Union’s Horizon 2020 research and innovation programme (Grant Agreement No. 899580); the German Federal Ministry of Education and Research (FKZ 13N14877 and 13N15956); and the Cluster of Excellence “Balance of the Microverse” (EXC 2051-Project No. 390713860). V.B. thanks for funding of this work also through the ProChance-career program of the Friedrich Schiller University Jena. A.P. thanks Aaron Z. Goldberg from National Research Council of Canada for fruitful discussions.


\begin{thebibliography}{144}%
\makeatletter
\providecommand \@ifxundefined [1]{%
 \@ifx{#1\undefined}
}%
\providecommand \@ifnum [1]{%
 \ifnum #1\expandafter \@firstoftwo
 \else \expandafter \@secondoftwo
 \fi
}%
\providecommand \@ifx [1]{%
 \ifx #1\expandafter \@firstoftwo
 \else \expandafter \@secondoftwo
 \fi
}%
\providecommand \natexlab [1]{#1}%
\providecommand \enquote  [1]{``#1''}%
\providecommand \bibnamefont  [1]{#1}%
\providecommand \bibfnamefont [1]{#1}%
\providecommand \citenamefont [1]{#1}%
\providecommand \href@noop [0]{\@secondoftwo}%
\providecommand \href [0]{\begingroup \@sanitize@url \@href}%
\providecommand \@href[1]{\@@startlink{#1}\@@href}%
\providecommand \@@href[1]{\endgroup#1\@@endlink}%
\providecommand \@sanitize@url [0]{\catcode `\\12\catcode `\$12\catcode
  `\&12\catcode `\#12\catcode `\^12\catcode `\_12\catcode `\%12\relax}%
\providecommand \@@startlink[1]{}%
\providecommand \@@endlink[0]{}%
\providecommand \url  [0]{\begingroup\@sanitize@url \@url }%
\providecommand \@url [1]{\endgroup\@href {#1}{\urlprefix }}%
\providecommand \urlprefix  [0]{URL }%
\providecommand \Eprint [0]{\href }%
\providecommand \doibase [0]{https://doi.org/}%
\providecommand \selectlanguage [0]{\@gobble}%
\providecommand \bibinfo  [0]{\@secondoftwo}%
\providecommand \bibfield  [0]{\@secondoftwo}%
\providecommand \translation [1]{[#1]}%
\providecommand \BibitemOpen [0]{}%
\providecommand \bibitemStop [0]{}%
\providecommand \bibitemNoStop [0]{.\EOS\space}%
\providecommand \EOS [0]{\spacefactor3000\relax}%
\providecommand \BibitemShut  [1]{\csname bibitem#1\endcsname}%
\let\auto@bib@innerbib\@empty
\bibitem [{\citenamefont {Nesse}\ and\ \citenamefont
  {Nesse}(2004)}]{nesse2004introduction}%
  \BibitemOpen
  \bibfield  {author} {\bibinfo {author} {\bibfnamefont {W.}~\bibnamefont
  {Nesse}}\ and\ \bibinfo {author} {\bibfnamefont {P.}~\bibnamefont {Nesse}},\
  }\href@noop {} {\emph {\bibinfo {title} {Introduction to Optical
  Mineralogy}}}\ (\bibinfo  {publisher} {Oxford University Press},\ \bibinfo
  {year} {2004})\BibitemShut {NoStop}%
\bibitem [{\citenamefont {Kaminsky}\ \emph {et~al.}(2004)\citenamefont
  {Kaminsky}, \citenamefont {Claborn},\ and\ \citenamefont {Kahr}}]{B201314M}%
  \BibitemOpen
  \bibfield  {author} {\bibinfo {author} {\bibfnamefont {W.}~\bibnamefont
  {Kaminsky}}, \bibinfo {author} {\bibfnamefont {K.}~\bibnamefont {Claborn}},\
  and\ \bibinfo {author} {\bibfnamefont {B.}~\bibnamefont {Kahr}},\ }\href
  {https://doi.org/10.1039/B201314M} {\bibfield  {journal} {\bibinfo  {journal}
  {Chem. Soc. Rev.}\ }\textbf {\bibinfo {volume} {33}},\ \bibinfo {pages} {514}
  (\bibinfo {year} {2004})}\BibitemShut {NoStop}%
\bibitem [{\citenamefont {Theeten}\ and\ \citenamefont
  {Aspnes}(1981)}]{doi:10.1146/annurev.ms.11.080181.000525}%
  \BibitemOpen
  \bibfield  {author} {\bibinfo {author} {\bibfnamefont {J.~B.}\ \bibnamefont
  {Theeten}}\ and\ \bibinfo {author} {\bibfnamefont {D.~E.}\ \bibnamefont
  {Aspnes}},\ }\href {https://doi.org/10.1146/annurev.ms.11.080181.000525}
  {\bibfield  {journal} {\bibinfo  {journal} {Annual Review of Materials
  Science}\ }\textbf {\bibinfo {volume} {11}},\ \bibinfo {pages} {97} (\bibinfo
  {year} {1981})},\ \Eprint
  {https://arxiv.org/abs/https://doi.org/10.1146/annurev.ms.11.080181.000525}
  {https://doi.org/10.1146/annurev.ms.11.080181.000525} \BibitemShut {NoStop}%
\bibitem [{\citenamefont {Losurdo}\ \emph {et~al.}(2009)\citenamefont
  {Losurdo}, \citenamefont {Bergmair}, \citenamefont {Bruno}, \citenamefont
  {Cattelan}, \citenamefont {Cobet}, \citenamefont {de~Martino}, \citenamefont
  {Fleischer}, \citenamefont {Dohcevic-Mitrovic}, \citenamefont {Esser},
  \citenamefont {Galliet}, \citenamefont {Gajic}, \citenamefont {Hemzal},
  \citenamefont {Hingerl}, \citenamefont {Humlicek}, \citenamefont
  {Ossikovski}, \citenamefont {Popovic},\ and\ \citenamefont
  {Saxl}}]{Losurdo2009}%
  \BibitemOpen
  \bibfield  {author} {\bibinfo {author} {\bibfnamefont {M.}~\bibnamefont
  {Losurdo}}, \bibinfo {author} {\bibfnamefont {M.}~\bibnamefont {Bergmair}},
  \bibinfo {author} {\bibfnamefont {G.}~\bibnamefont {Bruno}}, \bibinfo
  {author} {\bibfnamefont {D.}~\bibnamefont {Cattelan}}, \bibinfo {author}
  {\bibfnamefont {C.}~\bibnamefont {Cobet}}, \bibinfo {author} {\bibfnamefont
  {A.}~\bibnamefont {de~Martino}}, \bibinfo {author} {\bibfnamefont
  {K.}~\bibnamefont {Fleischer}}, \bibinfo {author} {\bibfnamefont
  {Z.}~\bibnamefont {Dohcevic-Mitrovic}}, \bibinfo {author} {\bibfnamefont
  {N.}~\bibnamefont {Esser}}, \bibinfo {author} {\bibfnamefont
  {M.}~\bibnamefont {Galliet}}, \bibinfo {author} {\bibfnamefont
  {R.}~\bibnamefont {Gajic}}, \bibinfo {author} {\bibfnamefont
  {D.}~\bibnamefont {Hemzal}}, \bibinfo {author} {\bibfnamefont
  {K.}~\bibnamefont {Hingerl}}, \bibinfo {author} {\bibfnamefont
  {J.}~\bibnamefont {Humlicek}}, \bibinfo {author} {\bibfnamefont
  {R.}~\bibnamefont {Ossikovski}}, \bibinfo {author} {\bibfnamefont {Z.~V.}\
  \bibnamefont {Popovic}},\ and\ \bibinfo {author} {\bibfnamefont
  {O.}~\bibnamefont {Saxl}},\ }\href
  {https://doi.org/10.1007/s11051-009-9662-6} {\bibfield  {journal} {\bibinfo
  {journal} {Journal of Nanoparticle Research}\ }\textbf {\bibinfo {volume}
  {11}},\ \bibinfo {pages} {1521} (\bibinfo {year} {2009})}\BibitemShut
  {NoStop}%
\bibitem [{\citenamefont {Andreou}\ and\ \citenamefont
  {Kalayjian}(2002)}]{1178170}%
  \BibitemOpen
  \bibfield  {author} {\bibinfo {author} {\bibfnamefont {A.}~\bibnamefont
  {Andreou}}\ and\ \bibinfo {author} {\bibfnamefont {Z.}~\bibnamefont
  {Kalayjian}},\ }\href {https://doi.org/10.1109/JSEN.2003.807946} {\bibfield
  {journal} {\bibinfo  {journal} {IEEE Sensors Journal}\ }\textbf {\bibinfo
  {volume} {2}},\ \bibinfo {pages} {566} (\bibinfo {year} {2002})}\BibitemShut
  {NoStop}%
\bibitem [{\citenamefont {Tinbergen}(1996)}]{tinbergen_1996}%
  \BibitemOpen
  \bibfield  {author} {\bibinfo {author} {\bibfnamefont {J.}~\bibnamefont
  {Tinbergen}},\ }\bibinfo {title} {Polarization in astronomy},\ in\ \href
  {https://doi.org/10.1017/CBO9780511525100.004} {\emph {\bibinfo {booktitle}
  {Astronomical Polarimetry}}}\ (\bibinfo  {publisher} {Cambridge University
  Press},\ \bibinfo {year} {1996})\ p.\ \bibinfo {pages} {27–44}\BibitemShut
  {NoStop}%
\bibitem [{\citenamefont {Snik}\ and\ \citenamefont {Keller}(2013)}]{Snik2013}%
  \BibitemOpen
  \bibfield  {author} {\bibinfo {author} {\bibfnamefont {F.}~\bibnamefont
  {Snik}}\ and\ \bibinfo {author} {\bibfnamefont {C.~U.}\ \bibnamefont
  {Keller}},\ }\bibinfo {title} {Astronomical polarimetry: Polarized views of
  stars and planets},\ in\ \href {https://doi.org/10.1007/978-94-007-5618-2_4}
  {\emph {\bibinfo {booktitle} {Planets, Stars and Stellar Systems: Volume 2:
  Astronomical Techniques, Software, and Data}}},\ \bibinfo {editor} {edited
  by\ \bibinfo {editor} {\bibfnamefont {T.~D.}\ \bibnamefont {Oswalt}}\ and\
  \bibinfo {editor} {\bibfnamefont {H.~E.}\ \bibnamefont {Bond}}}\ (\bibinfo
  {publisher} {Springer Netherlands},\ \bibinfo {address} {Dordrecht},\
  \bibinfo {year} {2013})\ pp.\ \bibinfo {pages} {175--221}\BibitemShut
  {NoStop}%
\bibitem [{\citenamefont {He}\ \emph {et~al.}(2021)\citenamefont {He},
  \citenamefont {He}, \citenamefont {Chang}, \citenamefont {Chen},
  \citenamefont {Ma},\ and\ \citenamefont {Booth}}]{He2021}%
  \BibitemOpen
  \bibfield  {author} {\bibinfo {author} {\bibfnamefont {C.}~\bibnamefont
  {He}}, \bibinfo {author} {\bibfnamefont {H.}~\bibnamefont {He}}, \bibinfo
  {author} {\bibfnamefont {J.}~\bibnamefont {Chang}}, \bibinfo {author}
  {\bibfnamefont {B.}~\bibnamefont {Chen}}, \bibinfo {author} {\bibfnamefont
  {H.}~\bibnamefont {Ma}},\ and\ \bibinfo {author} {\bibfnamefont {M.~J.}\
  \bibnamefont {Booth}},\ }\href {https://doi.org/10.1038/s41377-021-00639-x}
  {\bibfield  {journal} {\bibinfo  {journal} {Light: Science {\&}
  Applications}\ }\textbf {\bibinfo {volume} {10}},\ \bibinfo {pages} {194}
  (\bibinfo {year} {2021})}\BibitemShut {NoStop}%
\bibitem [{\citenamefont {Vitkin}\ \emph {et~al.}(2015)\citenamefont {Vitkin},
  \citenamefont {Ghosh},\ and\ \citenamefont
  {Martino}}]{doi:https://doi.org/10.1002/9781119011804.ch7}%
  \BibitemOpen
  \bibfield  {author} {\bibinfo {author} {\bibfnamefont {A.}~\bibnamefont
  {Vitkin}}, \bibinfo {author} {\bibfnamefont {N.}~\bibnamefont {Ghosh}},\ and\
  \bibinfo {author} {\bibfnamefont {A.~d.}\ \bibnamefont {Martino}},\ }\bibinfo
  {title} {Tissue polarimetry},\ in\ \href
  {https://doi.org/https://doi.org/10.1002/9781119011804.ch7} {\emph {\bibinfo
  {booktitle} {Photonics}}}\ (\bibinfo  {publisher} {John Wiley \& Sons, Ltd},\
  \bibinfo {year} {2015})\ Chap.~\bibinfo {chapter} {7}, pp.\ \bibinfo {pages}
  {239--321},\ \Eprint
  {https://arxiv.org/abs/https://onlinelibrary.wiley.com/doi/pdf/10.1002/9781119011804.ch7}
  {https://onlinelibrary.wiley.com/doi/pdf/10.1002/9781119011804.ch7}
  \BibitemShut {NoStop}%
\bibitem [{\citenamefont {Ghosh}\ and\ \citenamefont
  {Vitkin}(2011)}]{10.1117/1.3652896}%
  \BibitemOpen
  \bibfield  {author} {\bibinfo {author} {\bibfnamefont {N.}~\bibnamefont
  {Ghosh}}\ and\ \bibinfo {author} {\bibfnamefont {A.~I.}\ \bibnamefont
  {Vitkin}},\ }\href {https://doi.org/10.1117/1.3652896} {\bibfield  {journal}
  {\bibinfo  {journal} {Journal of Biomedical Optics}\ }\textbf {\bibinfo
  {volume} {16}},\ \bibinfo {pages} {110801} (\bibinfo {year}
  {2011})}\BibitemShut {NoStop}%
\bibitem [{\citenamefont {Ramella-Roman}\ \emph {et~al.}(2020)\citenamefont
  {Ramella-Roman}, \citenamefont {Saytashev},\ and\ \citenamefont
  {Piccini}}]{Ramella-Roman_2020}%
  \BibitemOpen
  \bibfield  {author} {\bibinfo {author} {\bibfnamefont {J.~C.}\ \bibnamefont
  {Ramella-Roman}}, \bibinfo {author} {\bibfnamefont {I.}~\bibnamefont
  {Saytashev}},\ and\ \bibinfo {author} {\bibfnamefont {M.}~\bibnamefont
  {Piccini}},\ }\href {https://doi.org/10.1088/2040-8986/abbf8a} {\bibfield
  {journal} {\bibinfo  {journal} {Journal of Optics}\ }\textbf {\bibinfo
  {volume} {22}},\ \bibinfo {pages} {123001} (\bibinfo {year}
  {2020})}\BibitemShut {NoStop}%
\bibitem [{\citenamefont {Gramatikov}(2014)}]{Gramatikov2014}%
  \BibitemOpen
  \bibfield  {author} {\bibinfo {author} {\bibfnamefont {B.~I.}\ \bibnamefont
  {Gramatikov}},\ }\href {https://doi.org/10.1186/1475-925X-13-52} {\bibfield
  {journal} {\bibinfo  {journal} {BioMedical Engineering OnLine}\ }\textbf
  {\bibinfo {volume} {13}},\ \bibinfo {pages} {52} (\bibinfo {year}
  {2014})}\BibitemShut {NoStop}%
\bibitem [{\citenamefont {Kok}\ \emph {et~al.}(2007)\citenamefont {Kok},
  \citenamefont {Munro}, \citenamefont {Nemoto}, \citenamefont {Ralph},
  \citenamefont {Dowling},\ and\ \citenamefont {Milburn}}]{RevModPhys.79.135}%
  \BibitemOpen
  \bibfield  {author} {\bibinfo {author} {\bibfnamefont {P.}~\bibnamefont
  {Kok}}, \bibinfo {author} {\bibfnamefont {W.~J.}\ \bibnamefont {Munro}},
  \bibinfo {author} {\bibfnamefont {K.}~\bibnamefont {Nemoto}}, \bibinfo
  {author} {\bibfnamefont {T.~C.}\ \bibnamefont {Ralph}}, \bibinfo {author}
  {\bibfnamefont {J.~P.}\ \bibnamefont {Dowling}},\ and\ \bibinfo {author}
  {\bibfnamefont {G.~J.}\ \bibnamefont {Milburn}},\ }\href
  {https://doi.org/10.1103/RevModPhys.79.135} {\bibfield  {journal} {\bibinfo
  {journal} {Rev. Mod. Phys.}\ }\textbf {\bibinfo {volume} {79}},\ \bibinfo
  {pages} {135} (\bibinfo {year} {2007})}\BibitemShut {NoStop}%
\bibitem [{\citenamefont {Hecht}(2017)}]{hecht2017optics}%
  \BibitemOpen
  \bibfield  {author} {\bibinfo {author} {\bibfnamefont {E.}~\bibnamefont
  {Hecht}},\ }\href@noop {} {\emph {\bibinfo {title} {Optics}}}\ (\bibinfo
  {publisher} {Pearson Education, Incorporated},\ \bibinfo {year}
  {2017})\BibitemShut {NoStop}%
\bibitem [{\citenamefont {Azzam}(2016)}]{Azzam:16}%
  \BibitemOpen
  \bibfield  {author} {\bibinfo {author} {\bibfnamefont {R.~M.~A.}\
  \bibnamefont {Azzam}},\ }\href {https://doi.org/10.1364/JOSAA.33.001396}
  {\bibfield  {journal} {\bibinfo  {journal} {J. Opt. Soc. Am. A}\ }\textbf
  {\bibinfo {volume} {33}},\ \bibinfo {pages} {1396} (\bibinfo {year}
  {2016})}\BibitemShut {NoStop}%
\bibitem [{\citenamefont {Bueno}(2000)}]{BUENO20003791}%
  \BibitemOpen
  \bibfield  {author} {\bibinfo {author} {\bibfnamefont {J.~M.}\ \bibnamefont
  {Bueno}},\ }\href
  {https://doi.org/https://doi.org/10.1016/S0042-6989(00)00220-0} {\bibfield
  {journal} {\bibinfo  {journal} {Vision Research}\ }\textbf {\bibinfo {volume}
  {40}},\ \bibinfo {pages} {3791} (\bibinfo {year} {2000})}\BibitemShut
  {NoStop}%
\bibitem [{\citenamefont {Knighton}\ \emph {et~al.}(2002)\citenamefont
  {Knighton}, \citenamefont {Huang},\ and\ \citenamefont
  {Greenfield}}]{knighton2002}%
  \BibitemOpen
  \bibfield  {author} {\bibinfo {author} {\bibfnamefont {R.~W.}\ \bibnamefont
  {Knighton}}, \bibinfo {author} {\bibfnamefont {X.-R.}\ \bibnamefont
  {Huang}},\ and\ \bibinfo {author} {\bibfnamefont {D.~S.}\ \bibnamefont
  {Greenfield}},\ }\href@noop {} {\bibfield  {journal} {\bibinfo  {journal}
  {Investigative Ophthalmology \& Visual Science}\ }\textbf {\bibinfo {volume}
  {43}},\ \bibinfo {pages} {383} (\bibinfo {year} {2002})},\ \Eprint
  {https://arxiv.org/abs/https://arvojournals.org/arvo/content\_public/journal/iovs/933222/7g0202000383.pdf}
  {https://arvojournals.org/arvo/content\_public/journal/iovs/933222/7g0202000383.pdf}
  \BibitemShut {NoStop}%
\bibitem [{\citenamefont {Benoit}\ \emph {et~al.}(2001)\citenamefont {Benoit},
  \citenamefont {Naoun}, \citenamefont {Louis-Dorr}, \citenamefont {Mala},\
  and\ \citenamefont {Raspiller}}]{Benoit:01}%
  \BibitemOpen
  \bibfield  {author} {\bibinfo {author} {\bibfnamefont {A.~M.}\ \bibnamefont
  {Benoit}}, \bibinfo {author} {\bibfnamefont {K.}~\bibnamefont {Naoun}},
  \bibinfo {author} {\bibfnamefont {V.}~\bibnamefont {Louis-Dorr}}, \bibinfo
  {author} {\bibfnamefont {L.}~\bibnamefont {Mala}},\ and\ \bibinfo {author}
  {\bibfnamefont {A.}~\bibnamefont {Raspiller}},\ }\href
  {https://doi.org/10.1364/AO.40.000565} {\bibfield  {journal} {\bibinfo
  {journal} {Appl. Opt.}\ }\textbf {\bibinfo {volume} {40}},\ \bibinfo {pages}
  {565} (\bibinfo {year} {2001})}\BibitemShut {NoStop}%
\bibitem [{\citenamefont {Huang}\ and\ \citenamefont
  {Knighton}(2003)}]{Huang:03}%
  \BibitemOpen
  \bibfield  {author} {\bibinfo {author} {\bibfnamefont {X.-R.}\ \bibnamefont
  {Huang}}\ and\ \bibinfo {author} {\bibfnamefont {R.~W.}\ \bibnamefont
  {Knighton}},\ }\href {https://doi.org/10.1364/AO.42.005726} {\bibfield
  {journal} {\bibinfo  {journal} {Appl. Opt.}\ }\textbf {\bibinfo {volume}
  {42}},\ \bibinfo {pages} {5726} (\bibinfo {year} {2003})}\BibitemShut
  {NoStop}%
\bibitem [{\citenamefont {Peres}\ and\ \citenamefont
  {Scudo}(2001{\natexlab{a}})}]{PhysRevLett.86.4160}%
  \BibitemOpen
  \bibfield  {author} {\bibinfo {author} {\bibfnamefont {A.}~\bibnamefont
  {Peres}}\ and\ \bibinfo {author} {\bibfnamefont {P.~F.}\ \bibnamefont
  {Scudo}},\ }\href {https://doi.org/10.1103/PhysRevLett.86.4160} {\bibfield
  {journal} {\bibinfo  {journal} {Phys. Rev. Lett.}\ }\textbf {\bibinfo
  {volume} {86}},\ \bibinfo {pages} {4160} (\bibinfo {year}
  {2001}{\natexlab{a}})}\BibitemShut {NoStop}%
\bibitem [{\citenamefont {Peres}\ and\ \citenamefont
  {Scudo}(2001{\natexlab{b}})}]{PhysRevLett.87.167901}%
  \BibitemOpen
  \bibfield  {author} {\bibinfo {author} {\bibfnamefont {A.}~\bibnamefont
  {Peres}}\ and\ \bibinfo {author} {\bibfnamefont {P.~F.}\ \bibnamefont
  {Scudo}},\ }\href {https://doi.org/10.1103/PhysRevLett.87.167901} {\bibfield
  {journal} {\bibinfo  {journal} {Phys. Rev. Lett.}\ }\textbf {\bibinfo
  {volume} {87}},\ \bibinfo {pages} {167901} (\bibinfo {year}
  {2001}{\natexlab{b}})}\BibitemShut {NoStop}%
\bibitem [{\citenamefont {Rezazadeh}\ \emph {et~al.}(2017)\citenamefont
  {Rezazadeh}, \citenamefont {Mani},\ and\ \citenamefont
  {Karimipour}}]{PhysRevA.96.022310}%
  \BibitemOpen
  \bibfield  {author} {\bibinfo {author} {\bibfnamefont {F.}~\bibnamefont
  {Rezazadeh}}, \bibinfo {author} {\bibfnamefont {A.}~\bibnamefont {Mani}},\
  and\ \bibinfo {author} {\bibfnamefont {V.}~\bibnamefont {Karimipour}},\
  }\href {https://doi.org/10.1103/PhysRevA.96.022310} {\bibfield  {journal}
  {\bibinfo  {journal} {Phys. Rev. A}\ }\textbf {\bibinfo {volume} {96}},\
  \bibinfo {pages} {022310} (\bibinfo {year} {2017})}\BibitemShut {NoStop}%
\bibitem [{\citenamefont {Bagan}\ \emph {et~al.}(2001)\citenamefont {Bagan},
  \citenamefont {Baig},\ and\ \citenamefont {Mu\~noz
  Tapia}}]{PhysRevLett.87.257903}%
  \BibitemOpen
  \bibfield  {author} {\bibinfo {author} {\bibfnamefont {E.}~\bibnamefont
  {Bagan}}, \bibinfo {author} {\bibfnamefont {M.}~\bibnamefont {Baig}},\ and\
  \bibinfo {author} {\bibfnamefont {R.}~\bibnamefont {Mu\~noz Tapia}},\ }\href
  {https://doi.org/10.1103/PhysRevLett.87.257903} {\bibfield  {journal}
  {\bibinfo  {journal} {Phys. Rev. Lett.}\ }\textbf {\bibinfo {volume} {87}},\
  \bibinfo {pages} {257903} (\bibinfo {year} {2001})}\BibitemShut {NoStop}%
\bibitem [{\citenamefont {Chiribella}\ \emph {et~al.}(2004)\citenamefont
  {Chiribella}, \citenamefont {D'Ariano}, \citenamefont {Perinotti},\ and\
  \citenamefont {Sacchi}}]{PhysRevLett.93.180503}%
  \BibitemOpen
  \bibfield  {author} {\bibinfo {author} {\bibfnamefont {G.}~\bibnamefont
  {Chiribella}}, \bibinfo {author} {\bibfnamefont {G.~M.}\ \bibnamefont
  {D'Ariano}}, \bibinfo {author} {\bibfnamefont {P.}~\bibnamefont
  {Perinotti}},\ and\ \bibinfo {author} {\bibfnamefont {M.~F.}\ \bibnamefont
  {Sacchi}},\ }\href {https://doi.org/10.1103/PhysRevLett.93.180503} {\bibfield
   {journal} {\bibinfo  {journal} {Phys. Rev. Lett.}\ }\textbf {\bibinfo
  {volume} {93}},\ \bibinfo {pages} {180503} (\bibinfo {year}
  {2004})}\BibitemShut {NoStop}%
\bibitem [{\citenamefont {Kolenderski}\ and\ \citenamefont
  {Demkowicz-Dobrzanski}(2008)}]{PhysRevA.78.052333}%
  \BibitemOpen
  \bibfield  {author} {\bibinfo {author} {\bibfnamefont {P.}~\bibnamefont
  {Kolenderski}}\ and\ \bibinfo {author} {\bibfnamefont {R.}~\bibnamefont
  {Demkowicz-Dobrzanski}},\ }\href {https://doi.org/10.1103/PhysRevA.78.052333}
  {\bibfield  {journal} {\bibinfo  {journal} {Phys. Rev. A}\ }\textbf {\bibinfo
  {volume} {78}},\ \bibinfo {pages} {052333} (\bibinfo {year}
  {2008})}\BibitemShut {NoStop}%
\bibitem [{\citenamefont {Safranek}\ \emph {et~al.}(2015)\citenamefont
  {Safranek}, \citenamefont {Ahmadi},\ and\ \citenamefont
  {Fuentes}}]{Safranek_2015}%
  \BibitemOpen
  \bibfield  {author} {\bibinfo {author} {\bibfnamefont {D.}~\bibnamefont
  {Safranek}}, \bibinfo {author} {\bibfnamefont {M.}~\bibnamefont {Ahmadi}},\
  and\ \bibinfo {author} {\bibfnamefont {I.}~\bibnamefont {Fuentes}},\ }\href
  {https://doi.org/10.1088/1367-2630/17/3/033012} {\bibfield  {journal}
  {\bibinfo  {journal} {New Journal of Physics}\ }\textbf {\bibinfo {volume}
  {17}},\ \bibinfo {pages} {033012} (\bibinfo {year} {2015})}\BibitemShut
  {NoStop}%
\bibitem [{\citenamefont {Koochakie}\ \emph {et~al.}(2021)\citenamefont
  {Koochakie}, \citenamefont {Jannesary},\ and\ \citenamefont
  {Karimipour}}]{Koochakie2021}%
  \BibitemOpen
  \bibfield  {author} {\bibinfo {author} {\bibfnamefont {M.~M.~R.}\
  \bibnamefont {Koochakie}}, \bibinfo {author} {\bibfnamefont {V.}~\bibnamefont
  {Jannesary}},\ and\ \bibinfo {author} {\bibfnamefont {V.}~\bibnamefont
  {Karimipour}},\ }\href {https://doi.org/10.1007/s11128-021-03243-5}
  {\bibfield  {journal} {\bibinfo  {journal} {Quantum Information Processing}\
  }\textbf {\bibinfo {volume} {20}},\ \bibinfo {pages} {329} (\bibinfo {year}
  {2021})}\BibitemShut {NoStop}%
\bibitem [{\citenamefont {Aiello}\ \emph {et~al.}(2007)\citenamefont {Aiello},
  \citenamefont {Puentes},\ and\ \citenamefont
  {Woerdman}}]{PhysRevA.76.032323}%
  \BibitemOpen
  \bibfield  {author} {\bibinfo {author} {\bibfnamefont {A.}~\bibnamefont
  {Aiello}}, \bibinfo {author} {\bibfnamefont {G.}~\bibnamefont {Puentes}},\
  and\ \bibinfo {author} {\bibfnamefont {J.~P.}\ \bibnamefont {Woerdman}},\
  }\href {https://doi.org/10.1103/PhysRevA.76.032323} {\bibfield  {journal}
  {\bibinfo  {journal} {Phys. Rev. A}\ }\textbf {\bibinfo {volume} {76}},\
  \bibinfo {pages} {032323} (\bibinfo {year} {2007})}\BibitemShut {NoStop}%
\bibitem [{\citenamefont {Yoon}\ \emph {et~al.}(2020)\citenamefont {Yoon},
  \citenamefont {Lee}, \citenamefont {Rockstuhl}, \citenamefont {Lee},\ and\
  \citenamefont {Lee}}]{Yoon_2020}%
  \BibitemOpen
  \bibfield  {author} {\bibinfo {author} {\bibfnamefont {S.-J.}\ \bibnamefont
  {Yoon}}, \bibinfo {author} {\bibfnamefont {J.-S.}\ \bibnamefont {Lee}},
  \bibinfo {author} {\bibfnamefont {C.}~\bibnamefont {Rockstuhl}}, \bibinfo
  {author} {\bibfnamefont {C.}~\bibnamefont {Lee}},\ and\ \bibinfo {author}
  {\bibfnamefont {K.-G.}\ \bibnamefont {Lee}},\ }\href
  {https://doi.org/10.1088/1681-7575/ab8801} {\bibfield  {journal} {\bibinfo
  {journal} {Metrologia}\ }\textbf {\bibinfo {volume} {57}},\ \bibinfo {pages}
  {045008} (\bibinfo {year} {2020})}\BibitemShut {NoStop}%
\bibitem [{\citenamefont {Rudnicki}\ \emph {et~al.}(2020)\citenamefont
  {Rudnicki}, \citenamefont {S\'{a}nchez-Soto}, \citenamefont {Leuchs},\ and\
  \citenamefont {Boyd}}]{Rudnicki:20}%
  \BibitemOpen
  \bibfield  {author} {\bibinfo {author} {\bibfnamefont {{\L}.}~\bibnamefont
  {Rudnicki}}, \bibinfo {author} {\bibfnamefont {L.~L.}\ \bibnamefont
  {S\'{a}nchez-Soto}}, \bibinfo {author} {\bibfnamefont {G.}~\bibnamefont
  {Leuchs}},\ and\ \bibinfo {author} {\bibfnamefont {R.~W.}\ \bibnamefont
  {Boyd}},\ }\href {https://doi.org/10.1364/OL.392955} {\bibfield  {journal}
  {\bibinfo  {journal} {Opt. Lett.}\ }\textbf {\bibinfo {volume} {45}},\
  \bibinfo {pages} {4607} (\bibinfo {year} {2020})}\BibitemShut {NoStop}%
\bibitem [{\citenamefont {Jarzyna}(2021)}]{jarzyna2021quantum}%
  \BibitemOpen
  \bibfield  {author} {\bibinfo {author} {\bibfnamefont {M.}~\bibnamefont
  {Jarzyna}},\ }\href@noop {} {\bibinfo {title} {Quantum limits to polarization
  measurement of classical light}} (\bibinfo {year} {2021}),\ \Eprint
  {https://arxiv.org/abs/2112.05578} {arXiv:2112.05578 [quant-ph]} \BibitemShut
  {NoStop}%
\bibitem [{\citenamefont {Hannonen}\ \emph {et~al.}(2020)\citenamefont
  {Hannonen}, \citenamefont {Hoenders}, \citenamefont {Els\"{a}sser},
  \citenamefont {Friberg},\ and\ \citenamefont {Set\"{a}l\"{a}}}]{Hannonen:20}%
  \BibitemOpen
  \bibfield  {author} {\bibinfo {author} {\bibfnamefont {A.}~\bibnamefont
  {Hannonen}}, \bibinfo {author} {\bibfnamefont {B.~J.}\ \bibnamefont
  {Hoenders}}, \bibinfo {author} {\bibfnamefont {W.}~\bibnamefont
  {Els\"{a}sser}}, \bibinfo {author} {\bibfnamefont {A.~T.}\ \bibnamefont
  {Friberg}},\ and\ \bibinfo {author} {\bibfnamefont {T.}~\bibnamefont
  {Set\"{a}l\"{a}}},\ }\href {https://doi.org/10.1364/JOSAA.385851} {\bibfield
  {journal} {\bibinfo  {journal} {J. Opt. Soc. Am. A}\ }\textbf {\bibinfo
  {volume} {37}},\ \bibinfo {pages} {714} (\bibinfo {year} {2020})}\BibitemShut
  {NoStop}%
\bibitem [{\citenamefont {Magnitskiy}\ \emph {et~al.}(2020)\citenamefont
  {Magnitskiy}, \citenamefont {Agapov},\ and\ \citenamefont
  {Chirkin}}]{Magnitskiy:20}%
  \BibitemOpen
  \bibfield  {author} {\bibinfo {author} {\bibfnamefont {S.}~\bibnamefont
  {Magnitskiy}}, \bibinfo {author} {\bibfnamefont {D.}~\bibnamefont {Agapov}},\
  and\ \bibinfo {author} {\bibfnamefont {A.}~\bibnamefont {Chirkin}},\ }\href
  {https://doi.org/10.1364/OL.387234} {\bibfield  {journal} {\bibinfo
  {journal} {Opt. Lett.}\ }\textbf {\bibinfo {volume} {45}},\ \bibinfo {pages}
  {3641} (\bibinfo {year} {2020})}\BibitemShut {NoStop}%
\bibitem [{\citenamefont {Magnitskiy}\ \emph {et~al.}(2022)\citenamefont
  {Magnitskiy}, \citenamefont {Agapov},\ and\ \citenamefont
  {Chirkin}}]{Magnitskiy:22}%
  \BibitemOpen
  \bibfield  {author} {\bibinfo {author} {\bibfnamefont {S.}~\bibnamefont
  {Magnitskiy}}, \bibinfo {author} {\bibfnamefont {D.}~\bibnamefont {Agapov}},\
  and\ \bibinfo {author} {\bibfnamefont {A.}~\bibnamefont {Chirkin}},\ }\href
  {https://doi.org/10.1364/OL.450206} {\bibfield  {journal} {\bibinfo
  {journal} {Opt. Lett.}\ }\textbf {\bibinfo {volume} {47}},\ \bibinfo {pages}
  {754} (\bibinfo {year} {2022})}\BibitemShut {NoStop}%
\bibitem [{\citenamefont {Restuccia}\ \emph {et~al.}(2022)\citenamefont
  {Restuccia}, \citenamefont {Gibson}, \citenamefont {Cronin},\ and\
  \citenamefont {Padgett}}]{Restuccia:2022}%
  \BibitemOpen
  \bibfield  {author} {\bibinfo {author} {\bibfnamefont {S.}~\bibnamefont
  {Restuccia}}, \bibinfo {author} {\bibfnamefont {G.~M.}\ \bibnamefont
  {Gibson}}, \bibinfo {author} {\bibfnamefont {L.}~\bibnamefont {Cronin}},\
  and\ \bibinfo {author} {\bibfnamefont {M.~J.}\ \bibnamefont {Padgett}},\
  }\href@noop {} {\bibfield  {journal} {\bibinfo  {journal} {Phys. Rev. A}\
  }\textbf {\bibinfo {volume} {106}},\ \bibinfo {pages} {062601} (\bibinfo
  {year} {2022})}\BibitemShut {NoStop}%
\bibitem [{\citenamefont {Vega}\ \emph {et~al.}(2021)\citenamefont {Vega},
  \citenamefont {Pertsch}, \citenamefont {Setzpfandt},\ and\ \citenamefont
  {Sukhorukov}}]{Vega:2021}%
  \BibitemOpen
  \bibfield  {author} {\bibinfo {author} {\bibfnamefont {A.}~\bibnamefont
  {Vega}}, \bibinfo {author} {\bibfnamefont {T.}~\bibnamefont {Pertsch}},
  \bibinfo {author} {\bibfnamefont {F.}~\bibnamefont {Setzpfandt}},\ and\
  \bibinfo {author} {\bibfnamefont {A.~A.}\ \bibnamefont {Sukhorukov}},\
  }\href@noop {} {\bibfield  {journal} {\bibinfo  {journal} {Phys. Rev. Appl.}\
  }\textbf {\bibinfo {volume} {16}},\ \bibinfo {pages} {064032} (\bibinfo
  {year} {2021})}\BibitemShut {NoStop}%
\bibitem [{\citenamefont {Pedram}\ \emph {et~al.}(2024)\citenamefont {Pedram},
  \citenamefont {Besaga}, \citenamefont {Setzpfandt},\ and\ \citenamefont
  {M\"{u}stecapl{\i}o\u{g}lu}}]{https://doi.org/10.1002/qute.202400059}%
  \BibitemOpen
  \bibfield  {author} {\bibinfo {author} {\bibfnamefont {A.}~\bibnamefont
  {Pedram}}, \bibinfo {author} {\bibfnamefont {V.~R.}\ \bibnamefont {Besaga}},
  \bibinfo {author} {\bibfnamefont {F.}~\bibnamefont {Setzpfandt}},\ and\
  \bibinfo {author} {\bibfnamefont {O.~E.}\ \bibnamefont
  {M\"{u}stecapl{\i}o\u{g}lu}},\ }\href
  {https://doi.org/https://doi.org/10.1002/qute.202400059} {\bibfield
  {journal} {\bibinfo  {journal} {Advanced Quantum Technologies}\ }\textbf
  {\bibinfo {volume} {7}},\ \bibinfo {pages} {2400059} (\bibinfo {year}
  {2024})},\ \Eprint
  {https://arxiv.org/abs/https://onlinelibrary.wiley.com/doi/pdf/10.1002/qute.202400059}
  {https://onlinelibrary.wiley.com/doi/pdf/10.1002/qute.202400059} \BibitemShut
  {NoStop}%
\bibitem [{\citenamefont {Goldberg}(2020)}]{PhysRevResearch.2.023038}%
  \BibitemOpen
  \bibfield  {author} {\bibinfo {author} {\bibfnamefont {A.~Z.}\ \bibnamefont
  {Goldberg}},\ }\href {https://doi.org/10.1103/PhysRevResearch.2.023038}
  {\bibfield  {journal} {\bibinfo  {journal} {Phys. Rev. Res.}\ }\textbf
  {\bibinfo {volume} {2}},\ \bibinfo {pages} {023038} (\bibinfo {year}
  {2020})}\BibitemShut {NoStop}%
\bibitem [{\citenamefont {Goldberg}\ \emph
  {et~al.}(2021{\natexlab{a}})\citenamefont {Goldberg}, \citenamefont {de~la
  Hoz}, \citenamefont {Bj\"{o}rk}, \citenamefont {Klimov}, \citenamefont
  {Grassl}, \citenamefont {Leuchs},\ and\ \citenamefont
  {S\'{a}nchez-Soto}}]{Goldberg:21}%
  \BibitemOpen
  \bibfield  {author} {\bibinfo {author} {\bibfnamefont {A.~Z.}\ \bibnamefont
  {Goldberg}}, \bibinfo {author} {\bibfnamefont {P.}~\bibnamefont {de~la Hoz}},
  \bibinfo {author} {\bibfnamefont {G.}~\bibnamefont {Bj\"{o}rk}}, \bibinfo
  {author} {\bibfnamefont {A.~B.}\ \bibnamefont {Klimov}}, \bibinfo {author}
  {\bibfnamefont {M.}~\bibnamefont {Grassl}}, \bibinfo {author} {\bibfnamefont
  {G.}~\bibnamefont {Leuchs}},\ and\ \bibinfo {author} {\bibfnamefont {L.~L.}\
  \bibnamefont {S\'{a}nchez-Soto}},\ }\href
  {https://doi.org/10.1364/AOP.404175} {\bibfield  {journal} {\bibinfo
  {journal} {Adv. Opt. Photon.}\ }\textbf {\bibinfo {volume} {13}},\ \bibinfo
  {pages} {1} (\bibinfo {year} {2021}{\natexlab{a}})}\BibitemShut {NoStop}%
\bibitem [{\citenamefont {Goldberg}(2021)}]{Goldberg_thesis}%
  \BibitemOpen
  \bibfield  {author} {\bibinfo {author} {\bibfnamefont {A.~Z.}\ \bibnamefont
  {Goldberg}},\ }\emph {\bibinfo {title} {Disquisitions on Quantum-Enhanced
  Polarimetry}},\ \href@noop {} {Ph.D. thesis} (\bibinfo {year}
  {2021})\BibitemShut {NoStop}%
\bibitem [{\citenamefont {Goldberg}(2022)}]{GOLDBERG2022185}%
  \BibitemOpen
  \bibfield  {author} {\bibinfo {author} {\bibfnamefont {A.~Z.}\ \bibnamefont
  {Goldberg}}\ }(\bibinfo  {publisher} {Elsevier},\ \bibinfo {year} {2022})\
  pp.\ \bibinfo {pages} {185--274}\BibitemShut {NoStop}%
\bibitem [{\citenamefont {Chryssomalakos}\ and\ \citenamefont
  {Hern\'andez-Coronado}(2017)}]{PhysRevA.95.052125}%
  \BibitemOpen
  \bibfield  {author} {\bibinfo {author} {\bibfnamefont {C.}~\bibnamefont
  {Chryssomalakos}}\ and\ \bibinfo {author} {\bibfnamefont {H.}~\bibnamefont
  {Hern\'andez-Coronado}},\ }\href {https://doi.org/10.1103/PhysRevA.95.052125}
  {\bibfield  {journal} {\bibinfo  {journal} {Phys. Rev. A}\ }\textbf {\bibinfo
  {volume} {95}},\ \bibinfo {pages} {052125} (\bibinfo {year}
  {2017})}\BibitemShut {NoStop}%
\bibitem [{\citenamefont {Bouchard}\ \emph {et~al.}(2017)\citenamefont
  {Bouchard}, \citenamefont {de~la Hoz}, \citenamefont {Bj\"{o}rk},
  \citenamefont {Boyd}, \citenamefont {Grassl}, \citenamefont {Hradil},
  \citenamefont {Karimi}, \citenamefont {Klimov}, \citenamefont {Leuchs},
  \citenamefont {\v{R}eh\'{a}\v{c}ek},\ and\ \citenamefont
  {S\'{a}nchez-Soto}}]{Bouchard:17}%
  \BibitemOpen
  \bibfield  {author} {\bibinfo {author} {\bibfnamefont {F.}~\bibnamefont
  {Bouchard}}, \bibinfo {author} {\bibfnamefont {P.}~\bibnamefont {de~la Hoz}},
  \bibinfo {author} {\bibfnamefont {G.}~\bibnamefont {Bj\"{o}rk}}, \bibinfo
  {author} {\bibfnamefont {R.~W.}\ \bibnamefont {Boyd}}, \bibinfo {author}
  {\bibfnamefont {M.}~\bibnamefont {Grassl}}, \bibinfo {author} {\bibfnamefont
  {Z.}~\bibnamefont {Hradil}}, \bibinfo {author} {\bibfnamefont
  {E.}~\bibnamefont {Karimi}}, \bibinfo {author} {\bibfnamefont {A.~B.}\
  \bibnamefont {Klimov}}, \bibinfo {author} {\bibfnamefont {G.}~\bibnamefont
  {Leuchs}}, \bibinfo {author} {\bibfnamefont {J.}~\bibnamefont
  {\v{R}eh\'{a}\v{c}ek}},\ and\ \bibinfo {author} {\bibfnamefont {L.~L.}\
  \bibnamefont {S\'{a}nchez-Soto}},\ }\href
  {https://doi.org/10.1364/OPTICA.4.001429} {\bibfield  {journal} {\bibinfo
  {journal} {Optica}\ }\textbf {\bibinfo {volume} {4}},\ \bibinfo {pages}
  {1429} (\bibinfo {year} {2017})}\BibitemShut {NoStop}%
\bibitem [{\citenamefont {Goldberg}\ and\ \citenamefont
  {James}(2018)}]{PhysRevA.98.032113}%
  \BibitemOpen
  \bibfield  {author} {\bibinfo {author} {\bibfnamefont {A.~Z.}\ \bibnamefont
  {Goldberg}}\ and\ \bibinfo {author} {\bibfnamefont {D.~F.~V.}\ \bibnamefont
  {James}},\ }\href {https://doi.org/10.1103/PhysRevA.98.032113} {\bibfield
  {journal} {\bibinfo  {journal} {Phys. Rev. A}\ }\textbf {\bibinfo {volume}
  {98}},\ \bibinfo {pages} {032113} (\bibinfo {year} {2018})}\BibitemShut
  {NoStop}%
\bibitem [{\citenamefont {Martin}\ \emph {et~al.}(2020)\citenamefont {Martin},
  \citenamefont {Weigert},\ and\ \citenamefont
  {Giraud}}]{Martin2020optimaldetectionof}%
  \BibitemOpen
  \bibfield  {author} {\bibinfo {author} {\bibfnamefont {J.}~\bibnamefont
  {Martin}}, \bibinfo {author} {\bibfnamefont {S.}~\bibnamefont {Weigert}},\
  and\ \bibinfo {author} {\bibfnamefont {O.}~\bibnamefont {Giraud}},\ }\href
  {https://doi.org/10.22331/q-2020-06-22-285} {\bibfield  {journal} {\bibinfo
  {journal} {{Quantum}}\ }\textbf {\bibinfo {volume} {4}},\ \bibinfo {pages}
  {285} (\bibinfo {year} {2020})}\BibitemShut {NoStop}%
\bibitem [{\citenamefont {Goldberg}\ \emph
  {et~al.}(2021{\natexlab{b}})\citenamefont {Goldberg}, \citenamefont {Klimov},
  \citenamefont {Leuchs},\ and\ \citenamefont
  {Sánchez-Soto}}]{ZGoldberg_2021}%
  \BibitemOpen
  \bibfield  {author} {\bibinfo {author} {\bibfnamefont {A.~Z.}\ \bibnamefont
  {Goldberg}}, \bibinfo {author} {\bibfnamefont {A.~B.}\ \bibnamefont
  {Klimov}}, \bibinfo {author} {\bibfnamefont {G.}~\bibnamefont {Leuchs}},\
  and\ \bibinfo {author} {\bibfnamefont {L.~L.}\ \bibnamefont
  {Sánchez-Soto}},\ }\href {https://doi.org/10.1088/2515-7647/abeb54}
  {\bibfield  {journal} {\bibinfo  {journal} {Journal of Physics: Photonics}\
  }\textbf {\bibinfo {volume} {3}},\ \bibinfo {pages} {022008} (\bibinfo {year}
  {2021}{\natexlab{b}})}\BibitemShut {NoStop}%
\bibitem [{\citenamefont {Eriksson}\ \emph {et~al.}(2023)\citenamefont
  {Eriksson}, \citenamefont {Goldberg}, \citenamefont {Hiekkam\"aki},
  \citenamefont {Bouchard}, \citenamefont {Rehacek}, \citenamefont {Hradil},
  \citenamefont {Leuchs}, \citenamefont {Fickler},\ and\ \citenamefont
  {S\'anchez-Soto}}]{PhysRevApplied.20.024052}%
  \BibitemOpen
  \bibfield  {author} {\bibinfo {author} {\bibfnamefont {M.}~\bibnamefont
  {Eriksson}}, \bibinfo {author} {\bibfnamefont {A.}~\bibnamefont {Goldberg}},
  \bibinfo {author} {\bibfnamefont {M.}~\bibnamefont {Hiekkam\"aki}}, \bibinfo
  {author} {\bibfnamefont {F.}~\bibnamefont {Bouchard}}, \bibinfo {author}
  {\bibfnamefont {J.}~\bibnamefont {Rehacek}}, \bibinfo {author} {\bibfnamefont
  {Z.}~\bibnamefont {Hradil}}, \bibinfo {author} {\bibfnamefont
  {G.}~\bibnamefont {Leuchs}}, \bibinfo {author} {\bibfnamefont
  {R.}~\bibnamefont {Fickler}},\ and\ \bibinfo {author} {\bibfnamefont
  {L.}~\bibnamefont {S\'anchez-Soto}},\ }\href
  {https://doi.org/10.1103/PhysRevApplied.20.024052} {\bibfield  {journal}
  {\bibinfo  {journal} {Phys. Rev. Appl.}\ }\textbf {\bibinfo {volume} {20}},\
  \bibinfo {pages} {024052} (\bibinfo {year} {2023})}\BibitemShut {NoStop}%
\bibitem [{\citenamefont {Adesso}\ \emph {et~al.}(2009)\citenamefont {Adesso},
  \citenamefont {Dell'Anno}, \citenamefont {De~Siena}, \citenamefont
  {Illuminati},\ and\ \citenamefont {Souza}}]{PhysRevA.79.040305}%
  \BibitemOpen
  \bibfield  {author} {\bibinfo {author} {\bibfnamefont {G.}~\bibnamefont
  {Adesso}}, \bibinfo {author} {\bibfnamefont {F.}~\bibnamefont {Dell'Anno}},
  \bibinfo {author} {\bibfnamefont {S.}~\bibnamefont {De~Siena}}, \bibinfo
  {author} {\bibfnamefont {F.}~\bibnamefont {Illuminati}},\ and\ \bibinfo
  {author} {\bibfnamefont {L.~A.~M.}\ \bibnamefont {Souza}},\ }\href
  {https://doi.org/10.1103/PhysRevA.79.040305} {\bibfield  {journal} {\bibinfo
  {journal} {Phys. Rev. A}\ }\textbf {\bibinfo {volume} {79}},\ \bibinfo
  {pages} {040305} (\bibinfo {year} {2009})}\BibitemShut {NoStop}%
\bibitem [{\citenamefont {Alipour}\ \emph {et~al.}(2014)\citenamefont
  {Alipour}, \citenamefont {Mehboudi},\ and\ \citenamefont
  {Rezakhani}}]{PhysRevLett.112.120405}%
  \BibitemOpen
  \bibfield  {author} {\bibinfo {author} {\bibfnamefont {S.}~\bibnamefont
  {Alipour}}, \bibinfo {author} {\bibfnamefont {M.}~\bibnamefont {Mehboudi}},\
  and\ \bibinfo {author} {\bibfnamefont {A.~T.}\ \bibnamefont {Rezakhani}},\
  }\href {https://doi.org/10.1103/PhysRevLett.112.120405} {\bibfield  {journal}
  {\bibinfo  {journal} {Phys. Rev. Lett.}\ }\textbf {\bibinfo {volume} {112}},\
  \bibinfo {pages} {120405} (\bibinfo {year} {2014})}\BibitemShut {NoStop}%
\bibitem [{\citenamefont {Brambilla}\ \emph {et~al.}(2008)\citenamefont
  {Brambilla}, \citenamefont {Caspani}, \citenamefont {Jedrkiewicz},
  \citenamefont {Lugiato},\ and\ \citenamefont {Gatti}}]{PhysRevA.77.053807}%
  \BibitemOpen
  \bibfield  {author} {\bibinfo {author} {\bibfnamefont {E.}~\bibnamefont
  {Brambilla}}, \bibinfo {author} {\bibfnamefont {L.}~\bibnamefont {Caspani}},
  \bibinfo {author} {\bibfnamefont {O.}~\bibnamefont {Jedrkiewicz}}, \bibinfo
  {author} {\bibfnamefont {L.~A.}\ \bibnamefont {Lugiato}},\ and\ \bibinfo
  {author} {\bibfnamefont {A.}~\bibnamefont {Gatti}},\ }\href
  {https://doi.org/10.1103/PhysRevA.77.053807} {\bibfield  {journal} {\bibinfo
  {journal} {Phys. Rev. A}\ }\textbf {\bibinfo {volume} {77}},\ \bibinfo
  {pages} {053807} (\bibinfo {year} {2008})}\BibitemShut {NoStop}%
\bibitem [{\citenamefont {Crowley}\ \emph {et~al.}(2014)\citenamefont
  {Crowley}, \citenamefont {Datta}, \citenamefont {Barbieri},\ and\
  \citenamefont {Walmsley}}]{PhysRevA.89.023845}%
  \BibitemOpen
  \bibfield  {author} {\bibinfo {author} {\bibfnamefont {P.~J.~D.}\
  \bibnamefont {Crowley}}, \bibinfo {author} {\bibfnamefont {A.}~\bibnamefont
  {Datta}}, \bibinfo {author} {\bibfnamefont {M.}~\bibnamefont {Barbieri}},\
  and\ \bibinfo {author} {\bibfnamefont {I.~A.}\ \bibnamefont {Walmsley}},\
  }\href {https://doi.org/10.1103/PhysRevA.89.023845} {\bibfield  {journal}
  {\bibinfo  {journal} {Phys. Rev. A}\ }\textbf {\bibinfo {volume} {89}},\
  \bibinfo {pages} {023845} (\bibinfo {year} {2014})}\BibitemShut {NoStop}%
\bibitem [{\citenamefont {Hayat}\ \emph {et~al.}(1999)\citenamefont {Hayat},
  \citenamefont {Joobeur},\ and\ \citenamefont {Saleh}}]{Hayat:99}%
  \BibitemOpen
  \bibfield  {author} {\bibinfo {author} {\bibfnamefont {M.~M.}\ \bibnamefont
  {Hayat}}, \bibinfo {author} {\bibfnamefont {A.}~\bibnamefont {Joobeur}},\
  and\ \bibinfo {author} {\bibfnamefont {B.~E.~A.}\ \bibnamefont {Saleh}},\
  }\href {https://doi.org/10.1364/JOSAA.16.000348} {\bibfield  {journal}
  {\bibinfo  {journal} {J. Opt. Soc. Am. A}\ }\textbf {\bibinfo {volume}
  {16}},\ \bibinfo {pages} {348} (\bibinfo {year} {1999})}\BibitemShut
  {NoStop}%
\bibitem [{\citenamefont {Heidmann}\ \emph {et~al.}(1987)\citenamefont
  {Heidmann}, \citenamefont {Horowicz}, \citenamefont {Reynaud}, \citenamefont
  {Giacobino}, \citenamefont {Fabre},\ and\ \citenamefont
  {Camy}}]{PhysRevLett.59.2555}%
  \BibitemOpen
  \bibfield  {author} {\bibinfo {author} {\bibfnamefont {A.}~\bibnamefont
  {Heidmann}}, \bibinfo {author} {\bibfnamefont {R.~J.}\ \bibnamefont
  {Horowicz}}, \bibinfo {author} {\bibfnamefont {S.}~\bibnamefont {Reynaud}},
  \bibinfo {author} {\bibfnamefont {E.}~\bibnamefont {Giacobino}}, \bibinfo
  {author} {\bibfnamefont {C.}~\bibnamefont {Fabre}},\ and\ \bibinfo {author}
  {\bibfnamefont {G.}~\bibnamefont {Camy}},\ }\href
  {https://doi.org/10.1103/PhysRevLett.59.2555} {\bibfield  {journal} {\bibinfo
   {journal} {Phys. Rev. Lett.}\ }\textbf {\bibinfo {volume} {59}},\ \bibinfo
  {pages} {2555} (\bibinfo {year} {1987})}\BibitemShut {NoStop}%
\bibitem [{\citenamefont {Ioannou}\ \emph {et~al.}(2021)\citenamefont
  {Ioannou}, \citenamefont {Nair}, \citenamefont {Fernandez-Corbaton},
  \citenamefont {Gu}, \citenamefont {Rockstuhl},\ and\ \citenamefont
  {Lee}}]{PhysRevA.104.052615}%
  \BibitemOpen
  \bibfield  {author} {\bibinfo {author} {\bibfnamefont {C.}~\bibnamefont
  {Ioannou}}, \bibinfo {author} {\bibfnamefont {R.}~\bibnamefont {Nair}},
  \bibinfo {author} {\bibfnamefont {I.}~\bibnamefont {Fernandez-Corbaton}},
  \bibinfo {author} {\bibfnamefont {M.}~\bibnamefont {Gu}}, \bibinfo {author}
  {\bibfnamefont {C.}~\bibnamefont {Rockstuhl}},\ and\ \bibinfo {author}
  {\bibfnamefont {C.}~\bibnamefont {Lee}},\ }\href
  {https://doi.org/10.1103/PhysRevA.104.052615} {\bibfield  {journal} {\bibinfo
   {journal} {Phys. Rev. A}\ }\textbf {\bibinfo {volume} {104}},\ \bibinfo
  {pages} {052615} (\bibinfo {year} {2021})}\BibitemShut {NoStop}%
\bibitem [{\citenamefont {Jakeman}\ and\ \citenamefont
  {Rarity}(1986)}]{JAKEMAN1986219}%
  \BibitemOpen
  \bibfield  {author} {\bibinfo {author} {\bibfnamefont {E.}~\bibnamefont
  {Jakeman}}\ and\ \bibinfo {author} {\bibfnamefont {J.}~\bibnamefont
  {Rarity}},\ }\href
  {https://doi.org/https://doi.org/10.1016/0030-4018(86)90288-9} {\bibfield
  {journal} {\bibinfo  {journal} {Optics Communications}\ }\textbf {\bibinfo
  {volume} {59}},\ \bibinfo {pages} {219} (\bibinfo {year} {1986})}\BibitemShut
  {NoStop}%
\bibitem [{\citenamefont {Losero}\ \emph {et~al.}(2018)\citenamefont {Losero},
  \citenamefont {Ruo-Berchera}, \citenamefont {Meda}, \citenamefont {Avella},\
  and\ \citenamefont {Genovese}}]{Losero2018}%
  \BibitemOpen
  \bibfield  {author} {\bibinfo {author} {\bibfnamefont {E.}~\bibnamefont
  {Losero}}, \bibinfo {author} {\bibfnamefont {I.}~\bibnamefont
  {Ruo-Berchera}}, \bibinfo {author} {\bibfnamefont {A.}~\bibnamefont {Meda}},
  \bibinfo {author} {\bibfnamefont {A.}~\bibnamefont {Avella}},\ and\ \bibinfo
  {author} {\bibfnamefont {M.}~\bibnamefont {Genovese}},\ }\href
  {https://doi.org/10.1038/s41598-018-25501-w} {\bibfield  {journal} {\bibinfo
  {journal} {Scientific Reports}\ }\textbf {\bibinfo {volume} {8}},\ \bibinfo
  {pages} {7431} (\bibinfo {year} {2018})}\BibitemShut {NoStop}%
\bibitem [{\citenamefont {Monras}\ and\ \citenamefont
  {Paris}(2007)}]{PhysRevLett.98.160401}%
  \BibitemOpen
  \bibfield  {author} {\bibinfo {author} {\bibfnamefont {A.}~\bibnamefont
  {Monras}}\ and\ \bibinfo {author} {\bibfnamefont {M.~G.~A.}\ \bibnamefont
  {Paris}},\ }\href {https://doi.org/10.1103/PhysRevLett.98.160401} {\bibfield
  {journal} {\bibinfo  {journal} {Phys. Rev. Lett.}\ }\textbf {\bibinfo
  {volume} {98}},\ \bibinfo {pages} {160401} (\bibinfo {year}
  {2007})}\BibitemShut {NoStop}%
\bibitem [{\citenamefont {Nair}(2018)}]{PhysRevLett.121.230801}%
  \BibitemOpen
  \bibfield  {author} {\bibinfo {author} {\bibfnamefont {R.}~\bibnamefont
  {Nair}},\ }\href {https://doi.org/10.1103/PhysRevLett.121.230801} {\bibfield
  {journal} {\bibinfo  {journal} {Phys. Rev. Lett.}\ }\textbf {\bibinfo
  {volume} {121}},\ \bibinfo {pages} {230801} (\bibinfo {year}
  {2018})}\BibitemShut {NoStop}%
\bibitem [{\citenamefont {Atkinson}\ \emph {et~al.}(2021)\citenamefont
  {Atkinson}, \citenamefont {Allen}, \citenamefont {Ferranti}, \citenamefont
  {McMillan},\ and\ \citenamefont {Matthews}}]{PhysRevApplied.16.044031}%
  \BibitemOpen
  \bibfield  {author} {\bibinfo {author} {\bibfnamefont {G.}~\bibnamefont
  {Atkinson}}, \bibinfo {author} {\bibfnamefont {E.}~\bibnamefont {Allen}},
  \bibinfo {author} {\bibfnamefont {G.}~\bibnamefont {Ferranti}}, \bibinfo
  {author} {\bibfnamefont {A.}~\bibnamefont {McMillan}},\ and\ \bibinfo
  {author} {\bibfnamefont {J.}~\bibnamefont {Matthews}},\ }\href
  {https://doi.org/10.1103/PhysRevApplied.16.044031} {\bibfield  {journal}
  {\bibinfo  {journal} {Phys. Rev. Appl.}\ }\textbf {\bibinfo {volume} {16}},\
  \bibinfo {pages} {044031} (\bibinfo {year} {2021})}\BibitemShut {NoStop}%
\bibitem [{\citenamefont {Sasaki}\ \emph {et~al.}(2002)\citenamefont {Sasaki},
  \citenamefont {Ban},\ and\ \citenamefont {Barnett}}]{PhysRevA.66.022308}%
  \BibitemOpen
  \bibfield  {author} {\bibinfo {author} {\bibfnamefont {M.}~\bibnamefont
  {Sasaki}}, \bibinfo {author} {\bibfnamefont {M.}~\bibnamefont {Ban}},\ and\
  \bibinfo {author} {\bibfnamefont {S.~M.}\ \bibnamefont {Barnett}},\ }\href
  {https://doi.org/10.1103/PhysRevA.66.022308} {\bibfield  {journal} {\bibinfo
  {journal} {Phys. Rev. A}\ }\textbf {\bibinfo {volume} {66}},\ \bibinfo
  {pages} {022308} (\bibinfo {year} {2002})}\BibitemShut {NoStop}%
\bibitem [{\citenamefont {Collins}\ and\ \citenamefont
  {Stephens}(2015)}]{PhysRevA.92.032324}%
  \BibitemOpen
  \bibfield  {author} {\bibinfo {author} {\bibfnamefont {D.}~\bibnamefont
  {Collins}}\ and\ \bibinfo {author} {\bibfnamefont {J.}~\bibnamefont
  {Stephens}},\ }\href {https://doi.org/10.1103/PhysRevA.92.032324} {\bibfield
  {journal} {\bibinfo  {journal} {Phys. Rev. A}\ }\textbf {\bibinfo {volume}
  {92}},\ \bibinfo {pages} {032324} (\bibinfo {year} {2015})}\BibitemShut
  {NoStop}%
\bibitem [{\citenamefont {Frey}\ \emph {et~al.}(2011)\citenamefont {Frey},
  \citenamefont {Collins},\ and\ \citenamefont {Gerlach}}]{Frey_2011}%
  \BibitemOpen
  \bibfield  {author} {\bibinfo {author} {\bibfnamefont {M.}~\bibnamefont
  {Frey}}, \bibinfo {author} {\bibfnamefont {D.}~\bibnamefont {Collins}},\ and\
  \bibinfo {author} {\bibfnamefont {K.}~\bibnamefont {Gerlach}},\ }\href
  {https://doi.org/10.1088/1751-8113/44/20/205306} {\bibfield  {journal}
  {\bibinfo  {journal} {Journal of Physics A: Mathematical and Theoretical}\
  }\textbf {\bibinfo {volume} {44}},\ \bibinfo {pages} {205306} (\bibinfo
  {year} {2011})}\BibitemShut {NoStop}%
\bibitem [{\citenamefont {Hotta}\ \emph {et~al.}(2005)\citenamefont {Hotta},
  \citenamefont {Karasawa},\ and\ \citenamefont {Ozawa}}]{PhysRevA.72.052334}%
  \BibitemOpen
  \bibfield  {author} {\bibinfo {author} {\bibfnamefont {M.}~\bibnamefont
  {Hotta}}, \bibinfo {author} {\bibfnamefont {T.}~\bibnamefont {Karasawa}},\
  and\ \bibinfo {author} {\bibfnamefont {M.}~\bibnamefont {Ozawa}},\ }\href
  {https://doi.org/10.1103/PhysRevA.72.052334} {\bibfield  {journal} {\bibinfo
  {journal} {Phys. Rev. A}\ }\textbf {\bibinfo {volume} {72}},\ \bibinfo
  {pages} {052334} (\bibinfo {year} {2005})}\BibitemShut {NoStop}%
\bibitem [{\citenamefont {Ji}\ \emph {et~al.}(2008)\citenamefont {Ji},
  \citenamefont {Wang}, \citenamefont {Duan}, \citenamefont {Feng},\ and\
  \citenamefont {Ying}}]{4655455}%
  \BibitemOpen
  \bibfield  {author} {\bibinfo {author} {\bibfnamefont {Z.}~\bibnamefont
  {Ji}}, \bibinfo {author} {\bibfnamefont {G.}~\bibnamefont {Wang}}, \bibinfo
  {author} {\bibfnamefont {R.}~\bibnamefont {Duan}}, \bibinfo {author}
  {\bibfnamefont {Y.}~\bibnamefont {Feng}},\ and\ \bibinfo {author}
  {\bibfnamefont {M.}~\bibnamefont {Ying}},\ }\href
  {https://doi.org/10.1109/TIT.2008.929940} {\bibfield  {journal} {\bibinfo
  {journal} {IEEE Transactions on Information Theory}\ }\textbf {\bibinfo
  {volume} {54}},\ \bibinfo {pages} {5172} (\bibinfo {year}
  {2008})}\BibitemShut {NoStop}%
\bibitem [{\citenamefont {Kessler}\ \emph {et~al.}(2014)\citenamefont
  {Kessler}, \citenamefont {Lovchinsky}, \citenamefont {Sushkov},\ and\
  \citenamefont {Lukin}}]{PhysRevLett.112.150802}%
  \BibitemOpen
  \bibfield  {author} {\bibinfo {author} {\bibfnamefont {E.~M.}\ \bibnamefont
  {Kessler}}, \bibinfo {author} {\bibfnamefont {I.}~\bibnamefont {Lovchinsky}},
  \bibinfo {author} {\bibfnamefont {A.~O.}\ \bibnamefont {Sushkov}},\ and\
  \bibinfo {author} {\bibfnamefont {M.~D.}\ \bibnamefont {Lukin}},\ }\href
  {https://doi.org/10.1103/PhysRevLett.112.150802} {\bibfield  {journal}
  {\bibinfo  {journal} {Phys. Rev. Lett.}\ }\textbf {\bibinfo {volume} {112}},\
  \bibinfo {pages} {150802} (\bibinfo {year} {2014})}\BibitemShut {NoStop}%
\bibitem [{\citenamefont {Zhou}\ \emph {et~al.}(2018)\citenamefont {Zhou},
  \citenamefont {Zhang}, \citenamefont {Preskill},\ and\ \citenamefont
  {Jiang}}]{Zhou2018}%
  \BibitemOpen
  \bibfield  {author} {\bibinfo {author} {\bibfnamefont {S.}~\bibnamefont
  {Zhou}}, \bibinfo {author} {\bibfnamefont {M.}~\bibnamefont {Zhang}},
  \bibinfo {author} {\bibfnamefont {J.}~\bibnamefont {Preskill}},\ and\
  \bibinfo {author} {\bibfnamefont {L.}~\bibnamefont {Jiang}},\ }\href
  {https://doi.org/10.1038/s41467-017-02510-3} {\bibfield  {journal} {\bibinfo
  {journal} {Nature Communications}\ }\textbf {\bibinfo {volume} {9}},\
  \bibinfo {pages} {78} (\bibinfo {year} {2018})}\BibitemShut {NoStop}%
\bibitem [{\citenamefont {Albarelli}\ \emph {et~al.}(2018)\citenamefont
  {Albarelli}, \citenamefont {Rossi}, \citenamefont {Tamascelli},\ and\
  \citenamefont {Genoni}}]{Albarelli2018restoringheisenberg}%
  \BibitemOpen
  \bibfield  {author} {\bibinfo {author} {\bibfnamefont {F.}~\bibnamefont
  {Albarelli}}, \bibinfo {author} {\bibfnamefont {M.~A.~C.}\ \bibnamefont
  {Rossi}}, \bibinfo {author} {\bibfnamefont {D.}~\bibnamefont {Tamascelli}},\
  and\ \bibinfo {author} {\bibfnamefont {M.~G.}\ \bibnamefont {Genoni}},\
  }\href {https://doi.org/10.22331/q-2018-12-03-110} {\bibfield  {journal}
  {\bibinfo  {journal} {{Quantum}}\ }\textbf {\bibinfo {volume} {2}},\ \bibinfo
  {pages} {110} (\bibinfo {year} {2018})}\BibitemShut {NoStop}%
\bibitem [{\citenamefont {Tratzmiller}\ \emph {et~al.}(2020)\citenamefont
  {Tratzmiller}, \citenamefont {Chen}, \citenamefont {Schwartz}, \citenamefont
  {Huelga},\ and\ \citenamefont {Plenio}}]{PhysRevA.101.032347}%
  \BibitemOpen
  \bibfield  {author} {\bibinfo {author} {\bibfnamefont {B.}~\bibnamefont
  {Tratzmiller}}, \bibinfo {author} {\bibfnamefont {Q.}~\bibnamefont {Chen}},
  \bibinfo {author} {\bibfnamefont {I.}~\bibnamefont {Schwartz}}, \bibinfo
  {author} {\bibfnamefont {S.~F.}\ \bibnamefont {Huelga}},\ and\ \bibinfo
  {author} {\bibfnamefont {M.~B.}\ \bibnamefont {Plenio}},\ }\href
  {https://doi.org/10.1103/PhysRevA.101.032347} {\bibfield  {journal} {\bibinfo
   {journal} {Phys. Rev. A}\ }\textbf {\bibinfo {volume} {101}},\ \bibinfo
  {pages} {032347} (\bibinfo {year} {2020})}\BibitemShut {NoStop}%
\bibitem [{\citenamefont {Taylor}\ and\ \citenamefont
  {Bowen}(2016)}]{TAYLOR20161}%
  \BibitemOpen
  \bibfield  {author} {\bibinfo {author} {\bibfnamefont {M.~A.}\ \bibnamefont
  {Taylor}}\ and\ \bibinfo {author} {\bibfnamefont {W.~P.}\ \bibnamefont
  {Bowen}},\ }\href
  {https://doi.org/https://doi.org/10.1016/j.physrep.2015.12.002} {\bibfield
  {journal} {\bibinfo  {journal} {Physics Reports}\ }\textbf {\bibinfo {volume}
  {615}},\ \bibinfo {pages} {1} (\bibinfo {year} {2016})},\ \bibinfo {note}
  {quantum metrology and its application in biology}\BibitemShut {NoStop}%
\bibitem [{\citenamefont {Pedram}\ \emph {et~al.}(2022)\citenamefont {Pedram},
  \citenamefont {M\"{u}stecapl{\i}o\u{g}lu},\ and\ \citenamefont
  {Kominis}}]{PhysRevResearch.4.033060}%
  \BibitemOpen
  \bibfield  {author} {\bibinfo {author} {\bibfnamefont {A.}~\bibnamefont
  {Pedram}}, \bibinfo {author} {\bibfnamefont {O.~E.}\ \bibnamefont
  {M\"{u}stecapl{\i}o\u{g}lu}},\ and\ \bibinfo {author} {\bibfnamefont {I.~K.}\
  \bibnamefont {Kominis}},\ }\href
  {https://doi.org/10.1103/PhysRevResearch.4.033060} {\bibfield  {journal}
  {\bibinfo  {journal} {Phys. Rev. Res.}\ }\textbf {\bibinfo {volume} {4}},\
  \bibinfo {pages} {033060} (\bibinfo {year} {2022})}\BibitemShut {NoStop}%
\bibitem [{\citenamefont {Lu}\ and\ \citenamefont {Chipman}(1996)}]{Lu:96}%
  \BibitemOpen
  \bibfield  {author} {\bibinfo {author} {\bibfnamefont {S.-Y.}\ \bibnamefont
  {Lu}}\ and\ \bibinfo {author} {\bibfnamefont {R.~A.}\ \bibnamefont
  {Chipman}},\ }\href {https://doi.org/10.1364/JOSAA.13.001106} {\bibfield
  {journal} {\bibinfo  {journal} {J. Opt. Soc. Am. A}\ }\textbf {\bibinfo
  {volume} {13}},\ \bibinfo {pages} {1106} (\bibinfo {year}
  {1996})}\BibitemShut {NoStop}%
\bibitem [{\citenamefont {Ghosh}\ \emph {et~al.}(2009)\citenamefont {Ghosh},
  \citenamefont {Wood}, \citenamefont {Li}, \citenamefont {Weisel},
  \citenamefont {Wilson}, \citenamefont {Li},\ and\ \citenamefont
  {Vitkin}}]{https://doi.org/10.1002/jbio.200810040}%
  \BibitemOpen
  \bibfield  {author} {\bibinfo {author} {\bibfnamefont {N.}~\bibnamefont
  {Ghosh}}, \bibinfo {author} {\bibfnamefont {M.~F.~G.}\ \bibnamefont {Wood}},
  \bibinfo {author} {\bibfnamefont {S.-h.}\ \bibnamefont {Li}}, \bibinfo
  {author} {\bibfnamefont {R.~D.}\ \bibnamefont {Weisel}}, \bibinfo {author}
  {\bibfnamefont {B.~C.}\ \bibnamefont {Wilson}}, \bibinfo {author}
  {\bibfnamefont {R.-K.}\ \bibnamefont {Li}},\ and\ \bibinfo {author}
  {\bibfnamefont {I.~A.}\ \bibnamefont {Vitkin}},\ }\href
  {https://doi.org/https://doi.org/10.1002/jbio.200810040} {\bibfield
  {journal} {\bibinfo  {journal} {Journal of Biophotonics}\ }\textbf {\bibinfo
  {volume} {2}},\ \bibinfo {pages} {145} (\bibinfo {year} {2009})},\ \Eprint
  {https://arxiv.org/abs/https://onlinelibrary.wiley.com/doi/pdf/10.1002/jbio.200810040}
  {https://onlinelibrary.wiley.com/doi/pdf/10.1002/jbio.200810040} \BibitemShut
  {NoStop}%
\bibitem [{\citenamefont {Gao}(2021)}]{GAO2021106735}%
  \BibitemOpen
  \bibfield  {author} {\bibinfo {author} {\bibfnamefont {W.}~\bibnamefont
  {Gao}},\ }\href
  {https://doi.org/https://doi.org/10.1016/j.optlaseng.2021.106735} {\bibfield
  {journal} {\bibinfo  {journal} {Optics and Lasers in Engineering}\ }\textbf
  {\bibinfo {volume} {147}},\ \bibinfo {pages} {106735} (\bibinfo {year}
  {2021})}\BibitemShut {NoStop}%
\bibitem [{\citenamefont {Shi}\ \emph {et~al.}(2016)\citenamefont {Shi},
  \citenamefont {Galvez},\ and\ \citenamefont {Alfano}}]{Shi:2016}%
  \BibitemOpen
  \bibfield  {author} {\bibinfo {author} {\bibfnamefont {L.}~\bibnamefont
  {Shi}}, \bibinfo {author} {\bibfnamefont {E.~J.}\ \bibnamefont {Galvez}},\
  and\ \bibinfo {author} {\bibfnamefont {R.~R.}\ \bibnamefont {Alfano}},\
  }\href {https://doi.org/10.1038/srep37714} {\bibfield  {journal} {\bibinfo
  {journal} {Scientific Reports}\ }\textbf {\bibinfo {volume} {6}},\ \bibinfo
  {pages} {37714} (\bibinfo {year} {2016})}\BibitemShut {NoStop}%
\bibitem [{\citenamefont {Huang}\ and\ \citenamefont
  {Zeng}(2020)}]{Huang_2020}%
  \BibitemOpen
  \bibfield  {author} {\bibinfo {author} {\bibfnamefont {P.}~\bibnamefont
  {Huang}}\ and\ \bibinfo {author} {\bibfnamefont {G.}~\bibnamefont {Zeng}},\
  }\href {https://doi.org/10.1088/1367-2630/ab82bc} {\bibfield  {journal}
  {\bibinfo  {journal} {New Journal of Physics}\ }\textbf {\bibinfo {volume}
  {22}},\ \bibinfo {pages} {053021} (\bibinfo {year} {2020})}\BibitemShut
  {NoStop}%
\bibitem [{\citenamefont {Lum}\ \emph {et~al.}(2021)\citenamefont {Lum},
  \citenamefont {Mazurek}, \citenamefont {Mikhaylov}, \citenamefont
  {Parzuchowski}, \citenamefont {Wilson}, \citenamefont {Jimenez},
  \citenamefont {Gerrits}, \citenamefont {Stevens}, \citenamefont {Cicerone},\
  and\ \citenamefont {Camp}}]{Lum:21}%
  \BibitemOpen
  \bibfield  {author} {\bibinfo {author} {\bibfnamefont {D.~J.}\ \bibnamefont
  {Lum}}, \bibinfo {author} {\bibfnamefont {M.~D.}\ \bibnamefont {Mazurek}},
  \bibinfo {author} {\bibfnamefont {A.}~\bibnamefont {Mikhaylov}}, \bibinfo
  {author} {\bibfnamefont {K.~M.}\ \bibnamefont {Parzuchowski}}, \bibinfo
  {author} {\bibfnamefont {R.~N.}\ \bibnamefont {Wilson}}, \bibinfo {author}
  {\bibfnamefont {R.}~\bibnamefont {Jimenez}}, \bibinfo {author} {\bibfnamefont
  {T.}~\bibnamefont {Gerrits}}, \bibinfo {author} {\bibfnamefont {M.~J.}\
  \bibnamefont {Stevens}}, \bibinfo {author} {\bibfnamefont {M.~T.}\
  \bibnamefont {Cicerone}},\ and\ \bibinfo {author} {\bibfnamefont {C.~H.}\
  \bibnamefont {Camp}},\ }\href {https://doi.org/10.1364/BOE.423743} {\bibfield
   {journal} {\bibinfo  {journal} {Biomed. Opt. Express}\ }\textbf {\bibinfo
  {volume} {12}},\ \bibinfo {pages} {3658} (\bibinfo {year}
  {2021})}\BibitemShut {NoStop}%
\bibitem [{\citenamefont {Mamani}\ \emph {et~al.}(2018)\citenamefont {Mamani},
  \citenamefont {Shi}, \citenamefont {Ahmed}, \citenamefont {Karnik},
  \citenamefont {Rodríguez-Contreras}, \citenamefont {Nolan},\ and\
  \citenamefont {Alfano}}]{https://doi.org/10.1002/jbio.201800096}%
  \BibitemOpen
  \bibfield  {author} {\bibinfo {author} {\bibfnamefont {S.}~\bibnamefont
  {Mamani}}, \bibinfo {author} {\bibfnamefont {L.}~\bibnamefont {Shi}},
  \bibinfo {author} {\bibfnamefont {T.}~\bibnamefont {Ahmed}}, \bibinfo
  {author} {\bibfnamefont {R.}~\bibnamefont {Karnik}}, \bibinfo {author}
  {\bibfnamefont {A.}~\bibnamefont {Rodríguez-Contreras}}, \bibinfo {author}
  {\bibfnamefont {D.}~\bibnamefont {Nolan}},\ and\ \bibinfo {author}
  {\bibfnamefont {R.}~\bibnamefont {Alfano}},\ }\href
  {https://doi.org/https://doi.org/10.1002/jbio.201800096} {\bibfield
  {journal} {\bibinfo  {journal} {Journal of Biophotonics}\ }\textbf {\bibinfo
  {volume} {11}},\ \bibinfo {pages} {e201800096} (\bibinfo {year} {2018})},\
  \Eprint
  {https://arxiv.org/abs/https://onlinelibrary.wiley.com/doi/pdf/10.1002/jbio.201800096}
  {https://onlinelibrary.wiley.com/doi/pdf/10.1002/jbio.201800096} \BibitemShut
  {NoStop}%
\bibitem [{\citenamefont {Klyshko}(1992)}]{KLYSHKO1992349}%
  \BibitemOpen
  \bibfield  {author} {\bibinfo {author} {\bibfnamefont {D.}~\bibnamefont
  {Klyshko}},\ }\href
  {https://doi.org/https://doi.org/10.1016/0375-9601(92)90837-C} {\bibfield
  {journal} {\bibinfo  {journal} {Physics Letters A}\ }\textbf {\bibinfo
  {volume} {163}},\ \bibinfo {pages} {349} (\bibinfo {year}
  {1992})}\BibitemShut {NoStop}%
\bibitem [{\citenamefont {Alodjants}\ and\ \citenamefont
  {Arakelian}(1999)}]{doi:10.1080/09500349908231279}%
  \BibitemOpen
  \bibfield  {author} {\bibinfo {author} {\bibfnamefont {A.~P.}\ \bibnamefont
  {Alodjants}}\ and\ \bibinfo {author} {\bibfnamefont {S.~M.}\ \bibnamefont
  {Arakelian}},\ }\href {https://doi.org/10.1080/09500349908231279} {\bibfield
  {journal} {\bibinfo  {journal} {Journal of Modern Optics}\ }\textbf {\bibinfo
  {volume} {46}},\ \bibinfo {pages} {475} (\bibinfo {year} {1999})},\ \Eprint
  {https://arxiv.org/abs/https://www.tandfonline.com/doi/pdf/10.1080/09500349908231279}
  {https://www.tandfonline.com/doi/pdf/10.1080/09500349908231279} \BibitemShut
  {NoStop}%
\bibitem [{\citenamefont {Luis}(2002)}]{PhysRevA.66.013806}%
  \BibitemOpen
  \bibfield  {author} {\bibinfo {author} {\bibfnamefont {A.}~\bibnamefont
  {Luis}},\ }\href {https://doi.org/10.1103/PhysRevA.66.013806} {\bibfield
  {journal} {\bibinfo  {journal} {Phys. Rev. A}\ }\textbf {\bibinfo {volume}
  {66}},\ \bibinfo {pages} {013806} (\bibinfo {year} {2002})}\BibitemShut
  {NoStop}%
\bibitem [{\citenamefont {Klimov}\ \emph {et~al.}(2005)\citenamefont {Klimov},
  \citenamefont {S\'anchez-Soto}, \citenamefont {Yustas}, \citenamefont
  {S\"oderholm},\ and\ \citenamefont {Bj\"ork}}]{PhysRevA.72.033813}%
  \BibitemOpen
  \bibfield  {author} {\bibinfo {author} {\bibfnamefont {A.~B.}\ \bibnamefont
  {Klimov}}, \bibinfo {author} {\bibfnamefont {L.~L.}\ \bibnamefont
  {S\'anchez-Soto}}, \bibinfo {author} {\bibfnamefont {E.~C.}\ \bibnamefont
  {Yustas}}, \bibinfo {author} {\bibfnamefont {J.}~\bibnamefont
  {S\"oderholm}},\ and\ \bibinfo {author} {\bibfnamefont {G.}~\bibnamefont
  {Bj\"ork}},\ }\href {https://doi.org/10.1103/PhysRevA.72.033813} {\bibfield
  {journal} {\bibinfo  {journal} {Phys. Rev. A}\ }\textbf {\bibinfo {volume}
  {72}},\ \bibinfo {pages} {033813} (\bibinfo {year} {2005})}\BibitemShut
  {NoStop}%
\bibitem [{\citenamefont {Luis}(2007{\natexlab{a}})}]{LUIS2007173}%
  \BibitemOpen
  \bibfield  {author} {\bibinfo {author} {\bibfnamefont {A.}~\bibnamefont
  {Luis}},\ }\href
  {https://doi.org/https://doi.org/10.1016/j.optcom.2007.01.016} {\bibfield
  {journal} {\bibinfo  {journal} {Optics Communications}\ }\textbf {\bibinfo
  {volume} {273}},\ \bibinfo {pages} {173} (\bibinfo {year}
  {2007}{\natexlab{a}})}\BibitemShut {NoStop}%
\bibitem [{\citenamefont {Klimov}\ \emph {et~al.}(2010)\citenamefont {Klimov},
  \citenamefont {Bj\"ork}, \citenamefont {S\"oderholm}, \citenamefont {Madsen},
  \citenamefont {Lassen}, \citenamefont {Andersen}, \citenamefont {Heersink},
  \citenamefont {Dong}, \citenamefont {Marquardt}, \citenamefont {Leuchs},\
  and\ \citenamefont {S\'anchez-Soto}}]{PhysRevLett.105.153602}%
  \BibitemOpen
  \bibfield  {author} {\bibinfo {author} {\bibfnamefont {A.~B.}\ \bibnamefont
  {Klimov}}, \bibinfo {author} {\bibfnamefont {G.}~\bibnamefont {Bj\"ork}},
  \bibinfo {author} {\bibfnamefont {J.}~\bibnamefont {S\"oderholm}}, \bibinfo
  {author} {\bibfnamefont {L.~S.}\ \bibnamefont {Madsen}}, \bibinfo {author}
  {\bibfnamefont {M.}~\bibnamefont {Lassen}}, \bibinfo {author} {\bibfnamefont
  {U.~L.}\ \bibnamefont {Andersen}}, \bibinfo {author} {\bibfnamefont
  {J.}~\bibnamefont {Heersink}}, \bibinfo {author} {\bibfnamefont
  {R.}~\bibnamefont {Dong}}, \bibinfo {author} {\bibfnamefont {C.}~\bibnamefont
  {Marquardt}}, \bibinfo {author} {\bibfnamefont {G.}~\bibnamefont {Leuchs}},\
  and\ \bibinfo {author} {\bibfnamefont {L.~L.}\ \bibnamefont
  {S\'anchez-Soto}},\ }\href {https://doi.org/10.1103/PhysRevLett.105.153602}
  {\bibfield  {journal} {\bibinfo  {journal} {Phys. Rev. Lett.}\ }\textbf
  {\bibinfo {volume} {105}},\ \bibinfo {pages} {153602} (\bibinfo {year}
  {2010})}\BibitemShut {NoStop}%
\bibitem [{\citenamefont {Björk}\ \emph {et~al.}(2010)\citenamefont {Björk},
  \citenamefont {Söderholm}, \citenamefont {Sánchez-Soto}, \citenamefont
  {Klimov}, \citenamefont {Ghiu}, \citenamefont {Marian},\ and\ \citenamefont
  {Marian}}]{BJORK20104440}%
  \BibitemOpen
  \bibfield  {author} {\bibinfo {author} {\bibfnamefont {G.}~\bibnamefont
  {Björk}}, \bibinfo {author} {\bibfnamefont {J.}~\bibnamefont {Söderholm}},
  \bibinfo {author} {\bibfnamefont {L.~L.}\ \bibnamefont {Sánchez-Soto}},
  \bibinfo {author} {\bibfnamefont {A.~B.}\ \bibnamefont {Klimov}}, \bibinfo
  {author} {\bibfnamefont {I.}~\bibnamefont {Ghiu}}, \bibinfo {author}
  {\bibfnamefont {P.}~\bibnamefont {Marian}},\ and\ \bibinfo {author}
  {\bibfnamefont {T.~A.}\ \bibnamefont {Marian}},\ }\href
  {https://doi.org/https://doi.org/10.1016/j.optcom.2010.04.088} {\bibfield
  {journal} {\bibinfo  {journal} {Optics Communications}\ }\textbf {\bibinfo
  {volume} {283}},\ \bibinfo {pages} {4440} (\bibinfo {year} {2010})},\
  \bibinfo {note} {electromagnetic Coherence and Polarization}\BibitemShut
  {NoStop}%
\bibitem [{\citenamefont {Luis}(2007{\natexlab{b}})}]{PhysRevA.75.053806}%
  \BibitemOpen
  \bibfield  {author} {\bibinfo {author} {\bibfnamefont {A.}~\bibnamefont
  {Luis}},\ }\href {https://doi.org/10.1103/PhysRevA.75.053806} {\bibfield
  {journal} {\bibinfo  {journal} {Phys. Rev. A}\ }\textbf {\bibinfo {volume}
  {75}},\ \bibinfo {pages} {053806} (\bibinfo {year}
  {2007}{\natexlab{b}})}\BibitemShut {NoStop}%
\bibitem [{\citenamefont {Iskhakov}\ \emph {et~al.}(2011)\citenamefont
  {Iskhakov}, \citenamefont {Agafonov}, \citenamefont {Chekhova}, \citenamefont
  {Rytikov},\ and\ \citenamefont {Leuchs}}]{PhysRevA.84.045804}%
  \BibitemOpen
  \bibfield  {author} {\bibinfo {author} {\bibfnamefont {T.~S.}\ \bibnamefont
  {Iskhakov}}, \bibinfo {author} {\bibfnamefont {I.~N.}\ \bibnamefont
  {Agafonov}}, \bibinfo {author} {\bibfnamefont {M.~V.}\ \bibnamefont
  {Chekhova}}, \bibinfo {author} {\bibfnamefont {G.~O.}\ \bibnamefont
  {Rytikov}},\ and\ \bibinfo {author} {\bibfnamefont {G.}~\bibnamefont
  {Leuchs}},\ }\href {https://doi.org/10.1103/PhysRevA.84.045804} {\bibfield
  {journal} {\bibinfo  {journal} {Phys. Rev. A}\ }\textbf {\bibinfo {volume}
  {84}},\ \bibinfo {pages} {045804} (\bibinfo {year} {2011})}\BibitemShut
  {NoStop}%
\bibitem [{\citenamefont {Nogueira}\ \emph {et~al.}(2013)\citenamefont
  {Nogueira}, \citenamefont {Silva}, \citenamefont {Gon{\c{c}}alves},\ and\
  \citenamefont {Vasconcelos}}]{10.1103/PhysRevA.87.043821}%
  \BibitemOpen
  \bibfield  {author} {\bibinfo {author} {\bibfnamefont {K.}~\bibnamefont
  {Nogueira}}, \bibinfo {author} {\bibfnamefont {J.~B.~R.}\ \bibnamefont
  {Silva}}, \bibinfo {author} {\bibfnamefont {J.~R.}\ \bibnamefont
  {Gon{\c{c}}alves}},\ and\ \bibinfo {author} {\bibfnamefont {H.~M.}\
  \bibnamefont {Vasconcelos}},\ }\href
  {https://doi.org/10.1103/PhysRevA.87.043821} {\bibfield  {journal} {\bibinfo
  {journal} {Physical Review A}\ }\textbf {\bibinfo {volume} {87}},\ \bibinfo
  {pages} {043821} (\bibinfo {year} {2013})}\BibitemShut {NoStop}%
\bibitem [{\citenamefont {Kothe}\ \emph {et~al.}(2013)\citenamefont {Kothe},
  \citenamefont {Madsen}, \citenamefont {Andersen},\ and\ \citenamefont
  {Bj\"ork}}]{PhysRevA.87.043814}%
  \BibitemOpen
  \bibfield  {author} {\bibinfo {author} {\bibfnamefont {C.}~\bibnamefont
  {Kothe}}, \bibinfo {author} {\bibfnamefont {L.~S.}\ \bibnamefont {Madsen}},
  \bibinfo {author} {\bibfnamefont {U.~L.}\ \bibnamefont {Andersen}},\ and\
  \bibinfo {author} {\bibfnamefont {G.}~\bibnamefont {Bj\"ork}},\ }\href
  {https://doi.org/10.1103/PhysRevA.87.043814} {\bibfield  {journal} {\bibinfo
  {journal} {Phys. Rev. A}\ }\textbf {\bibinfo {volume} {87}},\ \bibinfo
  {pages} {043814} (\bibinfo {year} {2013})}\BibitemShut {NoStop}%
\bibitem [{\citenamefont {Sánchez-Soto}\ \emph {et~al.}(2013)\citenamefont
  {Sánchez-Soto}, \citenamefont {Klimov}, \citenamefont {de~la Hoz},\ and\
  \citenamefont {Leuchs}}]{Sanchez-Soto_2013}%
  \BibitemOpen
  \bibfield  {author} {\bibinfo {author} {\bibfnamefont {L.~L.}\ \bibnamefont
  {Sánchez-Soto}}, \bibinfo {author} {\bibfnamefont {A.~B.}\ \bibnamefont
  {Klimov}}, \bibinfo {author} {\bibfnamefont {P.}~\bibnamefont {de~la Hoz}},\
  and\ \bibinfo {author} {\bibfnamefont {G.}~\bibnamefont {Leuchs}},\ }\href
  {https://doi.org/10.1088/0953-4075/46/10/104011} {\bibfield  {journal}
  {\bibinfo  {journal} {Journal of Physics B: Atomic, Molecular and Optical
  Physics}\ }\textbf {\bibinfo {volume} {46}},\ \bibinfo {pages} {104011}
  (\bibinfo {year} {2013})}\BibitemShut {NoStop}%
\bibitem [{\citenamefont {Morio}\ and\ \citenamefont
  {Goudail}(2004)}]{Morio:04}%
  \BibitemOpen
  \bibfield  {author} {\bibinfo {author} {\bibfnamefont {J.}~\bibnamefont
  {Morio}}\ and\ \bibinfo {author} {\bibfnamefont {F.}~\bibnamefont
  {Goudail}},\ }\href {https://doi.org/10.1364/OL.29.002234} {\bibfield
  {journal} {\bibinfo  {journal} {Opt. Lett.}\ }\textbf {\bibinfo {volume}
  {29}},\ \bibinfo {pages} {2234} (\bibinfo {year} {2004})}\BibitemShut
  {NoStop}%
\bibitem [{\citenamefont {Gil}\ \emph {et~al.}(2013)\citenamefont {Gil},
  \citenamefont {Jos\'{e}},\ and\ \citenamefont {Ossikovski}}]{Gil:13}%
  \BibitemOpen
  \bibfield  {author} {\bibinfo {author} {\bibfnamefont {J.~J.}\ \bibnamefont
  {Gil}}, \bibinfo {author} {\bibfnamefont {I.~S.}\ \bibnamefont {Jos\'{e}}},\
  and\ \bibinfo {author} {\bibfnamefont {R.}~\bibnamefont {Ossikovski}},\
  }\href {https://doi.org/10.1364/JOSAA.30.000032} {\bibfield  {journal}
  {\bibinfo  {journal} {J. Opt. Soc. Am. A}\ }\textbf {\bibinfo {volume}
  {30}},\ \bibinfo {pages} {32} (\bibinfo {year} {2013})}\BibitemShut {NoStop}%
\bibitem [{\citenamefont {Ossikovski}\ \emph {et~al.}(2007)\citenamefont
  {Ossikovski}, \citenamefont {Martino},\ and\ \citenamefont
  {Guyot}}]{Ossikovski:07}%
  \BibitemOpen
  \bibfield  {author} {\bibinfo {author} {\bibfnamefont {R.}~\bibnamefont
  {Ossikovski}}, \bibinfo {author} {\bibfnamefont {A.~D.}\ \bibnamefont
  {Martino}},\ and\ \bibinfo {author} {\bibfnamefont {S.}~\bibnamefont
  {Guyot}},\ }\href {https://doi.org/10.1364/OL.32.000689} {\bibfield
  {journal} {\bibinfo  {journal} {Opt. Lett.}\ }\textbf {\bibinfo {volume}
  {32}},\ \bibinfo {pages} {689} (\bibinfo {year} {2007})}\BibitemShut
  {NoStop}%
\bibitem [{\citenamefont {Ossikovski}(2012)}]{Ossikovski:12}%
  \BibitemOpen
  \bibfield  {author} {\bibinfo {author} {\bibfnamefont {R.}~\bibnamefont
  {Ossikovski}},\ }\href {https://doi.org/10.1364/OL.37.000220} {\bibfield
  {journal} {\bibinfo  {journal} {Opt. Lett.}\ }\textbf {\bibinfo {volume}
  {37}},\ \bibinfo {pages} {220} (\bibinfo {year} {2012})}\BibitemShut
  {NoStop}%
\bibitem [{\citenamefont {Ortega-Quijano}\ and\ \citenamefont
  {Arce-Diego}(2011)}]{Ortega-Quijano:11}%
  \BibitemOpen
  \bibfield  {author} {\bibinfo {author} {\bibfnamefont {N.}~\bibnamefont
  {Ortega-Quijano}}\ and\ \bibinfo {author} {\bibfnamefont {J.~L.}\
  \bibnamefont {Arce-Diego}},\ }\href {https://doi.org/10.1364/OL.36.001942}
  {\bibfield  {journal} {\bibinfo  {journal} {Opt. Lett.}\ }\textbf {\bibinfo
  {volume} {36}},\ \bibinfo {pages} {1942} (\bibinfo {year}
  {2011})}\BibitemShut {NoStop}%
\bibitem [{\citenamefont {Ossikovski}\ \emph {et~al.}(2008)\citenamefont
  {Ossikovski}, \citenamefont {Anastasiadou}, \citenamefont {Ben~Hatit},
  \citenamefont {Garcia-Caurel},\ and\ \citenamefont
  {De~Martino}}]{https://doi.org/10.1002/pssa.200777793}%
  \BibitemOpen
  \bibfield  {author} {\bibinfo {author} {\bibfnamefont {R.}~\bibnamefont
  {Ossikovski}}, \bibinfo {author} {\bibfnamefont {M.}~\bibnamefont
  {Anastasiadou}}, \bibinfo {author} {\bibfnamefont {S.}~\bibnamefont
  {Ben~Hatit}}, \bibinfo {author} {\bibfnamefont {E.}~\bibnamefont
  {Garcia-Caurel}},\ and\ \bibinfo {author} {\bibfnamefont {A.}~\bibnamefont
  {De~Martino}},\ }\href
  {https://doi.org/https://doi.org/10.1002/pssa.200777793} {\bibfield
  {journal} {\bibinfo  {journal} {physica status solidi (a)}\ }\textbf
  {\bibinfo {volume} {205}},\ \bibinfo {pages} {720} (\bibinfo {year}
  {2008})},\ \Eprint
  {https://arxiv.org/abs/https://onlinelibrary.wiley.com/doi/pdf/10.1002/pssa.200777793}
  {https://onlinelibrary.wiley.com/doi/pdf/10.1002/pssa.200777793} \BibitemShut
  {NoStop}%
\bibitem [{\citenamefont {Jones}(2003)}]{https://doi.org/10.1002/mrm.10331}%
  \BibitemOpen
  \bibfield  {author} {\bibinfo {author} {\bibfnamefont {D.~K.}\ \bibnamefont
  {Jones}},\ }\href {https://doi.org/https://doi.org/10.1002/mrm.10331}
  {\bibfield  {journal} {\bibinfo  {journal} {Magnetic Resonance in Medicine}\
  }\textbf {\bibinfo {volume} {49}},\ \bibinfo {pages} {7} (\bibinfo {year}
  {2003})},\ \Eprint
  {https://arxiv.org/abs/https://onlinelibrary.wiley.com/doi/pdf/10.1002/mrm.10331}
  {https://onlinelibrary.wiley.com/doi/pdf/10.1002/mrm.10331} \BibitemShut
  {NoStop}%
\bibitem [{\citenamefont {Bancelin}\ \emph {et~al.}(2014)\citenamefont
  {Bancelin}, \citenamefont {Nazac}, \citenamefont {Ibrahim}, \citenamefont
  {Dokl\'{a}dal}, \citenamefont {Decenci\`{e}re}, \citenamefont {Teig},
  \citenamefont {Haddad}, \citenamefont {Fernandez}, \citenamefont
  {Schanne-Klein},\ and\ \citenamefont {Martino}}]{Bancelin:14}%
  \BibitemOpen
  \bibfield  {author} {\bibinfo {author} {\bibfnamefont {S.}~\bibnamefont
  {Bancelin}}, \bibinfo {author} {\bibfnamefont {A.}~\bibnamefont {Nazac}},
  \bibinfo {author} {\bibfnamefont {B.~H.}\ \bibnamefont {Ibrahim}}, \bibinfo
  {author} {\bibfnamefont {P.}~\bibnamefont {Dokl\'{a}dal}}, \bibinfo {author}
  {\bibfnamefont {E.}~\bibnamefont {Decenci\`{e}re}}, \bibinfo {author}
  {\bibfnamefont {B.}~\bibnamefont {Teig}}, \bibinfo {author} {\bibfnamefont
  {H.}~\bibnamefont {Haddad}}, \bibinfo {author} {\bibfnamefont
  {H.}~\bibnamefont {Fernandez}}, \bibinfo {author} {\bibfnamefont {M.-C.}\
  \bibnamefont {Schanne-Klein}},\ and\ \bibinfo {author} {\bibfnamefont
  {A.~D.}\ \bibnamefont {Martino}},\ }\href
  {https://doi.org/10.1364/OE.22.022561} {\bibfield  {journal} {\bibinfo
  {journal} {Opt. Express}\ }\textbf {\bibinfo {volume} {22}},\ \bibinfo
  {pages} {22561} (\bibinfo {year} {2014})}\BibitemShut {NoStop}%
\bibitem [{\citenamefont {Menzel}\ \emph {et~al.}(2017)\citenamefont {Menzel},
  \citenamefont {Reckfort}, \citenamefont {Weigand}, \citenamefont {K\"{o}se},
  \citenamefont {Amunts},\ and\ \citenamefont {Axer}}]{Menzel:17}%
  \BibitemOpen
  \bibfield  {author} {\bibinfo {author} {\bibfnamefont {M.}~\bibnamefont
  {Menzel}}, \bibinfo {author} {\bibfnamefont {J.}~\bibnamefont {Reckfort}},
  \bibinfo {author} {\bibfnamefont {D.}~\bibnamefont {Weigand}}, \bibinfo
  {author} {\bibfnamefont {H.}~\bibnamefont {K\"{o}se}}, \bibinfo {author}
  {\bibfnamefont {K.}~\bibnamefont {Amunts}},\ and\ \bibinfo {author}
  {\bibfnamefont {M.}~\bibnamefont {Axer}},\ }\href
  {https://doi.org/10.1364/BOE.8.003163} {\bibfield  {journal} {\bibinfo
  {journal} {Biomed. Opt. Express}\ }\textbf {\bibinfo {volume} {8}},\ \bibinfo
  {pages} {3163} (\bibinfo {year} {2017})}\BibitemShut {NoStop}%
\bibitem [{\citenamefont {Schmitz}\ \emph {et~al.}(2020)\citenamefont
  {Schmitz}, \citenamefont {Lippert}, \citenamefont {Amunts},\ and\
  \citenamefont {Axer}}]{10.1117/12.2548935}%
  \BibitemOpen
  \bibfield  {author} {\bibinfo {author} {\bibfnamefont {D.}~\bibnamefont
  {Schmitz}}, \bibinfo {author} {\bibfnamefont {T.}~\bibnamefont {Lippert}},
  \bibinfo {author} {\bibfnamefont {K.}~\bibnamefont {Amunts}},\ and\ \bibinfo
  {author} {\bibfnamefont {M.}~\bibnamefont {Axer}},\ }in\ \href
  {https://doi.org/10.1117/12.2548935} {\emph {\bibinfo {booktitle} {Medical
  Imaging 2020: Physics of Medical Imaging}}},\ Vol.\ \bibinfo {volume}
  {11312},\ \bibinfo {editor} {edited by\ \bibinfo {editor} {\bibfnamefont
  {G.-H.}\ \bibnamefont {Chen}}\ and\ \bibinfo {editor} {\bibfnamefont
  {H.}~\bibnamefont {Bosmans}}},\ \bibinfo {organization} {International
  Society for Optics and Photonics}\ (\bibinfo  {publisher} {SPIE},\ \bibinfo
  {year} {2020})\ p.\ \bibinfo {pages} {1131239}\BibitemShut {NoStop}%
\bibitem [{\citenamefont {Chang}\ \emph {et~al.}(2023)\citenamefont {Chang},
  \citenamefont {Yang}, \citenamefont {Novoseltseva}, \citenamefont
  {Abdelhakeem}, \citenamefont {Hyman}, \citenamefont {Fu}, \citenamefont {Li},
  \citenamefont {Chen}, \citenamefont {Augustinack}, \citenamefont {Magnain},
  \citenamefont {Fischl}, \citenamefont {Mckee}, \citenamefont {Boas},
  \citenamefont {Chen},\ and\ \citenamefont
  {Wang}}]{https://doi.org/10.1002/advs.202303381}%
  \BibitemOpen
  \bibfield  {author} {\bibinfo {author} {\bibfnamefont {S.}~\bibnamefont
  {Chang}}, \bibinfo {author} {\bibfnamefont {J.}~\bibnamefont {Yang}},
  \bibinfo {author} {\bibfnamefont {A.}~\bibnamefont {Novoseltseva}}, \bibinfo
  {author} {\bibfnamefont {A.}~\bibnamefont {Abdelhakeem}}, \bibinfo {author}
  {\bibfnamefont {M.}~\bibnamefont {Hyman}}, \bibinfo {author} {\bibfnamefont
  {X.}~\bibnamefont {Fu}}, \bibinfo {author} {\bibfnamefont {C.}~\bibnamefont
  {Li}}, \bibinfo {author} {\bibfnamefont {S.-C.}\ \bibnamefont {Chen}},
  \bibinfo {author} {\bibfnamefont {J.~C.}\ \bibnamefont {Augustinack}},
  \bibinfo {author} {\bibfnamefont {C.}~\bibnamefont {Magnain}}, \bibinfo
  {author} {\bibfnamefont {B.}~\bibnamefont {Fischl}}, \bibinfo {author}
  {\bibfnamefont {A.~C.}\ \bibnamefont {Mckee}}, \bibinfo {author}
  {\bibfnamefont {D.~A.}\ \bibnamefont {Boas}}, \bibinfo {author}
  {\bibfnamefont {I.~A.}\ \bibnamefont {Chen}},\ and\ \bibinfo {author}
  {\bibfnamefont {H.}~\bibnamefont {Wang}},\ }\href
  {https://doi.org/https://doi.org/10.1002/advs.202303381} {\bibfield
  {journal} {\bibinfo  {journal} {Advanced Science}\ }\textbf {\bibinfo
  {volume} {10}},\ \bibinfo {pages} {2303381} (\bibinfo {year} {2023})},\
  \Eprint
  {https://arxiv.org/abs/https://onlinelibrary.wiley.com/doi/pdf/10.1002/advs.202303381}
  {https://onlinelibrary.wiley.com/doi/pdf/10.1002/advs.202303381} \BibitemShut
  {NoStop}%
\bibitem [{\citenamefont {Klimov}\ \emph {et~al.}(2006)\citenamefont {Klimov},
  \citenamefont {Romero},\ and\ \citenamefont {S\'{a}nchez-Soto}}]{Klimov:06}%
  \BibitemOpen
  \bibfield  {author} {\bibinfo {author} {\bibfnamefont {A.~B.}\ \bibnamefont
  {Klimov}}, \bibinfo {author} {\bibfnamefont {J.~L.}\ \bibnamefont {Romero}},\
  and\ \bibinfo {author} {\bibfnamefont {L.~L.}\ \bibnamefont
  {S\'{a}nchez-Soto}},\ }\href {https://doi.org/10.1364/JOSAB.23.000126}
  {\bibfield  {journal} {\bibinfo  {journal} {J. Opt. Soc. Am. B}\ }\textbf
  {\bibinfo {volume} {23}},\ \bibinfo {pages} {126} (\bibinfo {year}
  {2006})}\BibitemShut {NoStop}%
\bibitem [{\citenamefont {Klimov}\ \emph {et~al.}(2008)\citenamefont {Klimov},
  \citenamefont {Romero}, \citenamefont {S\'anchez-Soto}, \citenamefont
  {Messina},\ and\ \citenamefont {Napoli}}]{PhysRevA.77.033853}%
  \BibitemOpen
  \bibfield  {author} {\bibinfo {author} {\bibfnamefont {A.~B.}\ \bibnamefont
  {Klimov}}, \bibinfo {author} {\bibfnamefont {J.~L.}\ \bibnamefont {Romero}},
  \bibinfo {author} {\bibfnamefont {L.~L.}\ \bibnamefont {S\'anchez-Soto}},
  \bibinfo {author} {\bibfnamefont {A.}~\bibnamefont {Messina}},\ and\ \bibinfo
  {author} {\bibfnamefont {A.}~\bibnamefont {Napoli}},\ }\href
  {https://doi.org/10.1103/PhysRevA.77.033853} {\bibfield  {journal} {\bibinfo
  {journal} {Phys. Rev. A}\ }\textbf {\bibinfo {volume} {77}},\ \bibinfo
  {pages} {033853} (\bibinfo {year} {2008})}\BibitemShut {NoStop}%
\bibitem [{\citenamefont {Rivas}\ and\ \citenamefont
  {Luis}(2013)}]{PhysRevA.88.052120}%
  \BibitemOpen
  \bibfield  {author} {\bibinfo {author} {\bibfnamefont {A.}~\bibnamefont
  {Rivas}}\ and\ \bibinfo {author} {\bibfnamefont {A.}~\bibnamefont {Luis}},\
  }\href {https://doi.org/10.1103/PhysRevA.88.052120} {\bibfield  {journal}
  {\bibinfo  {journal} {Phys. Rev. A}\ }\textbf {\bibinfo {volume} {88}},\
  \bibinfo {pages} {052120} (\bibinfo {year} {2013})}\BibitemShut {NoStop}%
\bibitem [{\citenamefont {Denis}\ and\ \citenamefont
  {Martin}(2022)}]{PhysRevResearch.4.013178}%
  \BibitemOpen
  \bibfield  {author} {\bibinfo {author} {\bibfnamefont {J.}~\bibnamefont
  {Denis}}\ and\ \bibinfo {author} {\bibfnamefont {J.}~\bibnamefont {Martin}},\
  }\href {https://doi.org/10.1103/PhysRevResearch.4.013178} {\bibfield
  {journal} {\bibinfo  {journal} {Phys. Rev. Res.}\ }\textbf {\bibinfo {volume}
  {4}},\ \bibinfo {pages} {013178} (\bibinfo {year} {2022})}\BibitemShut
  {NoStop}%
\bibitem [{\citenamefont {Helstrom}(1967)}]{HELSTROM1967101}%
  \BibitemOpen
  \bibfield  {author} {\bibinfo {author} {\bibfnamefont {C.}~\bibnamefont
  {Helstrom}},\ }\href
  {https://doi.org/https://doi.org/10.1016/0375-9601(67)90366-0} {\bibfield
  {journal} {\bibinfo  {journal} {Physics Letters A}\ }\textbf {\bibinfo
  {volume} {25}},\ \bibinfo {pages} {101} (\bibinfo {year} {1967})}\BibitemShut
  {NoStop}%
\bibitem [{\citenamefont {Helstrom}(1968)}]{1054108}%
  \BibitemOpen
  \bibfield  {author} {\bibinfo {author} {\bibfnamefont {C.}~\bibnamefont
  {Helstrom}},\ }\href {https://doi.org/10.1109/TIT.1968.1054108} {\bibfield
  {journal} {\bibinfo  {journal} {IEEE Transactions on Information Theory}\
  }\textbf {\bibinfo {volume} {14}},\ \bibinfo {pages} {234} (\bibinfo {year}
  {1968})}\BibitemShut {NoStop}%
\bibitem [{\citenamefont {Helstrom}(1969)}]{Helstrom1969}%
  \BibitemOpen
  \bibfield  {author} {\bibinfo {author} {\bibfnamefont {C.~W.}\ \bibnamefont
  {Helstrom}},\ }\href {https://doi.org/10.1007/BF01007479} {\bibfield
  {journal} {\bibinfo  {journal} {Journal of Statistical Physics}\ }\textbf
  {\bibinfo {volume} {1}},\ \bibinfo {pages} {231} (\bibinfo {year}
  {1969})}\BibitemShut {NoStop}%
\bibitem [{\citenamefont {Braunstein}\ and\ \citenamefont
  {Caves}(1994)}]{PhysRevLett.72.3439}%
  \BibitemOpen
  \bibfield  {author} {\bibinfo {author} {\bibfnamefont {S.~L.}\ \bibnamefont
  {Braunstein}}\ and\ \bibinfo {author} {\bibfnamefont {C.~M.}\ \bibnamefont
  {Caves}},\ }\href {https://doi.org/10.1103/PhysRevLett.72.3439} {\bibfield
  {journal} {\bibinfo  {journal} {Phys. Rev. Lett.}\ }\textbf {\bibinfo
  {volume} {72}},\ \bibinfo {pages} {3439} (\bibinfo {year}
  {1994})}\BibitemShut {NoStop}%
\bibitem [{\citenamefont {PARIS}(2009)}]{doi:10.1142/S0219749909004839}%
  \BibitemOpen
  \bibfield  {author} {\bibinfo {author} {\bibfnamefont {M.~G.~A.}\
  \bibnamefont {PARIS}},\ }\href {https://doi.org/10.1142/S0219749909004839}
  {\bibfield  {journal} {\bibinfo  {journal} {International Journal of Quantum
  Information}\ }\textbf {\bibinfo {volume} {07}},\ \bibinfo {pages} {125}
  (\bibinfo {year} {2009})},\ \Eprint
  {https://arxiv.org/abs/https://doi.org/10.1142/S0219749909004839}
  {https://doi.org/10.1142/S0219749909004839} \BibitemShut {NoStop}%
\bibitem [{\citenamefont {Sidhu}\ and\ \citenamefont
  {Kok}(2020)}]{doi:10.1116/1.5119961}%
  \BibitemOpen
  \bibfield  {author} {\bibinfo {author} {\bibfnamefont {J.~S.}\ \bibnamefont
  {Sidhu}}\ and\ \bibinfo {author} {\bibfnamefont {P.}~\bibnamefont {Kok}},\
  }\href {https://doi.org/10.1116/1.5119961} {\bibfield  {journal} {\bibinfo
  {journal} {AVS Quantum Science}\ }\textbf {\bibinfo {volume} {2}},\ \bibinfo
  {pages} {014701} (\bibinfo {year} {2020})},\ \Eprint
  {https://arxiv.org/abs/https://doi.org/10.1116/1.5119961}
  {https://doi.org/10.1116/1.5119961} \BibitemShut {NoStop}%
\bibitem [{\citenamefont {Johansson}\ \emph {et~al.}(2012)\citenamefont
  {Johansson}, \citenamefont {Nation},\ and\ \citenamefont
  {Nori}}]{JOHANSSON20121760}%
  \BibitemOpen
  \bibfield  {author} {\bibinfo {author} {\bibfnamefont {J.}~\bibnamefont
  {Johansson}}, \bibinfo {author} {\bibfnamefont {P.}~\bibnamefont {Nation}},\
  and\ \bibinfo {author} {\bibfnamefont {F.}~\bibnamefont {Nori}},\ }\href
  {https://doi.org/https://doi.org/10.1016/j.cpc.2012.02.021} {\bibfield
  {journal} {\bibinfo  {journal} {Computer Physics Communications}\ }\textbf
  {\bibinfo {volume} {183}},\ \bibinfo {pages} {1760} (\bibinfo {year}
  {2012})}\BibitemShut {NoStop}%
\bibitem [{\citenamefont {Johansson}\ \emph {et~al.}(2013)\citenamefont
  {Johansson}, \citenamefont {Nation},\ and\ \citenamefont
  {Nori}}]{JOHANSSON20131234}%
  \BibitemOpen
  \bibfield  {author} {\bibinfo {author} {\bibfnamefont {J.}~\bibnamefont
  {Johansson}}, \bibinfo {author} {\bibfnamefont {P.}~\bibnamefont {Nation}},\
  and\ \bibinfo {author} {\bibfnamefont {F.}~\bibnamefont {Nori}},\ }\href
  {https://doi.org/https://doi.org/10.1016/j.cpc.2012.11.019} {\bibfield
  {journal} {\bibinfo  {journal} {Computer Physics Communications}\ }\textbf
  {\bibinfo {volume} {184}},\ \bibinfo {pages} {1234} (\bibinfo {year}
  {2013})}\BibitemShut {NoStop}%
\bibitem [{\citenamefont {Zhang}\ \emph {et~al.}(2022)\citenamefont {Zhang},
  \citenamefont {Yu}, \citenamefont {Yuan}, \citenamefont {Wang}, \citenamefont
  {Demkowicz-Dobrza\ifmmode~\acute{n}\else \'{n}\fi{}ski},\ and\ \citenamefont
  {Liu}}]{PhysRevResearch.4.043057}%
  \BibitemOpen
  \bibfield  {author} {\bibinfo {author} {\bibfnamefont {M.}~\bibnamefont
  {Zhang}}, \bibinfo {author} {\bibfnamefont {H.-M.}\ \bibnamefont {Yu}},
  \bibinfo {author} {\bibfnamefont {H.}~\bibnamefont {Yuan}}, \bibinfo {author}
  {\bibfnamefont {X.}~\bibnamefont {Wang}}, \bibinfo {author} {\bibfnamefont
  {R.}~\bibnamefont {Demkowicz-Dobrza\ifmmode~\acute{n}\else \'{n}\fi{}ski}},\
  and\ \bibinfo {author} {\bibfnamefont {J.}~\bibnamefont {Liu}},\ }\href
  {https://doi.org/10.1103/PhysRevResearch.4.043057} {\bibfield  {journal}
  {\bibinfo  {journal} {Phys. Rev. Res.}\ }\textbf {\bibinfo {volume} {4}},\
  \bibinfo {pages} {043057} (\bibinfo {year} {2022})}\BibitemShut {NoStop}%
\bibitem [{\citenamefont {Belsley}\ and\ \citenamefont
  {Matthews}(2022)}]{Belsley:2022}%
  \BibitemOpen
  \bibfield  {author} {\bibinfo {author} {\bibfnamefont {A.}~\bibnamefont
  {Belsley}}\ and\ \bibinfo {author} {\bibfnamefont {J.~C.~F.}\ \bibnamefont
  {Matthews}},\ }\href {https://doi.org/10.1063/5.0122769} {\bibfield
  {journal} {\bibinfo  {journal} {Applied Physics Letters}\ }\textbf {\bibinfo
  {volume} {121}},\ \bibinfo {pages} {184001} (\bibinfo {year} {2022})},\
  \Eprint
  {https://arxiv.org/abs/https://pubs.aip.org/aip/apl/article-pdf/doi/10.1063/5.0122769/16487852/184001\_1\_online.pdf}
  {https://pubs.aip.org/aip/apl/article-pdf/doi/10.1063/5.0122769/16487852/184001\_1\_online.pdf}
  \BibitemShut {NoStop}%
\bibitem [{\citenamefont {Wolfgramm}\ \emph {et~al.}(2013)\citenamefont
  {Wolfgramm}, \citenamefont {Vitelli}, \citenamefont {Beduini}, \citenamefont
  {Godbout},\ and\ \citenamefont {Mitchell}}]{Wolfgramm:2013}%
  \BibitemOpen
  \bibfield  {author} {\bibinfo {author} {\bibfnamefont {F.}~\bibnamefont
  {Wolfgramm}}, \bibinfo {author} {\bibfnamefont {C.}~\bibnamefont {Vitelli}},
  \bibinfo {author} {\bibfnamefont {F.~A.}\ \bibnamefont {Beduini}}, \bibinfo
  {author} {\bibfnamefont {N.}~\bibnamefont {Godbout}},\ and\ \bibinfo {author}
  {\bibfnamefont {M.~W.}\ \bibnamefont {Mitchell}},\ }\href
  {https://doi.org/10.1038/nphoton.2012.300} {\bibfield  {journal} {\bibinfo
  {journal} {Nature Photonics}\ }\textbf {\bibinfo {volume} {7}},\ \bibinfo
  {pages} {28} (\bibinfo {year} {2013})}\BibitemShut {NoStop}%
\bibitem [{\citenamefont {Peng}\ and\ \citenamefont {Zhao}(2023)}]{Peng:2023}%
  \BibitemOpen
  \bibfield  {author} {\bibinfo {author} {\bibfnamefont {Y.}~\bibnamefont
  {Peng}}\ and\ \bibinfo {author} {\bibfnamefont {Y.}~\bibnamefont {Zhao}},\
  }\href {https://doi.org/https://doi.org/10.1016/j.snb.2023.133616} {\bibfield
   {journal} {\bibinfo  {journal} {Sensors and Actuators B: Chemical}\ }\textbf
  {\bibinfo {volume} {383}},\ \bibinfo {pages} {133616} (\bibinfo {year}
  {2023})}\BibitemShut {NoStop}%
\bibitem [{\citenamefont {Yurke}\ \emph {et~al.}(1986)\citenamefont {Yurke},
  \citenamefont {McCall},\ and\ \citenamefont {Klauder}}]{Yurke:1986}%
  \BibitemOpen
  \bibfield  {author} {\bibinfo {author} {\bibfnamefont {B.}~\bibnamefont
  {Yurke}}, \bibinfo {author} {\bibfnamefont {S.~L.}\ \bibnamefont {McCall}},\
  and\ \bibinfo {author} {\bibfnamefont {J.~R.}\ \bibnamefont {Klauder}},\
  }\href {https://doi.org/10.1103/PhysRevA.33.4033} {\bibfield  {journal}
  {\bibinfo  {journal} {Phys. Rev. A}\ }\textbf {\bibinfo {volume} {33}},\
  \bibinfo {pages} {4033} (\bibinfo {year} {1986})}\BibitemShut {NoStop}%
\bibitem [{\citenamefont {Kuzmich}\ and\ \citenamefont
  {Mandel}(1998)}]{Kuzmich:1998}%
  \BibitemOpen
  \bibfield  {author} {\bibinfo {author} {\bibfnamefont {A.}~\bibnamefont
  {Kuzmich}}\ and\ \bibinfo {author} {\bibfnamefont {L.}~\bibnamefont
  {Mandel}},\ }\href {https://doi.org/10.1088/1355-5111/10/3/008} {\bibfield
  {journal} {\bibinfo  {journal} {Quantum and Semiclassical Optics: Journal of
  the European Optical Society Part B}\ }\textbf {\bibinfo {volume} {10}},\
  \bibinfo {pages} {493} (\bibinfo {year} {1998})}\BibitemShut {NoStop}%
\bibitem [{\citenamefont {Sanaka}\ \emph {et~al.}(2004)\citenamefont {Sanaka},
  \citenamefont {Jennewein}, \citenamefont {Pan}, \citenamefont {Resch},\ and\
  \citenamefont {Zeilinger}}]{Sanaka:2004}%
  \BibitemOpen
  \bibfield  {author} {\bibinfo {author} {\bibfnamefont {K.}~\bibnamefont
  {Sanaka}}, \bibinfo {author} {\bibfnamefont {T.}~\bibnamefont {Jennewein}},
  \bibinfo {author} {\bibfnamefont {J.-W.}\ \bibnamefont {Pan}}, \bibinfo
  {author} {\bibfnamefont {K.}~\bibnamefont {Resch}},\ and\ \bibinfo {author}
  {\bibfnamefont {A.}~\bibnamefont {Zeilinger}},\ }\href
  {https://doi.org/10.1103/PhysRevLett.92.017902} {\bibfield  {journal}
  {\bibinfo  {journal} {Phys. Rev. Lett.}\ }\textbf {\bibinfo {volume} {92}},\
  \bibinfo {pages} {017902} (\bibinfo {year} {2004})}\BibitemShut {NoStop}%
\bibitem [{\citenamefont {Israel}\ \emph {et~al.}(2012)\citenamefont {Israel},
  \citenamefont {Afek}, \citenamefont {Rosen}, \citenamefont {Ambar},\ and\
  \citenamefont {Silberberg}}]{Israel:2012}%
  \BibitemOpen
  \bibfield  {author} {\bibinfo {author} {\bibfnamefont {Y.}~\bibnamefont
  {Israel}}, \bibinfo {author} {\bibfnamefont {I.}~\bibnamefont {Afek}},
  \bibinfo {author} {\bibfnamefont {S.}~\bibnamefont {Rosen}}, \bibinfo
  {author} {\bibfnamefont {O.}~\bibnamefont {Ambar}},\ and\ \bibinfo {author}
  {\bibfnamefont {Y.}~\bibnamefont {Silberberg}},\ }\href
  {https://doi.org/10.1103/PhysRevA.85.022115} {\bibfield  {journal} {\bibinfo
  {journal} {Phys. Rev. A}\ }\textbf {\bibinfo {volume} {85}},\ \bibinfo
  {pages} {022115} (\bibinfo {year} {2012})}\BibitemShut {NoStop}%
\bibitem [{\citenamefont {Israel}\ \emph {et~al.}(2014)\citenamefont {Israel},
  \citenamefont {Rosen},\ and\ \citenamefont {Silberberg}}]{Israel:2014}%
  \BibitemOpen
  \bibfield  {author} {\bibinfo {author} {\bibfnamefont {Y.}~\bibnamefont
  {Israel}}, \bibinfo {author} {\bibfnamefont {S.}~\bibnamefont {Rosen}},\ and\
  \bibinfo {author} {\bibfnamefont {Y.}~\bibnamefont {Silberberg}},\ }\href
  {https://doi.org/10.1103/PhysRevLett.112.103604} {\bibfield  {journal}
  {\bibinfo  {journal} {Phys. Rev. Lett.}\ }\textbf {\bibinfo {volume} {112}},\
  \bibinfo {pages} {103604} (\bibinfo {year} {2014})}\BibitemShut {NoStop}%
\bibitem [{\citenamefont {Hong}\ \emph {et~al.}(1987)\citenamefont {Hong},
  \citenamefont {Ou},\ and\ \citenamefont {Mandel}}]{Hong:1987}%
  \BibitemOpen
  \bibfield  {author} {\bibinfo {author} {\bibfnamefont {C.~K.}\ \bibnamefont
  {Hong}}, \bibinfo {author} {\bibfnamefont {Z.~Y.}\ \bibnamefont {Ou}},\ and\
  \bibinfo {author} {\bibfnamefont {L.}~\bibnamefont {Mandel}},\ }\href
  {https://doi.org/10.1103/PhysRevLett.59.2044} {\bibfield  {journal} {\bibinfo
   {journal} {Phys. Rev. Lett.}\ }\textbf {\bibinfo {volume} {59}},\ \bibinfo
  {pages} {2044} (\bibinfo {year} {1987})}\BibitemShut {NoStop}%
\bibitem [{\citenamefont {Adamson}\ \emph {et~al.}(2007)\citenamefont
  {Adamson}, \citenamefont {Shalm}, \citenamefont {Mitchell},\ and\
  \citenamefont {Steinberg}}]{Adamson:2007}%
  \BibitemOpen
  \bibfield  {author} {\bibinfo {author} {\bibfnamefont {R.~B.~A.}\
  \bibnamefont {Adamson}}, \bibinfo {author} {\bibfnamefont {L.~K.}\
  \bibnamefont {Shalm}}, \bibinfo {author} {\bibfnamefont {M.~W.}\ \bibnamefont
  {Mitchell}},\ and\ \bibinfo {author} {\bibfnamefont {A.~M.}\ \bibnamefont
  {Steinberg}},\ }\href {https://doi.org/10.1103/PhysRevLett.98.043601}
  {\bibfield  {journal} {\bibinfo  {journal} {Phys. Rev. Lett.}\ }\textbf
  {\bibinfo {volume} {98}},\ \bibinfo {pages} {043601} (\bibinfo {year}
  {2007})}\BibitemShut {NoStop}%
\bibitem [{\citenamefont {Wolfgramm}\ \emph {et~al.}(2010)\citenamefont
  {Wolfgramm}, \citenamefont {Cer\`{e}},\ and\ \citenamefont
  {Mitchell}}]{Wolfgramm:2010}%
  \BibitemOpen
  \bibfield  {author} {\bibinfo {author} {\bibfnamefont {F.}~\bibnamefont
  {Wolfgramm}}, \bibinfo {author} {\bibfnamefont {A.}~\bibnamefont
  {Cer\`{e}}},\ and\ \bibinfo {author} {\bibfnamefont {M.~W.}\ \bibnamefont
  {Mitchell}},\ }\href {https://doi.org/10.1364/JOSAB.27.000A25} {\bibfield
  {journal} {\bibinfo  {journal} {J. Opt. Soc. Am. B}\ }\textbf {\bibinfo
  {volume} {27}},\ \bibinfo {pages} {A25} (\bibinfo {year} {2010})}\BibitemShut
  {NoStop}%
\bibitem [{\citenamefont {James}\ \emph {et~al.}(2001)\citenamefont {James},
  \citenamefont {Kwiat}, \citenamefont {Munro},\ and\ \citenamefont
  {White}}]{James:2001}%
  \BibitemOpen
  \bibfield  {author} {\bibinfo {author} {\bibfnamefont {D.~F.~V.}\
  \bibnamefont {James}}, \bibinfo {author} {\bibfnamefont {P.~G.}\ \bibnamefont
  {Kwiat}}, \bibinfo {author} {\bibfnamefont {W.~J.}\ \bibnamefont {Munro}},\
  and\ \bibinfo {author} {\bibfnamefont {A.~G.}\ \bibnamefont {White}},\
  }\href@noop {} {\bibfield  {journal} {\bibinfo  {journal} {Phys. Rev. A}\
  }\textbf {\bibinfo {volume} {64}},\ \bibinfo {pages} {052312} (\bibinfo
  {year} {2001})}\BibitemShut {NoStop}%
\bibitem [{\citenamefont {Bengtsson}\ and\ \citenamefont
  {Zyczkowski}(2006)}]{Bengtsson_Zyczkowski_2006}%
  \BibitemOpen
  \bibfield  {author} {\bibinfo {author} {\bibfnamefont {I.}~\bibnamefont
  {Bengtsson}}\ and\ \bibinfo {author} {\bibfnamefont {K.}~\bibnamefont
  {Zyczkowski}},\ }\href@noop {} {\emph {\bibinfo {title} {Geometry of Quantum
  States: An Introduction to Quantum Entanglement}}}\ (\bibinfo  {publisher}
  {Cambridge University Press},\ \bibinfo {year} {2006})\BibitemShut {NoStop}%
\bibitem [{\citenamefont {Giraud}\ \emph {et~al.}(2015)\citenamefont {Giraud},
  \citenamefont {Braun}, \citenamefont {Baguette}, \citenamefont {Bastin},\
  and\ \citenamefont {Martin}}]{PhysRevLett.114.080401}%
  \BibitemOpen
  \bibfield  {author} {\bibinfo {author} {\bibfnamefont {O.}~\bibnamefont
  {Giraud}}, \bibinfo {author} {\bibfnamefont {D.}~\bibnamefont {Braun}},
  \bibinfo {author} {\bibfnamefont {D.}~\bibnamefont {Baguette}}, \bibinfo
  {author} {\bibfnamefont {T.}~\bibnamefont {Bastin}},\ and\ \bibinfo {author}
  {\bibfnamefont {J.}~\bibnamefont {Martin}},\ }\href
  {https://doi.org/10.1103/PhysRevLett.114.080401} {\bibfield  {journal}
  {\bibinfo  {journal} {Phys. Rev. Lett.}\ }\textbf {\bibinfo {volume} {114}},\
  \bibinfo {pages} {080401} (\bibinfo {year} {2015})}\BibitemShut {NoStop}%
\bibitem [{\citenamefont {Majorana}(1932)}]{Majorana1932}%
  \BibitemOpen
  \bibfield  {author} {\bibinfo {author} {\bibfnamefont {E.}~\bibnamefont
  {Majorana}},\ }\href {https://doi.org/10.1007/BF02960953} {\bibfield
  {journal} {\bibinfo  {journal} {Il Nuovo Cimento (1924-1942)}\ }\textbf
  {\bibinfo {volume} {9}},\ \bibinfo {pages} {43} (\bibinfo {year}
  {1932})}\BibitemShut {NoStop}%
\bibitem [{\citenamefont
  {Schwinger}(1977)}]{https://doi.org/10.1111/j.2164-0947.1977.tb02957.x}%
  \BibitemOpen
  \bibfield  {author} {\bibinfo {author} {\bibfnamefont {J.}~\bibnamefont
  {Schwinger}},\ }\href
  {https://doi.org/https://doi.org/10.1111/j.2164-0947.1977.tb02957.x}
  {\bibfield  {journal} {\bibinfo  {journal} {Transactions of the New York
  Academy of Sciences}\ }\textbf {\bibinfo {volume} {38}},\ \bibinfo {pages}
  {170} (\bibinfo {year} {1977})},\ \Eprint
  {https://arxiv.org/abs/https://nyaspubs.onlinelibrary.wiley.com/doi/pdf/10.1111/j.2164-0947.1977.tb02957.x}
  {https://nyaspubs.onlinelibrary.wiley.com/doi/pdf/10.1111/j.2164-0947.1977.tb02957.x}
  \BibitemShut {NoStop}%
\bibitem [{\citenamefont {Liu}\ and\ \citenamefont
  {Fu}(2016)}]{PhysRevA.94.022123}%
  \BibitemOpen
  \bibfield  {author} {\bibinfo {author} {\bibfnamefont {H.~D.}\ \bibnamefont
  {Liu}}\ and\ \bibinfo {author} {\bibfnamefont {L.~B.}\ \bibnamefont {Fu}},\
  }\href {https://doi.org/10.1103/PhysRevA.94.022123} {\bibfield  {journal}
  {\bibinfo  {journal} {Phys. Rev. A}\ }\textbf {\bibinfo {volume} {94}},\
  \bibinfo {pages} {022123} (\bibinfo {year} {2016})}\BibitemShut {NoStop}%
\bibitem [{\citenamefont {Lee}(1988)}]{Lee_1988}%
  \BibitemOpen
  \bibfield  {author} {\bibinfo {author} {\bibfnamefont {C.~T.}\ \bibnamefont
  {Lee}},\ }\href {https://doi.org/10.1088/0305-4470/21/19/013} {\bibfield
  {journal} {\bibinfo  {journal} {Journal of Physics A: Mathematical and
  General}\ }\textbf {\bibinfo {volume} {21}},\ \bibinfo {pages} {3749}
  (\bibinfo {year} {1988})}\BibitemShut {NoStop}%
\bibitem [{\citenamefont
  {Bargmann}(1961)}]{https://doi.org/10.1002/cpa.3160140303}%
  \BibitemOpen
  \bibfield  {author} {\bibinfo {author} {\bibfnamefont {V.}~\bibnamefont
  {Bargmann}},\ }\href {https://doi.org/https://doi.org/10.1002/cpa.3160140303}
  {\bibfield  {journal} {\bibinfo  {journal} {Communications on Pure and
  Applied Mathematics}\ }\textbf {\bibinfo {volume} {14}},\ \bibinfo {pages}
  {187} (\bibinfo {year} {1961})},\ \Eprint
  {https://arxiv.org/abs/https://onlinelibrary.wiley.com/doi/pdf/10.1002/cpa.3160140303}
  {https://onlinelibrary.wiley.com/doi/pdf/10.1002/cpa.3160140303} \BibitemShut
  {NoStop}%
\bibitem [{\citenamefont {Gazeau}(2009)}]{gazeau2009coherent}%
  \BibitemOpen
  \bibfield  {author} {\bibinfo {author} {\bibfnamefont {J.}~\bibnamefont
  {Gazeau}},\ }\href@noop {} {\emph {\bibinfo {title} {Coherent States in
  Quantum Physics}}}\ (\bibinfo  {publisher} {Wiley},\ \bibinfo {year}
  {2009})\BibitemShut {NoStop}%
\bibitem [{\citenamefont {Goldberg}\ \emph {et~al.}(2020)\citenamefont
  {Goldberg}, \citenamefont {Klimov}, \citenamefont {Grassl}, \citenamefont
  {Leuchs},\ and\ \citenamefont {Sánchez-Soto}}]{10.1116/5.0025819}%
  \BibitemOpen
  \bibfield  {author} {\bibinfo {author} {\bibfnamefont {A.~Z.}\ \bibnamefont
  {Goldberg}}, \bibinfo {author} {\bibfnamefont {A.~B.}\ \bibnamefont
  {Klimov}}, \bibinfo {author} {\bibfnamefont {M.}~\bibnamefont {Grassl}},
  \bibinfo {author} {\bibfnamefont {G.}~\bibnamefont {Leuchs}},\ and\ \bibinfo
  {author} {\bibfnamefont {L.~L.}\ \bibnamefont {Sánchez-Soto}},\ }\href
  {https://doi.org/10.1116/5.0025819} {\bibfield  {journal} {\bibinfo
  {journal} {AVS Quantum Science}\ }\textbf {\bibinfo {volume} {2}},\ \bibinfo
  {pages} {044701} (\bibinfo {year} {2020})}\BibitemShut {NoStop}%
\bibitem [{\citenamefont {Hall}(1994)}]{HALL1994103}%
  \BibitemOpen
  \bibfield  {author} {\bibinfo {author} {\bibfnamefont {B.}~\bibnamefont
  {Hall}},\ }\href {https://doi.org/https://doi.org/10.1006/jfan.1994.1064}
  {\bibfield  {journal} {\bibinfo  {journal} {Journal of Functional Analysis}\
  }\textbf {\bibinfo {volume} {122}},\ \bibinfo {pages} {103} (\bibinfo {year}
  {1994})}\BibitemShut {NoStop}%
\bibitem [{\citenamefont {Lieb}\ and\ \citenamefont
  {Solovej}(2014)}]{10.1007/s11511-014-0113-6}%
  \BibitemOpen
  \bibfield  {author} {\bibinfo {author} {\bibfnamefont {E.~H.}\ \bibnamefont
  {Lieb}}\ and\ \bibinfo {author} {\bibfnamefont {J.~P.}\ \bibnamefont
  {Solovej}},\ }\href {https://doi.org/10.1007/s11511-014-0113-6} {\bibfield
  {journal} {\bibinfo  {journal} {Acta Mathematica}\ }\textbf {\bibinfo
  {volume} {212}},\ \bibinfo {pages} {379 } (\bibinfo {year}
  {2014})}\BibitemShut {NoStop}%
\bibitem [{\citenamefont {Lieb}(1978)}]{Lieb1978}%
  \BibitemOpen
  \bibfield  {author} {\bibinfo {author} {\bibfnamefont {E.~H.}\ \bibnamefont
  {Lieb}},\ }\href {https://doi.org/10.1007/BF01940328} {\bibfield  {journal}
  {\bibinfo  {journal} {Communications in Mathematical Physics}\ }\textbf
  {\bibinfo {volume} {62}},\ \bibinfo {pages} {35} (\bibinfo {year}
  {1978})}\BibitemShut {NoStop}%
\bibitem [{\citenamefont {Schupp}(1999)}]{Schupp1999}%
  \BibitemOpen
  \bibfield  {author} {\bibinfo {author} {\bibfnamefont {P.}~\bibnamefont
  {Schupp}},\ }\href {https://doi.org/10.1007/s002200050734} {\bibfield
  {journal} {\bibinfo  {journal} {Communications in Mathematical Physics}\
  }\textbf {\bibinfo {volume} {207}},\ \bibinfo {pages} {481} (\bibinfo {year}
  {1999})}\BibitemShut {NoStop}%
\bibitem [{\citenamefont {Baecklund}\ and\ \citenamefont
  {Bengtsson}(2014)}]{Baecklund_2014}%
  \BibitemOpen
  \bibfield  {author} {\bibinfo {author} {\bibfnamefont {A.}~\bibnamefont
  {Baecklund}}\ and\ \bibinfo {author} {\bibfnamefont {I.}~\bibnamefont
  {Bengtsson}},\ }\href {https://doi.org/10.1088/0031-8949/2014/T163/014012}
  {\bibfield  {journal} {\bibinfo  {journal} {Physica Scripta}\ }\textbf
  {\bibinfo {volume} {2014}},\ \bibinfo {pages} {014012} (\bibinfo {year}
  {2014})}\BibitemShut {NoStop}%
\bibitem [{\citenamefont {Baecklund}(2013)}]{Baecklund_thesis}%
  \BibitemOpen
  \bibfield  {author} {\bibinfo {author} {\bibfnamefont {A.}~\bibnamefont
  {Baecklund}},\ }\emph {\bibinfo {title} {Maximization of the Wehrl Entropy in
  Finite Dimensions}},\ \href@noop {} {\bibinfo {type} {Master of science
  thesis}},\ \bibinfo  {school} {KTH Royal Institute of Technology} (\bibinfo
  {year} {2013})\BibitemShut {NoStop}%
\bibitem [{\citenamefont {Giraud}\ \emph {et~al.}(2010)\citenamefont {Giraud},
  \citenamefont {Braun},\ and\ \citenamefont {Braun}}]{Giraud_2010}%
  \BibitemOpen
  \bibfield  {author} {\bibinfo {author} {\bibfnamefont {O.}~\bibnamefont
  {Giraud}}, \bibinfo {author} {\bibfnamefont {P.}~\bibnamefont {Braun}},\ and\
  \bibinfo {author} {\bibfnamefont {D.}~\bibnamefont {Braun}},\ }\href
  {https://doi.org/10.1088/1367-2630/12/6/063005} {\bibfield  {journal}
  {\bibinfo  {journal} {New Journal of Physics}\ }\textbf {\bibinfo {volume}
  {12}},\ \bibinfo {pages} {063005} (\bibinfo {year} {2010})}\BibitemShut
  {NoStop}%
\bibitem [{\citenamefont {Aulbach}\ \emph {et~al.}(2010)\citenamefont
  {Aulbach}, \citenamefont {Markham},\ and\ \citenamefont
  {Murao}}]{Aulbach_2010}%
  \BibitemOpen
  \bibfield  {author} {\bibinfo {author} {\bibfnamefont {M.}~\bibnamefont
  {Aulbach}}, \bibinfo {author} {\bibfnamefont {D.}~\bibnamefont {Markham}},\
  and\ \bibinfo {author} {\bibfnamefont {M.}~\bibnamefont {Murao}},\ }\href
  {https://doi.org/10.1088/1367-2630/12/7/073025} {\bibfield  {journal}
  {\bibinfo  {journal} {New Journal of Physics}\ }\textbf {\bibinfo {volume}
  {12}},\ \bibinfo {pages} {073025} (\bibinfo {year} {2010})}\BibitemShut
  {NoStop}%
\bibitem [{\citenamefont {Baguette}\ \emph {et~al.}(2014)\citenamefont
  {Baguette}, \citenamefont {Bastin},\ and\ \citenamefont
  {Martin}}]{PhysRevA.90.032314}%
  \BibitemOpen
  \bibfield  {author} {\bibinfo {author} {\bibfnamefont {D.}~\bibnamefont
  {Baguette}}, \bibinfo {author} {\bibfnamefont {T.}~\bibnamefont {Bastin}},\
  and\ \bibinfo {author} {\bibfnamefont {J.}~\bibnamefont {Martin}},\ }\href
  {https://doi.org/10.1103/PhysRevA.90.032314} {\bibfield  {journal} {\bibinfo
  {journal} {Phys. Rev. A}\ }\textbf {\bibinfo {volume} {90}},\ \bibinfo
  {pages} {032314} (\bibinfo {year} {2014})}\BibitemShut {NoStop}%
\bibitem [{\citenamefont {H\"ubener}\ \emph {et~al.}(2009)\citenamefont
  {H\"ubener}, \citenamefont {Kleinmann}, \citenamefont {Wei}, \citenamefont
  {Gonz\'alez-Guill\'en},\ and\ \citenamefont {G\"uhne}}]{PhysRevA.80.032324}%
  \BibitemOpen
  \bibfield  {author} {\bibinfo {author} {\bibfnamefont {R.}~\bibnamefont
  {H\"ubener}}, \bibinfo {author} {\bibfnamefont {M.}~\bibnamefont
  {Kleinmann}}, \bibinfo {author} {\bibfnamefont {T.-C.}\ \bibnamefont {Wei}},
  \bibinfo {author} {\bibfnamefont {C.}~\bibnamefont {Gonz\'alez-Guill\'en}},\
  and\ \bibinfo {author} {\bibfnamefont {O.}~\bibnamefont {G\"uhne}},\ }\href
  {https://doi.org/10.1103/PhysRevA.80.032324} {\bibfield  {journal} {\bibinfo
  {journal} {Phys. Rev. A}\ }\textbf {\bibinfo {volume} {80}},\ \bibinfo
  {pages} {032324} (\bibinfo {year} {2009})}\BibitemShut {NoStop}%
\end{thebibliography}
%

\appendix*
\section{Majorana Constellations and Quantumness of Polarization States}
A quantum state describing a particle with angular momentum or spin $S$ is in a ($2S+1$)-dimensional Hilbert space $\mathcal{H}_S$. For spin $S$, we label $2S+1$ levels as $m=-S,...,S$. $\mathcal{H}_S$ is spanned by the set of basis ${\ket{S,m}}$, which is usually called the angular momentum basis or Dicke basis. One should note that for spin $S$ the space of states is $\mathcal{H}_S\in \mathbb{C}^{2S+1}$ but for angular momentum the space of bound states is the infinite sum of such spaces for every possible value that $S$ can take, $\mathcal{H}=\bigoplus_S \mathcal{H}_S$~\cite{Bengtsson_Zyczkowski_2006}. Additionally, $\mathcal{H}_S$ is the carrier of the irreducible representation (irrep) of spin $S$ of $\mathfrak{su}(2)$, and therefore any two states in $\mathcal{H}_S$ which only differ by an $\mathfrak{su}(2)$ rotation are physically equivalent. An arbitrary pure state describing such a system can be written as
\begin{equation}\label{spin_state}
  \ket{\Psi} = \sum_{m=-S}^{S} c_S \ket{S,m}.
\end{equation}
Using a method called the Majorana spin representation, one can construct an arbitrary pure state with angular momentum $S$ as a superposition of $2S$ spin-1/2 systems. Note that although the Hilbert space of $S$ spin-1/2 systems $\mathcal{H}\in \mathbb{C}^{2^S}$ is of dimension $2^S$, its symmetric subspace, is of dimension $2S+1$~\cite{Bengtsson_Zyczkowski_2006,PhysRevLett.114.080401}. Majorana showed that there exists a bijection between the pure spin state given in~\cref{spin_state} and roots of the polynomial $M(z)$, which is called the Majorana polynomial~\cite{Majorana1932,Bengtsson_Zyczkowski_2006}.
\begin{equation}\label{majorana_pol}
  M(z) = \sum_{k}^{2S} c_k \binom{2S}{k}^{1/2} z^{2S-k}.
\end{equation}
The roots $\xi_k$ of $M(z)$ in turn, have bijective correspondence to the points on $\mathcal{S}^2$. Therefore, one can uniquely represent the angular momentum state $\ket{\psi}$ in~\cref{spin_state} on the surface of the extended complex plane (Riemann sphere), up to the action of $\mathfrak{su}(2)$ rotations~\cite{Bengtsson_Zyczkowski_2006}. One can accomplish this task by the stereographic projection
\begin{equation}\label{str_proj}
  \text{tan}^{-1}(\frac{\theta_k}{2})e^{i\phi_k} = \xi_k,
\end{equation}
in which $\theta_k$ and $\phi_k$ are the polar and azimuthal coordinates for a point on the sphere. Within the symmetric subspace of the Hilbert space that we mentioned earlier, using Majorana representation one can decompose the angular momentum state in~\cref{spin_state} as a single product of individual spin-1/2 systems.
\begin{equation}\label{majorana_decom}
  \ket{\Psi} = \bigotimes_{k=1}^{2S}\ket{\psi_k},\quad \ket{\psi_k} = \text{cos}\frac{\theta_k}{2}\ket{0} + e^{i\phi_k}\text{sin}\frac{\theta_k}{2}\ket{1}.
\end{equation}
Schwinger showed that commutation relations of an angular momentum operator can be reduced to those of a two-dimensional harmonic oscillator and using a transformation one can map quantum mechanical spins onto bosonic fields~\cite{https://doi.org/10.1111/j.2164-0947.1977.tb02957.x}. This method is called Schwinger boson representation, which is a special case of Jordan-Schwinger map. Upon application of this method the Schwinger boson representation of the angular momentum operators takes the form given in~\cref{stokesops} for the Stokes operators. In light of this transformation, we can express the quantum polarization states expressed using Fock basis in~\cref{polarizationstate} using Dicke (angular momentum) basis. The optical state can be thought of as a collection of spin-1/2 states with $n_1 = S+m$ particles in spin-up and $n_2=S-m$ particles in spin-down states. It should be noted that, by construction, unlike the usual angular momentum addition rules, combining the mentioned spin up and down states will result in a symmetric state with $n_1+n_2=2S$ spin-1/2 particles with total spin $S$. The polarization state in Dicke basis is given by   
\begin{equation}\label{dicke_polarizationstate}
  \ket{n_1,n_2}\equiv \ket{S,m} =\frac{\hat{a}^{\dagger}\vphantom{a}{}^{(S+m)}\hat{b}^{\dagger}\vphantom{a}{}^{(S-m)}}{\sqrt{(S+m)!(S-m)!}}\ket{0,0}.
\end{equation}
Due to the isomorphism that is established by Schwinger boson representation between spin systems and two-mode bosonic systems, one can use Majorana's method to represent any pure polarization state with $n=2S$ photons as a collection of points on the surface of a Riemann sphere (Poincare sphere within the context of polarization optics). 
\begin{equation}\label{majorana_optical}
\begin{aligned}
\ket{\Psi} = \frac{1}{\mathcal{N}_n(U)}\prod_{k=1}^{n} \hat{a}^{\dagger}_{u_k} \ket{0,0},\\
\hat{a}^{\dagger}_{u_k} = \text{cos}(\frac{\theta_k}{2})\hat{a}^{\dagger} + e^{i\phi_k} \text{sin}(\frac{\theta_k}{2})\hat{b}^{\dagger}.
\end{aligned}
\end{equation}
In~\cref{majorana_optical}, $\mathcal{N}_n(U)$ is the normalization factor which is a function of $U\equiv \{ u_1, ..., u_{2S} \}$ and $u_k =(\theta_k,\phi_k)$ are the coordinates of the points on the Poincare sphere, which are called the constellations of the state. The normalization factor is given by~\cite{PhysRevA.94.022123,Lee_1988}
\begin{equation}\label{normalization_majorana}
\begin{aligned}
\mathcal{N}_n(U) = \Biggl[ \frac{(n+1)!}{2^n}\sum_{k=0}^{[n/2]}\frac{D_k^n}{(2k+1)!!} \Biggr]^{\frac{1}{2}},\\
D_k^n \equiv \sum_{i_1=1}^{n}\sum_{j_1>i_1}^{n}...\sum_{i_k>i_{k-1}}^{n*}\sum_{j_k>i_k}^{n*}(u_{i_1}.u_{j_1})...(u_{i_k}.u_{j_k}).
\end{aligned}
\end{equation}
In~\cref{normalization_majorana}, in the first line, $[\bullet]$ at the upper limit of the sum signifies the floor function and in the second line, $*$ indicates a restriction on the summations such that the values that the indices take in each term must all be different.

The methodology described so far, explains how to represent a general pure polarization state as a constellation of points on the surface of a sphere and its connection to the original method by Majorana developed for quantum angular momentum states. However, it is constructive to briefly describe another approach based on coherent states to make connections between the constellations, quasi-probability distribution and different notions of quantumness of polarization states. It is convenient to express the Bloch coherent (also known as spin coherent, atomic coherent and $\mathfrak{su}(2)$ coherent) state using Dicke basis.
\begin{equation}\label{bloch_coherent}
  \ket{z} = \hat{D}(\theta,\phi)\ket{S,-S}=\frac{1}{(1+|z|^2)^S}\sum_{-S}^{S}\binom{2S}{S+m}^{\frac{1}{2}}z^{S+m}\ket{S,m}.
\end{equation}
In~\cref{bloch_coherent}, $\hat{D}$ is the displacement operator on $\mathcal{S}^2$. Any pure state in Dicke basis can be written as~\cref{spin_state}. The stellar function (also known as Fock-Bargmann representation) of the state for $\mathcal{H}_S$ is defined using the overlap between the state~\cref{spin_state} and the Bloch coherent state~\cite{https://doi.org/10.1002/cpa.3160140303,gazeau2009coherent}.
\begin{equation}\label{bargmann}
  F(z) = (1+|z|^2)^S \braket{\Psi^*|z} = \sum_{m=-S}^{S} c_S \binom{2S}{S+m}^{\frac{1}{2}} z^{S+m}.
\end{equation}
In~\cref{bargmann} we used the following notation.
\begin{equation}\label{bargmann_notation}
  \bra{\Psi} = \sum_{m=-S}^{S} c_S \bra{S,m};\quad \ket{\Psi^*} = \sum_{m=-S}^{S} c_S^* \ket{S,m};\quad \bra{\Psi^*} = \sum_{m=-S}^{S} c_S \ket{S,m}.
\end{equation}
It is worth mentioning that the stellar function can also be defined in $\mathcal{H}_{\infty}$ (continuous-variable quantum systems) using the canonical coherent states~\cite{gazeau2009coherent,10.1116/5.0025819}. $F(z)$ is a holomorphic function which provides an analytic representation of a quantum state. The transformation in~\cref{bargmann} defines the Bargmann-Segal transform on the spin coherent states which maps a quantum state to the space of holomorphic functions, called the Segal–Bargmann space~\cite{https://doi.org/10.1002/cpa.3160140303,10.1116/5.0025819,HALL1994103,gazeau2009coherent}. The Bargmann representation is directly related to the Husimi Q-function, which is a quasi-probability distribution on the phase space. For the case of a pure state $\hat{\rho}=\ketbra{\Psi}{\Psi}$ and this relationship becomes 
\begin{equation}\label{husimi_bargmann}
  Q(z) = \braopket{z}{\hat{\rho}}{z} = (1+|z|^2)^S |F(z^*)|^2.
\end{equation}
Husimi function (unlike Wigner function) is non-negative definite and is bounded by $0\leq Q(z) \leq 1/\pi$. It is evident that the zeros of the Husimi Q-function are the complex conjugates of the zeros of the stellar function. The Q function also has a geometric interpretation on the manifold of quantum states. The Fubini-Study distance between states $\ket{\Psi}$ and $\ket{z}$ is related to the Q function by $\mathbb{D}_{FS}=\text{arccos}\sqrt{Q(z)}$. Using Husimi Q-function one can define Wehrl entropy on the sphere ($\mathcal{H}_S$) as
\begin{equation}\label{wehrl_entropy}
  S_W(\hat{\rho}) = -\frac{2S+1}{4\pi}\int_{\mathcal{S}^2} dz Q(z)\text{ln}Q(z).
\end{equation}
Unlike Von-Neumann entropy, Wehrl entropy doesn't vanish for pure states and is always positive. It can be interpreted as an information measure for a joint noisy measurement of position and momentum of the quantum system. Lieb and Solovej proved that Wehrl entropy on $\mathcal{H}_S$ is lower bounded by
\begin{equation}\label{wehrl_bound}
  S_W(\hat{\rho}) \geq \frac{2S}{2S+1}, 
\end{equation}
and the lower bound is attained only for the Bloch coherent states for which the Majorana constellations are concentrated at a single point on the Poincare sphere~\cite{10.1007/s11511-014-0113-6,10.1116/5.0025819}. The more spread out the points are on the sphere, the larger the value of the Wehrl entropy becomes. It is also proven that canonical coherent states minimize Wehrl entropy on the space of states defined on $\mathcal{H}_\infty$~\cite{Lieb1978}. Since the coherent states are known to be the most classical-like states, one can interpret the Wehrl entropy as a measure of non-classicality such that the larger the value for $S_W(\hat{\rho})$ is, more ``distant" the state becomes from the Bloch coherent states. In this sense, the states that maximize the Wehrl entropy, for which the Majorana constellations are distributed such that the points are as far away as possible from each other on the unit sphere, are the most quantum ones. For these states the Majorana points are distributed symmetrically on the Poincare sphere. To obtain these maximally non-classical states, for each $S$, one must solve an optimization problem. However, the problem of optimizing the value of $S_W(\hat{\rho})$ based on the positions of the Majorana points on $\mathcal{S}^2$ is extremely challenging, especially for larger values of $S$~\cite{Lee_1988,Schupp1999}. These states are obtained for a limited set of values of $S$, not by full-scale optimization, but by selecting a set of candidates for the optimal states based on the geometry of the constellations and checking if the selected states result in a local maxima for $S_W(\hat{\rho})$~\cite{Baecklund_2014,Baecklund_thesis}. We must make it clear that maximization of Wehrl entropy is not the only criterion for finding the ``maximally quantum" states in the literature. Depending on the problem under investigation, there are various procedures to find such states. Application of different extremum principles result in a different set of maximally quantum states for discrete variable systems, as opposed to the continuous variable case~\cite{10.1116/5.0025819}. The important theme underlying the methodology for finding such states is that all of these methods are based on identifying an suitable functional of quantum states and finding the maximally quantum states by optimizing that functional. For a more detailed explanation of such procedures the readers can refer to~\cite{10.1116/5.0025819,Giraud_2010,Aulbach_2010,PhysRevA.90.032314,PhysRevA.80.032324}.\\

One can verify that in the case of mentioned maximally quantum states, due to the fact that the constellations are symmetrically distributed on $\mathcal{S}^2$, the polarization measure defined in~\cref{dop} vanishes. Although these states are classically unpolarized, they contain hidden polarization which can be understood as a manifestation of the higher moments of the Stokes operators~\cite{Goldberg:21,PhysRevA.66.013806,PhysRevA.80.032324,Sanchez-Soto_2013}. One can expand a general polarization state in terms of irreducible tensors as
\begin{equation}\label{rho_tensor}
  \hat{\rho} = \sum_{K=0}^{2S}\sum_{q=-K}^{K}\rho_{Kq}\hat{T}_{Kq}^{(S)}.
\end{equation}
In~\cref{rho_tensor}, $\hat{T}_{Kq}^{(S)}$ are the irreducible tensors defined as~\cite{Goldberg:21,10.1116/5.0025819}
\begin{equation}\label{irr_tensor}
  \hat{T}_{Kq}^{(S)} = \sqrt{\frac{2K+1}{2S+1}}\sum_{m,m'=-S}^{S}C_{Sm,Kq}^{Sm'}\ketbra{S,m'}{S,m},
\end{equation}
in which $C_{Sm,Kq}^{Sm'}$ are the Clebsch–Gordan coefficients that couple a spin $S$ and a spin $K$ ($0\leq K \leq 2S$) to a total spin $S$. The tensors form an orthonormal basis $\text{Tr}[\hat{T}_{Kq}^{(S)}(\hat{T}_{K'q'}^{(S')})^{\dagger}]=\delta_{SS'}\delta_{KK'}\delta_{qq'}$ and are covariant under $\mathfrak{su}(2)$ transformations. The expansion coefficients $\rho_{Kq}=\text{Tr}(\hat{\rho}\hat{T}_{Kq}^{\dagger})$ are known as the state multipoles. Complete characterization of the state requires knowledge of all of these multipoles. One can use these multipoles to expand the Q function as~\cite{Goldberg:21,10.1116/5.0025819}
\begin{equation}\label{husimi_decomp}
  Q(z) = \sum_{K=0}^{2S}Q^{(K)},
\end{equation}
where $Q^{(K)}$ is the $K^{\text{th}}$ multipole of the Q function and contains information of the $K^{\text{th}}$ moment of the Stokes operators. It can be expressed as~\cite{Goldberg:21,10.1116/5.0025819}
\begin{equation}\label{husimi_polar}
\begin{aligned}
  Q^{(K)} = \sqrt{\frac{4\pi}{2S+1}}\sum_{q=-K}^{K}C_{SS,K0}^{SS}\rho_{Kq}Y^{*}_{Kq},\\
  C_{SS,K0}^{SS} = \frac{\sqrt{2S+1}(2S)!}{\sqrt{(2S-K)!(2S+1+K)!}},
\end{aligned}
\end{equation} 
in which $Y^{*}_{Kq}$ are the spherical harmonics and in the second line of~\cref{husimi_polar} the analytical form for the Clebsch–Gordan coefficient $C_{SS,K0}^{SS}$ is given. Since it is calculated using all of the state multipoles, the Husimi Q function (and therefore the Wehrl entropy), contains complete information of the state. One can utilize the information encoded in the Q function to define a distance based polarization measure which takes into account the contributions from all of the multipoles ~\cite{PhysRevA.66.013806}
\begin{equation}\label{q_dop}
  \mathbb{P}_{q} = 1 - \frac{1}{4\pi}\Sigma,\quad \Sigma = \frac{1}{\int d\Omega [Q(\theta,\phi)]^2} .
\end{equation}
In~\cref{q_dop}, the DOP of a state is defined as the distance between the Q function of that state and the Q function for the maximally unpolarized state ($Q_{\text{unpol}}=1/4\pi$). The DOP defined in~\cref{q_dop} is more suitable for quantum states of light compared to~\cref{dop} because, as explained previously, the higher order multipoles play a more significant role in the polarization behavior of the quantum states. However, in most of the cases, one can gain substantial amount of information on polarization characteristics of the state only considering a finite number of its multipoles. A useful quantity which encodes the polarization information of a quantum state up to its $M^{\text{th}}$ order is the $M^{\text{th}}$ order cumulative multipolar distribution defined below~\cite{Goldberg:21,Sanchez-Soto_2013,10.1116/5.0025819}.
\begin{equation}\label{cumulative_multipole}
  \mathcal{A}_M = \sum_{K=1}^{M}\sum_{q=-K}^{K}|\rho_{Kq}|^2.
\end{equation}
$\mathfrak{su}(2)$ coherent states maximize $\mathcal{A}_M$ for all orders $M$. The states that minimize $\mathcal{A}_M$ are called Kings of Quantumness (anticoherent states). A state is called $M$-anticoherent if for all $t\leq M$ and unit vectors $\mathbf{n}$ the moments of the Stokes vector are isotropic $\langle(\mathbf{\hat{S}.n})^t\rangle=c_t$ and independent of $\mathbf{n}$. It is shown that the states with the anticoherence order of 2 and higher are optimal for estimating all three rotation parameters~\cite{PhysRevA.95.052125,Bouchard:17,PhysRevA.78.052333,PhysRevA.98.032113,Martin2020optimaldetectionof,ZGoldberg_2021,PhysRevApplied.20.024052}.

\end{document}